\documentclass[fleqn,usenatbib]{mnras}

\usepackage{newtxtext,newtxmath}

\usepackage[T1]{fontenc}

\DeclareRobustCommand{\VAN}[3]{#2}
\let\VANthebibliography\thebibliography
\def\thebibliography{\DeclareRobustCommand{\VAN}[3]{##3}\VANthebibliography}

\usepackage{graphicx}	
\usepackage{amsmath}	
\usepackage{siunitx} 
\usepackage{chemformula} 
\usepackage{booktabs} 
\usepackage[normalem]{ulem} 
\usepackage{pdflscape} 
\usepackage{xcolor} 
\usepackage{colortbl} 

\title[Size and composition of comets]{A link between the size and composition of comets}

\author[J. E. Robinson et al.]{
James E. Robinson,$^{1}$\thanks{E-mail: james.robinson@ed.ac.uk}\footnotemark[2]
Uri Malamud,$^{2}\thanks{These authors contributed equally to this work. Robinson - observational data compilation ; Malamud - initiative and theory.}$
Cyrielle Opitom,$^{1}$
Hagai Perets,$^{2}$
and J\"{u}rgen Blum$^{3}$
\\
$^{1}$Institute for Astronomy, University of Edinburgh, Edinburgh EH9 3HJ, UK\\
$^{2}$Department of Physics, Technion - Israel Institute of Technology, Technion City, 3200003 Haifa, Israel\\
$^{3}$Institute for Geophysics and extraterrestrial Physics, Technische Universit\"{a}t Braunschweig, Mendelssohnstr. 3, D-38106, Braunschweig, Germany
}

\date{Accepted 2024 March 21. Received 2024 March 7; in original form 2023 October 17}

\pubyear{2024}

\begin{document}
\label{firstpage}
\pagerange{\pageref{firstpage}--\pageref{lastpage}}
\maketitle

\begin{abstract}
All cometary nuclei that formed in the early Solar System incorporated radionuclides and therefore were subject to internal radiogenic heating. Previous work predicts that if comets have a pebble-pile structure internal temperature build-up is enhanced due to very low thermal conductivity, leading to internal differentiation. An internal thermal gradient causes widespread sublimation and migration of either ice condensates, or gases released from amorphous ice hosts during their crystallisation. Overall, the models predict that the degree of differentiation and re-distribution of volatile species to a shallower near-surface layer depends primarily on nucleus size. Hence, we hypothesise that cometary activity should reveal a correlation between the abundance of volatile species and the size of the nucleus. To explore this hypothesis we have conducted a thorough literature search for measurements of the composition and size of cometary nuclei, compiling these into a unified database. 
We report a statistically significant correlation between the measured abundance of \ch{CO}/\ch{H2O} and the size of cometary nuclei.
We further recover the measured slope of abundance as a function of size, using a theoretical model based on our previous thermophysical models, invoking re-entrapment of outward migrating high volatility gases in the near-surface pristine amorphous ice layers. This model replicates the observed trend and supports the theory of internal differentiation of cometary nuclei by early radiogenic heating. We make our database available for future studies, and we advocate for collection of more measurements to allow more precise and statistically significant analyses to be conducted in the future.
\end{abstract}

\begin{keywords}
comets: general -- astronomical data bases: miscellaneous
\end{keywords}



\section{Introduction}\label{S:intro}
In a recent study of the long-term evolution of comets \citep{MalamudEtAl-2022}, it was found, through a combination of various empirical laboratory works, that pebble-made comets have an extremely low thermal conductivity. 
The result of which is that due to radiogenic heating, comets are able to attain higher temperatures during their evolution than those typically considered in past models. This implies that internal transport of various volatile species is ubiquitous in comets.

Comet nuclei consist of various volatile species. Water and probably also carbon dioxide are currently viewed as primary volatile species, that exist as amorphous solids \citep{rubinVolatilesH2OCO22023}, and can trap other high-volatility species during their incipient formation \citep{BarNunEtAl-1985,SimonEtAl-2023}. We cannot know for certain which high-volatility species exist inside the comet as pure ice condensates, and which ones were co-deposited within the amorphous ice hosts, or in what precise proportion. 
Regardless, as was envisaged in \cite{MalamudEtAl-2022}, due to the internal temperature gradient in the comet, volatiles must migrate towards the surface upon their direct sublimation or else upon the phase transitions of their amorphous ice hosts. While migrating through an intact matrix of amorphous ice, the volatiles may become sequentially entrapped in the amorphous ice again, even though local temperatures are otherwise still too high for their deposition as pure ices \citep{BarNunEtAl-1985,LauferEtAl-1987,BarNunEtAl-1987,BarNunEtAl-1988,CollinsEtAl-2003,KumiEtAl-2006,GalvezEtAl-2007,MateEtAl-2008,GalvezEtAl-2008,HerreroEtAl-2010,CarMackEtAl-2023}. Amorphous ice hosts can thus become highly enriched in entrapped gases, since they have a very large uptake (storage capacity) of high-volatility species, amounting to a few tens of \% of their own mass 
\citep[see][and references therein]{CarMackEtAl-2023}. Even if sequential deposition were to be ignored, hyper- and super-volatile species will still generally flow outwards and eventually freeze according to their respective deposition temperatures, forming an onion-like internal stratification \citep{DeSanctisEtAl-2001,ChoiEtAl-2002,Davidsson-2021}.

Not only might this process differentiate hyper- and super-volatile species from the bulk of the comet to a much narrower layer, closer to the surface, but the degree of differentiation could be strongly dependent on the comet's size. While \cite{MalamudEtAl-2022} have established a correlation between the degree of differentiation and several factors, such as the assumed composition, the time of formation, the pore-space permeability and the pebble size, the dependence on the nucleus size involves fewer uncertainties, as follows.

Internal temperatures are, generally, governed by the interplay between the rate at which internal heat is released, and the rate in which it can diffuse out, either through conduction, advection or radiative transport. The amount of radiogenic heat release depends on the radionuclide abundances which in turn depend on the comet's formation time (for short-lived radionuclides) and the assumed composition of refractories which sets the initial radiogenic abundances at $t=0$. Arguments were made in \cite{MalamudEtAl-2022} that for outer Solar System objects the abundances do not necessarily adhere to meteoritic levels, which are usually invoked in thermophysical models.

The effectiveness of advective flow and thus of differentiation, depends on the permeability of gas within the porous matrix in which it travels. This parameter is unconstrained in pebble media and has a potentially large range \citep{GundlachEtAl-2011a,GundlachEtAl-2020,SchweighartEtAl-2021,Guttler.2023}.
Radiative transport strongly depends on temperature as well as pebble size \citep{HuEtAl-2019,Bischoff.2021}. Finally, heat conduction out of the interior, which is the primary mode of heat transport, depends quadratically on the characteristic length scale of the object. Hence, all else being equal (radionuclide abundances, internal permeability, thermal conductivity, heat capacity etc.), there is no question that the size of the nucleus dictates how much heat the nucleus can retain. The larger the comet, the greater bulk migration of hyper- and super-volatiles we might expect, sweeping them outwards and depositing them in differentiated layers of either gas laden amorphous ice or as pure ice. The former is much more likely for hyper-volatiles, as discussed next.

The survival of pure ices of hyper-volatile materials is not impossible, however probably unlikely in most present-day observed comets. Since comets are expected to lose their hyper-volatile content either in the contemporary Kuiper Belt \citep{LisseEtAl-2021} or even in the primordial disc \citep{Davidsson-2021} relatively quickly, they must be emplaced onto distant Oort-cloud orbits early enough in the Solar System formation history in order to avoid this fate.
At least some fraction of the outer nucleus must be kept below the threshold temperature of incipient sublimation of such ices, unaffected by both external insolation and internal radiogenic heating. C/2016 R2 might be an example of a rare, hyper-volatile rich, yet water and dust poor comet, belonging to this category \citep{McKayEtAl-2019}. Even without early emplacement onto an Oort-cloud-like orbit, several super-volatile species as well as amorphous water and carbon dioxide ice would remain sufficiently cold and thus safe against insolation in the outer Solar System, as evident from both theory and observations \citep{PrialnikEtAl-1987,Jewitt-2009,LiEtAl-2020,ParhiPrialnik-2023}. If perturbed into the inner Solar System for the first time, more vigorous activity is expected, however erosion limits the penetration of a heat wave to the interior \citep{CapriaEtAl-2017}, thus shielding the inner nucleus. The bulk composition of the eroded surface reflects the early rather than contemporary orbital state.

Based on the aforementioned arguments, and in light of the elevated internal temperatures suggested by the pebble nucleus model of \cite{MalamudEtAl-2022}, the following predictions have instigated this study:\\

\noindent 1. Only extremely small nuclei cool effectively enough in order to prevent any internal differentiation, whereas increased nucleus size correlates with increased hyper- and super-volatile differentiation and concentration near the surface, triggered by internal radiogenic evolution. Thus, active comets will appear to have greater abundances of high-volatility species as a function of their size.\\

\noindent 2. More dynamically evolved short period active comets will, as a group, be more eroded and therefore expose deeper layers compared to long period comets, which might be reflected in their size-dependent volatile abundances.\\

\noindent 3. If prolonged activity of comets strictly requires gas laden amorphous ice, then small comets might outlive larger comets, because the latter might concentrate amorphous ice in a thinner outer layer. In turn, a testable prediction is that dynamically evolved active comets are smaller than their long period counterparts, because large inner Solar System comets have had more orbits over which to erode and become dormant.\\

In this paper we present the first evidence that comet observations partly support the above predictions, and in particular the size-composition dependence for the hyper-volatile CO. Future efforts are certainly needed in order to increase the quantity of currently available data and help verify our predictions. In what follows we first carefully describe the criteria for assembling our data set, consisting of comets for which both the size and the coma composition are reliably known (Section \ref{S:DataSet}). We then present the observational evidence in Section \ref{S:Findings}. A discussion of our findings is given in Section \ref{S:Discussion}. The paper is concluded in Section \ref{S:Conclusions}.

\section{The data set}\label{S:DataSet}
In this section, we describe the data we used for this study. We aimed to gather the largest amount of size and composition data as we could, given what was available in the literature at the time of writing.

Cometary activity allows us to measure the composition and abundance of gases in the coma that are being released from the nucleus via sublimation of volatile ices. This provides an opportunity to gain insight into the composition of the ices contained in cometary nuclei. Many radiative processes take place in cometary atmospheres, which can be observed across a range of wavelengths \citep{biverChemistryCometAtmospheres2022,bodewitsRadiativeProcessesDiagnostics2022}. Spectroscopic observations of comets have been used to measure the composition of their coma through detection of emission bands or lines from a variety of molecules. This complements much rarer data from direct mass spectroscopy measurements following flyby of a comet by a space mission. In this study, we used mass spectroscopy data only for comet 67P/C-G \citep{rubinElementalMolecularAbundances2019}. 
We have gathered measurements from a large number of sources covering a range of molecules, from relatively complex ones to small radicals.

The other key information for this study is the size of comet nuclei, which is difficult to measure. Indeed, comets far from the Sun are faint and challenging to observe because of their small sizes and low geometric albedo. As they move inwards, solar radiation increases and causes the formation of the coma surrounding and obscuring the nucleus. Comets are most often discovered/observed while active, as this is when they are brightest.

In this study we generally refer to two broad classes of comets: nearly isotropic comets (NIC), which possess a fairly uniform inclination distribution, long orbital periods and a Tisserand parameter $T<2$; and ecliptic comets (EC), also known as short period comets, which are sub-classified into Jupiter family comets (with a Tisserand parameter $2<T<3$ and periods below 20 years) and Chiron-type comets (with $T>3$ and periods in the range 20-200 years) \citep{Levison-1996}. 

\subsection{Comet Nuclei Sizes}\label{SS:DataSetSize}

Different methods are used to measure the size of a comet nucleus, from optical photometric observations of inactive comets, observations of thermal emission from the nucleus, or direct measurements by spacecrafts. 
These techniques have different levels of accuracy and usually rely on different assumptions. We briefly describe the techniques that were used for the objects in this study but refer the reader to \cite{lamySizesShapesAlbedos2004} and \cite{knightPhysicalSurfaceProperties2023} for a more complete description alongside the advantages/shortfalls of the different techniques. 

\subsubsection{Observing reflected light}
This is one of the most commonly used techniques to determine the size of cometary nuclei. When a comet is at large heliocentric distances ($\gtrsim 4$ au for short period comets, further away for other types of comets), the heating from the Sun is insufficient to efficiently drive sublimation of water ice and there is little to no coma obscuring the nucleus. The nucleus can then be observed directly, or its flux estimated once a small coma contribution is modelled and subtracted. The flux measured for the nucleus is then used to compute the nucleus size.
This technique presents some difficulties, as it relies on observing faint objects far from the Sun. With a limited spatial resolution, it can also be impossible to ascertain that the object is truly inactive vs having an unresolved coma. For most objects, when these parameters are unknown, assumptions also need to be made concerning the geometric albedo of the target and its phase curve properties, introducing uncertainties in the nucleus size obtained. 

\subsubsection{Observing thermal emission}
At longer wavelengths, the thermal emission of comet nuclei can be detected, which can be linked to the nucleus size. 
This generally requires observing at infrared wavelengths, often with a space telescope.
Similar to observations of reflected light at optical wavelengths, in cases where the object is active the coma contribution has to be modelled to retrieve the contribution of the nucleus.
Thermal modelling of the nucleus is then used to determine its size \cite[e.g.\ NEATM;][]{harrisThermalModelNearEarth1998}. 
In most cases, assumptions must be made on the rotation period of the object, its shape (often, the derived size is an effective radius, assuming a spherical nucleus), albedo, or thermal inertia.
Likewise, interferometric measurements can be made of the submillimetre continuum component of the thermal emission, from which nucleus size can also be determined through thermal modelling \citep[e.g.][]{altenhoffCoordinatedRadioContinuum1999,boissierEarthbasedDetectionMillimetric2011,boissierMillimetreContinuumObservations2013}.
At these wavelengths dust in the coma contributes much less to the total emission than in the IR and visible ranges, as such one expects such observations to be dominated by thermal emission from the nucleus.
However, the submillimetre flux from the nucleus is lower than the IR flux making these observations challenging except for bright and/or nearby comets.
Furthermore, in cases where thermal emission and visible photometry can be measured simultaneously it is possible to solve for the nucleus radius and albedo independently \citep{lamySizesShapesAlbedos2004}.

\subsubsection{Radar observations} 
Radar observations, where a burst of microwaves is sent towards the nucleus of a comet and the reflected echo measured, can accurately constrain the shape and size of comets and asteroids that pass very close to the Earth. However, this technique is limited by a relatively small number of comet nuclei with near-Earth orbits. 

\subsubsection{Space-based size measurements from flyby / rendezvous}
The most accurate determination of the shape and size of any small Solar System body is made by direct observation during a spacecraft flyby/rendezvous. Naturally only a handful of comets have been visited by a spacecraft. These accurate size measurements are invaluable, alongside detailed information on the shapes (e.g.\ bilobate) and terrain (topography) of comet nuclei.

\subsubsection{Other size-measuring techniques}
In this study, we have gathered comet size measurements determined using the techniques mentioned above. 
Other techniques have been used to measure nucleus sizes, but either the comets they targeted did not have composition information and were not useful for this study or they were judged to be less reliable. 
We investigated the size estimates inferred by \cite{jewittDestructionLongPeriodComets2022} using the water production rate and non-gravitational acceleration of long period comets.
The former model assumes the production rate is linked to the sublimating area and therefore nucleus size, and the latter model estimates the nucleus mass/size from the magnitude of non-gravitational accelerations on the comet orbit.
These methods are generally less accurate than those described above; they make use of simplified models of cometary activity and assume physical parameters such as active surface area and nucleus density.
Furthermore, photometric/thermal techniques generally provide an upper limit on nucleus size whereas the techniques of \cite{jewittDestructionLongPeriodComets2022} could either overestimate the size of hyperactive comets (when icy grains in the coma enhance activity) or underestimate the size depending on the accuracy of the model and choice of physical parameters.
At the time of writing this source provides the only available literature size estimates for a number of NICs (8 comets for which we also found compositional information) therefore we considered using these comets in our analysis. 
We found that our overall results were not significantly changed by the inclusion of sizes from \cite{jewittDestructionLongPeriodComets2022} therefore we elected not to include these sizes in the final results to avoid potential biases.
However, we welcome the efforts to broaden the dataset of comet sizes in this manner and perhaps future work can incorporate such size estimates into a more complete analysis.

\subsubsection{Selection criteria} 
\label{SS:size_selection_criteria}

In addition to searching the literature, we used the Small-Body Database Lookup tool\footnote{\url{https://ssd.jpl.nasa.gov/tools/sbdb_lookup.html}} to query all comets listed in the MPC comet list\footnote{\url{https://www.minorplanetcenter.net/iau/MPCORB/CometEls.txt}} (via \verb|astroquery|, accessed 23/06/2022) and checked the original references. 
Furthermore we searched for additional size measurements in the Properties of Comet Nuclei v2.0 PDS\footnote{\url{https://pds.nasa.gov/}} dataset \citep{barnesPropertiesCometNuclei2010}.
The studies used for our final selection of comet nucleus sizes are listed in Table \ref{tab:sizes}.

Many comets had multiple measurements from different sources and obtained with different techniques. Previous compilations of comet nuclei sizes have sometimes taken the average of multiple measurements and used their variation to estimate the uncertainty \citep[e.g.][for 46P and 96P]{combiSurveyWaterProduction2019}. In this work we chose to pick a single size for each comet, considering the reliability of each source and the methods used.
We do this as one would expect the measurement of nucleus size to be more frequently biased towards larger sizes due to additional signal from unresolved activity.
We applied the following guidelines to make our selection:
\begin{itemize}
    \item Use spacecraft rendezvous/flyby measurements (a direct measurement of size) where available.
    \item Choose the smallest measured radius as some studies only provided an upper limit on size.
    \item Prefer more modern sources which generally apply more well established techniques on mostly higher quality data.
    \item Select a measurement with a directly calculated uncertainty if available\footnote{When no uncertainty is provided we use the radius-uncertainty relation of the literature data to assign an uncertainty estimate (Figure \ref{fig:radius_uncertainty})}. For some of these objects the uncertainty on the radius is expressed as an upper and lower limit. In these cases, for simplicity, we have taken the mean of these values to obtain a single uncertainty. 
\end{itemize}

For further details about size selection for particular comet nuclei, please refer to Appendix \ref{Appendix:Radius_Sources}.

\subsubsection{Comet fragments} 

Comets have often been observed to split into multiple fragments due to tidal forces while passing too close to the Sun or a planet (e.g.\ the Kreutz sungrazers and the 1994 Shoemaker-Levy encounter with Jupiter) or for underdetermined reasons such as rotational spin-up, impacts or gas pressure \citep{Boehnhardt-2002}. As such in this study we must be wary of when the nucleus size was measured relative to the observation of its composition.
A famous example is comet 73P/Schwassmann-Wachmann 3, which fragmented in 1995 \citep[and possibly earlier as well - ][]{schullerCometSchwassmannWachmann1930}. 
This is one of the only cases where the composition of different fragments of a split comet could be measured separately. 
The compositions of fragments B and C have been measured by \cite{finkTaxonomicSurveyComet2009}, \cite{dellorussoEmergingTrendsComet2016} and \cite{lippiInvestigationOriginsComets2021} and were determined to be similar. 
However, in this work we decided to use only pre-fragmentation sizes as this is more representative of the initial size of the primordial comet nucleus.
We were able to find composition information of nuclei/fragments for a number of comets including 73P, 51P/Harrington, C/1996 B2 (Hyakutake) and C/2001 A2 (LINEAR), however we have excluded them from this study as they do not have reliable size determination prior to the splitting events.
Likewise for other known split comets \citep[listed in][]{boehnhardtSplitComets2004}, either no composition information was available or the size pre-splitting was unknown so they were not used in this study.

\subsection{Comet Compositions}\label{SS:DataSetComposition}

\subsubsection{Observed species}
Over the years, various techniques have been used to measure the compositions of comets by observing their atmosphere. \cite{biverChemistryCometAtmospheres2022} present an overview of the current stage of our knowledge about comet composition and how it is measured. The simplest measurements to obtain are made at optical wavelengths, using spectroscopy or narrow-band filters to measure the abundance of a set of radicals, such as \ch{CN}, \ch{C2}, \ch{C3}, \ch{OH}, \ch{NH} \citep[see for example][]{ahearnEnsemblePropertiesComets1995}. While databases containing comets observed at optical wavelengths are the largest available, the species they sample are what we call product or daughter species. They are not present as such in the nucleus ices but are instead produced in the coma by the photo-dissociation of larger molecules. Therefore they might not directly represent the composition of nucleus ices. 
Molecules produced directly by the sublimation of nucleus ices, called parent species, tend to emit at infrared and radio wavelengths (H$_2$O, CO, CO$_2$, HCN, NH$_3$, CH$_4$, ...). These molecules are harder to detect, and their abundances can typically only be measured for relatively bright comets. In this study, we have included both types of species, in order to gather the largest sample possible. We have also included abundances measured \textit{in situ} using mass spectroscopy in the coma of comet 67P/C-G by the ROSINA mass spectrometer onboard the Rosetta spacecraft \citep{rubinElementalMolecularAbundances2019}. 
We note that there are composition measurements of 67P made using other instruments on Rosetta \citep[e.g.][]{bockelee-morvanEvolutionCOCH2016,feldmanFUVSpectralSignatures2018,biverLongtermMonitoringOutgassing2019}, however we used the measurements of \cite{rubinElementalMolecularAbundances2019} as they report abundances for a large number of species together.

Comparing the abundance of species from different studies, sometimes derived from observations at different wavelengths and with different techniques, can be a challenge. Indeed, the size of the field of view and the model parameters used can have a significant effect on the abundances measured. 
For example, in hyperactive comets sublimation of icy grains in the coma can be a significant source of volatile gas compared to production from the nucleus alone \citep[e.g.\ 103P,][]{kelleyIceAggregatesComa2014}. This could in principle lead to changes in measured abundance depending on where in the coma the measurement was made.
However, the goal of this study is to find broad trends among comets rather than performing detailed comparison between a small number of comets. Given the limited number of targets for which we could find both composition and size measurements, we collected all the sources we could find for abundance measurements. We focused first on large-scale studies, and then complemented our database using works focused on individual comets. 

For observations of radicals at optical wavelengths, we considered mainly the following studies.
Among the largest available studies is the one by \cite{ahearnEnsemblePropertiesComets1995}. They published the results of a survey of 85 comets observed from the 1970s until 1992 using narrow-band photometry at optical wavelengths. They sample a range of ECs and NICs. This was later updated by \cite{schleicherEXTREMELYANOMALOUSMOLECULAR2008}. 
We queried that dataset from PDS \cite[Lowell Observatory Cometary Database - Production Rates,][]{osipLowellObservatoryCometary2003}. 
\cite{cochranThirtyYearsCometary2012} obtained abundances from optical spectroscopy of 130 comets from 1980 - 2008 while 
\cite{langland-shulaCometClassificationNew2011} observed 26 comets using the Kask double spectrograph at Lick observatory (primarily in the $300 - 600\ \si{nm}$ range).
Finally, \cite{finkTaxonomicSurveyComet2009} present abundances for 50 comets with significant enough detections to derive reliable production rates from observations made in the wavelength range $520 - 1040\ \si{nm}$ at the Catalina Site telescope \citep[]{Fink.1996}.
In this particular study, there are no exact uncertainties published, only a subjective quality grade. We thus converted the quality grade into an uncertainty using the suggested percentage errors.
Most of these studies contain measurements of the $\mathrm{Af\rho}$ parameter, a proxy for the dust production, in addition to the radical production rates. For completeness and as an estimate of the dust-to-gas ratio we have included $\mathrm{Af\rho}$ measurements in this study.

For observations of parent species, we focused mainly on the following studies. \cite{dellorussoEmergingTrendsComet2016} present high resolution IR spectroscopy of 30 comets observed between 1997 and 2013. 
We complemented this with data from \cite{lippiInvestigationOriginsComets2021}. When a target was available in both data sets, we used data from \cite{dellorussoEmergingTrendsComet2016} as it presents the largest dataset.
\cite{ootsuboAKARINEARINFRAREDSPECTROSCOPIC2012} present CO$_2$ production rates for a sample of 18 comets with the AKARI satellite, and \cite{reachSurveyCometaryCO22013} measured abundances of CO and CO$_2$ with the Spitzer space telescope for 23 comets.  
The production rates of \ch{CO} and \ch{CO2} were further complemented by an existing compilation by \cite{harringtonpintoSurveyCOCO2022}, which contains the results of \cite{ootsuboAKARINEARINFRAREDSPECTROSCOPIC2012} \& \cite{reachSurveyCometaryCO22013} alongside additional sources.
Table \ref{tab:composition data} summarises the number of comets included in the largest of these studies, as well as the species they measured. 

\begin{table*}
\begin{tabular}{lllll}
\hline
Source                                           & Method                & Wavelength & Number of Comets & Species                                                                                      \\ \hline
\cite{ahearnEnsemblePropertiesComets1995}        & Narrowband photometry & Visible    & 85               & \ch{CN}, \ch{C2}, \ch{C3}, \ch{NH}, $\mathrm{Af\rho}$, \ch{OH}                               \\
\cite{finkTaxonomicSurveyComet2009}              & Spectroscopy          & Visible    & 92 (50)          & \ch{C2}, \ch{NH2}, \ch{CN}, $\mathrm{Af\rho}$, \ch{H2O}                                      \\
\cite{langland-shulaCometClassificationNew2011}  & Spectroscopy          & Visible    & 26               & \ch{CN}, \ch{C2}, \ch{C3}, \ch{NH}, \ch{NH2}, $\mathrm{Af\rho}$                              \\
\cite{cochranThirtyYearsCometary2012}            & Spectroscopy          & Visible    & 130 (110)        & \ch{CN}, \ch{NH}, \ch{C3}, \ch{CH}, \ch{C2}, \ch{NH2}, \ch{OH}                               \\
\cite{ootsuboAKARINEARINFRAREDSPECTROSCOPIC2012} & Spectroscopy          & IR         & 18 (17)          & \ch{H2O}, \ch{CO2}, \ch{CO}                                                                  \\
\cite{reachSurveyCometaryCO22013}                & Spectroscopy          & IR         & 23 (20)          & \ch{OH}, \ch{CO2}, $\mathrm{Af\rho}$                                                         \\
\cite{dellorussoEmergingTrendsComet2016}         & Spectroscopy          & IR         & 30               & \ch{CH3OH}, \ch{HCN}, \ch{NH3}, \ch{H2CO}, \ch{C2H2}, \ch{C2H6}, \ch{CH4}, \ch{H2O}          \\
\cite{lippiInvestigationOriginsComets2021}       & Spectroscopy          & IR         & 20               & \ch{CH3OH}, \ch{HCN}, \ch{NH3}, \ch{H2CO}, \ch{C2H2}, \ch{C2H6}, \ch{CH4}, \ch{CO}, \ch{H2O} \\ \hline
\end{tabular}
\caption{Table of the main compositional surveys considered in this work.
The source is listed along with the observational method and wavelength.
The number of comets considered in each study is listed, where the number in brackets is the number for which some or all of the species were detected.
The species targeted by each survey are also listed, including $\mathrm{Af\rho}$ which is a proxy for dust production.
}
\label{tab:composition data}
\end{table*}

The rest of the data comes from publications focused on individual comets by 
\cite{biverSpectroscopicMonitoringComet1999,biverRadioObservationsComet2007,biverMolecularCompositionOutgassing2011,biverMolecularInvestigationsComets2011,biverAmmoniaOtherParent2012,biverMolecularCompositionComet2021,biverMolecularCompositionShortperiod2021,biverObservationsComet20202022,bockelee-morvanNewMoleculesFound2000,bockelee-morvanOutgassingCompositionComet2004,bockelee-morvanStudyDistantActivity2010,bockelee-morvanWaterHydrogenCyanide2022a,bodewitsCOLLISIONALEXCAVATIONASTEROID2011,bonevFirstCometObservations2021,dellorussoPostperihelionVolatileProduction2020,faggiQuantifyingEvolutionMolecular2019,moulaneMonitoringActivityComposition2018,opitomLongtermActivityOutburst2016,rothTaleTwoComets2018,rothProbingEvolutionaryHistory2020,rubinElementalMolecularAbundances2019}

\subsubsection{Selection criteria}
\label{SS:comp_selection_criteria}
To select the final abundance measurements presented in Section \ref{S:Findings}, we applied the following methodology: 
\begin{itemize}
\item We collected data from all sources and identified the available species. We selected only data sets that also had available production rates for \ch{OH}/\ch{H2O} or \ch{CN} (we considered abundances relative to water or CN when looking for correlation between composition and size).

\item For each data set, if multiple measurements were provided for a comet we calculated the mean heliocentric distance and date of the observations and the corresponding average for each production rate if required. We used a weighted average if and when the source provided the corresponding weights.

\item If multiple sources were available for a comet, we selected production rate ratios from the largest available dataset, to prioritise using larger homogeneous datasets.

\item For improved reliability, preference was given to sources who published production rates with an associated uncertainty.
\end{itemize}

		By selecting composition from a single source we attempt to avoid difficulties in combining abundance measurements from multiple observational techniques and/or observational circumstances.
Assessing the changes in compositional abundance as a function of methodology, viewing geometry and other circumstances such as outburst events for all comets analysed here is beyond the scope of this work.
We note that determining the true bulk abundance of species in a cometary nucleus from remote observations alone will always be a difficult problem, and we attempted to minimise these effects by selecting from larger homogeneous datasets where possible.

As mentioned above, we focused mostly on abundances of other species relative to water. However, observations at optical (and sometimes radio) wavelengths only provide measurements of the \ch{OH} production rate. Since OH is produced by the photo-dissociation of water, it is possible to convert between \ch{H2O} and \ch{OH} production rates. Several ways to do the conversion have been presented in the literature and we decided to use the conversion ratio of $\ch{OH} = 0.85 \times \ch{H2O}$ based on the photo-dissociation rate of water into OH and H \citep{harrisProductionOutflowVelocity2002}. As \ch{H2O} and \ch{OH} are hard to observe at optical wavelengths, abundance is sometimes reported relative to CN. We thus included these observations as well for completeness.

Most studies present composition measurements for a comet in a distinct time window, e.g.\ around its perihelion passage or the date range over which it was observable/observing time was available. If not provided in the original study, we determined the mean heliocentric distance and date of the observation for each comet from each source.
In most cases this accurately captures the mean epoch at which the comet composition was measured, however in certain long running observing campaigns the mean might not reflect the true range of observing conditions \citep[e.g.][]{ahearnEnsemblePropertiesComets1995}.

Whenever a source provided several measurements for the same target, and if a final summary table was not provided by the authors, we took a mean of the compositions for that source to get a single mixing ratio for each object (uncertainties were propagated forward when available), unless a large gap in heliocentric distance/time was present between the observations. In that case, the measurements closest to perihelion were selected. 
Some sources provide an upper and lower uncertainty estimate on composition/mixing ratio. For simplicity, with these measurements we took the mean value of the upper and lower limits to be the uncertainty.

As a sanity check we considered the range of heliocentric distances used to calculate each mean composition.
For the comet compositions where this range is large ($>1~ \si{au}$) we inspected the individual measurements to ensure they were consistent.
10P and 103P displayed significant variation in their \ch{CO2}/\ch{H2O} abundance as reported by \cite{reachSurveyCometaryCO22013}, therefore for these two comets we excluded the production rates measured at $>2 ~ \si{au}$ where ices more volatile than \ch{H2O} begin to dominate activity.
Likewise, there was large variation in the measurements of \ch{CO2}/\ch{H2O} for C/1995 O1 compiled by \cite{harringtonpintoSurveyCOCO2022}.
The reported observations spanned a wide range of heliocentric distances, with some taken while the comet was $>3\ \si{au}$.
As such, we selected the measurement with the lowest heliocentric distance ($r_h = 2.93\ \si{au}$) for our analysis.
Furthermore, we note here that \cite{cochranThirtyYearsCometary2012} provided average production rates with respect to \ch{CN} scaled to $1\ \si{au}$; as such we set all measurements 
from this source to heliocentric distance $r_h = 1\ \si{au}$ when incorporating their results into our dataset.

We have taken the steps described above, applying a consistent methodology when selecting which source to use for a given composition measurement, in order to utilise the wide range of literature measurements in a reliable manner.
We note that choice of source will greatly affect the outcome of an analysis such as ours.
For example there is significant variation in the \ch{CO2}/\ch{H2O} production rates of comets such as 19P between \cite{ootsuboAKARINEARINFRAREDSPECTROSCOPIC2012} \& \cite{reachSurveyCometaryCO22013}.
We make available the full data table with all comet composition and size measurements (and their literature sources) so that in all cases the provenance of the data is clear. 
Furthermore we hope that this data collection is of value to future studies investigating the size and/or compositions of comets.
A sample of the dataset is displayed in Table \ref{Appendix:composition_data_table} and the full dataset is available \href{https://doi.org/10.7488/ds/7723}{at this link}.

\section{Observational findings}\label{S:Findings}

\subsection{Raw data analysis}
\begin{figure*}
\begin{tabular}{ccc}
    \includegraphics[width=0.3\textwidth]{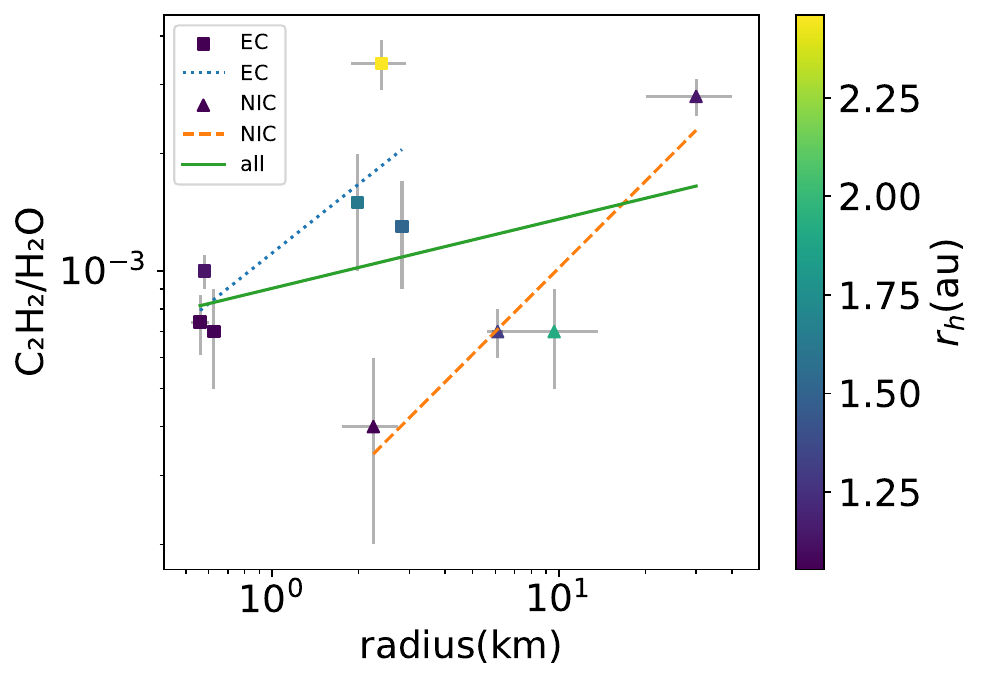} &
	\includegraphics[width=0.3\textwidth]{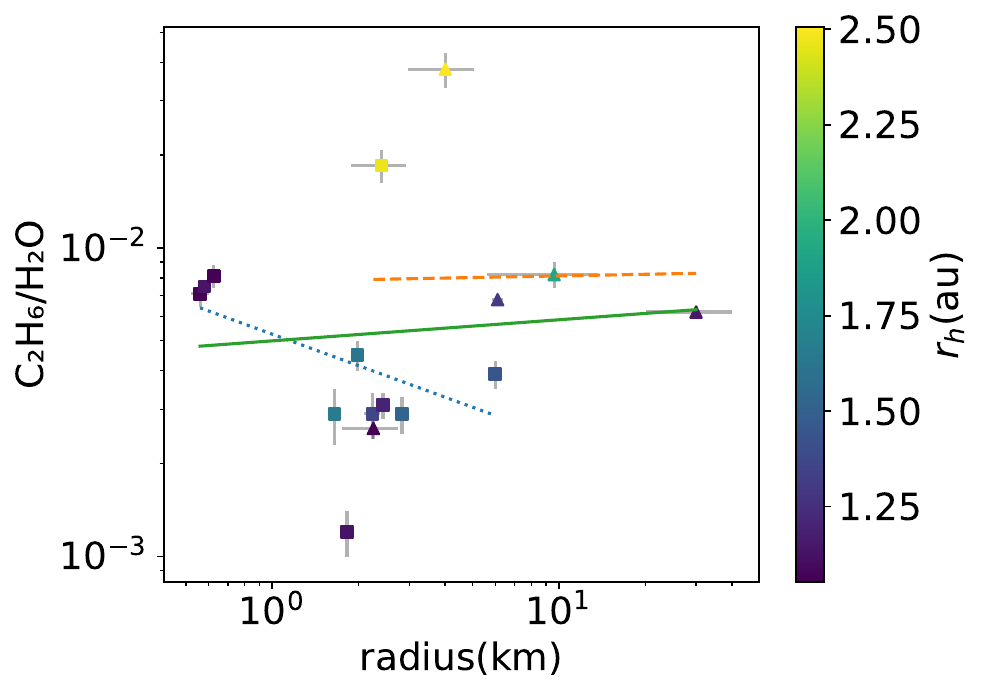} &
	\includegraphics[width=0.3\textwidth]{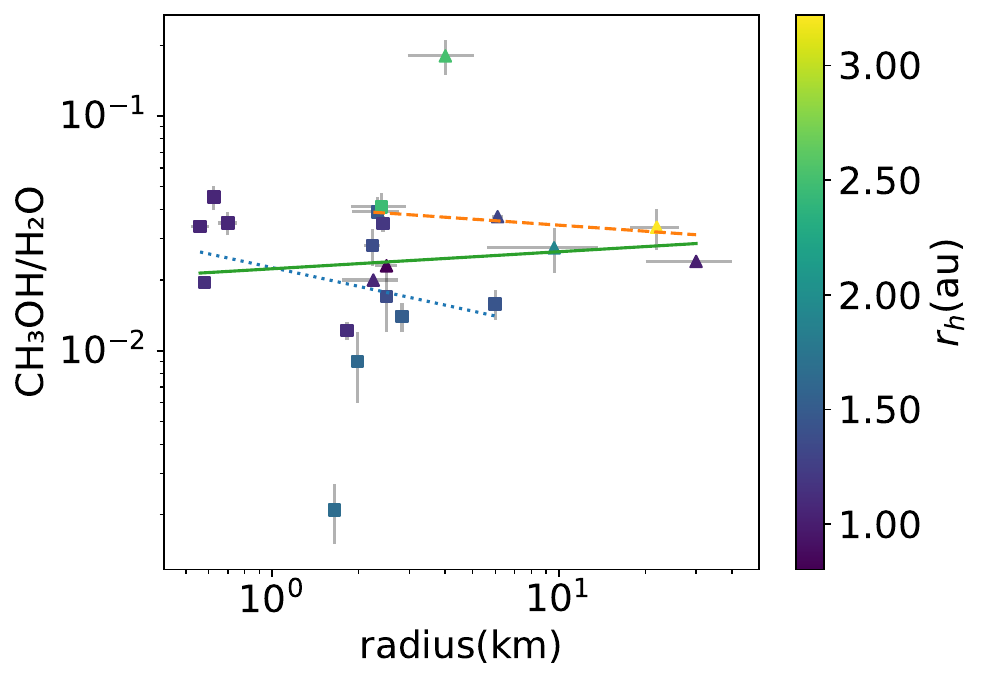} \\
	\includegraphics[width=0.3\textwidth]{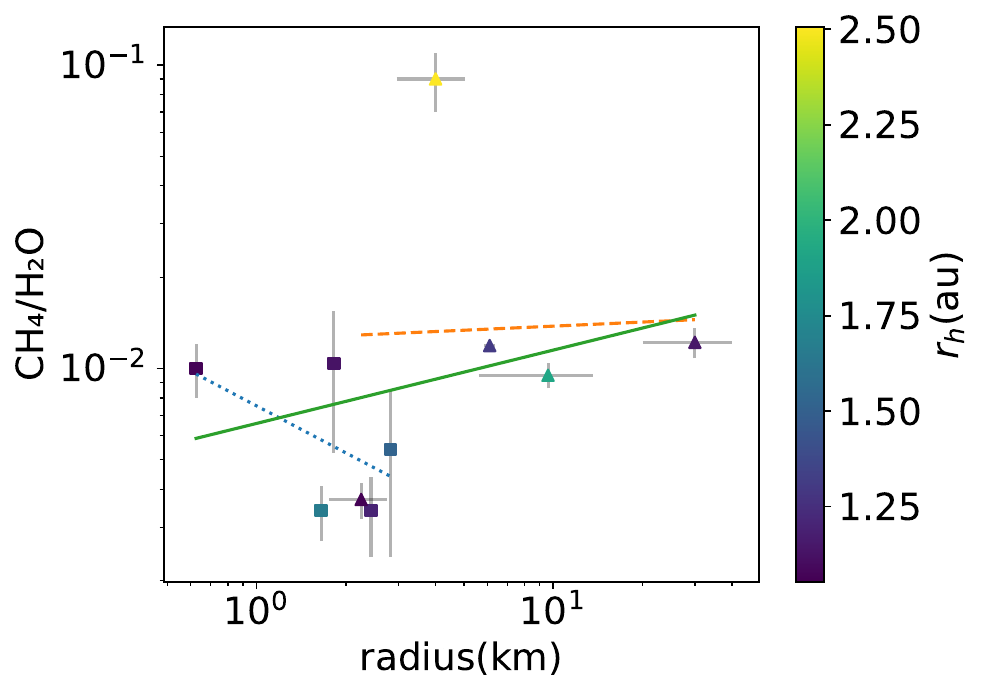} &   
	\includegraphics[width=0.3\textwidth]{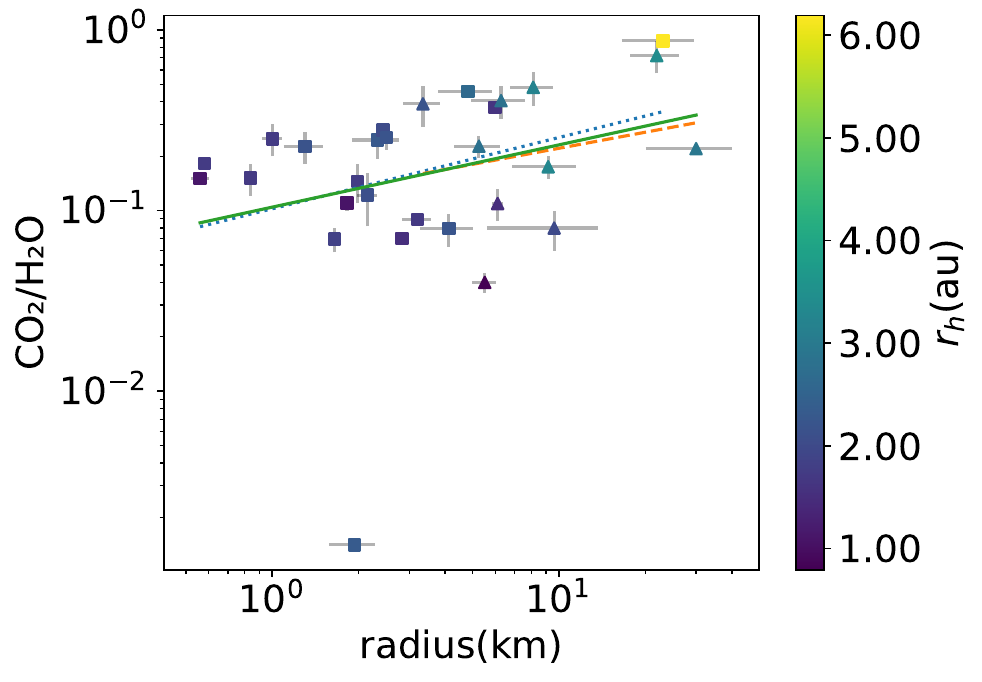} &
	\includegraphics[width=0.3\textwidth]{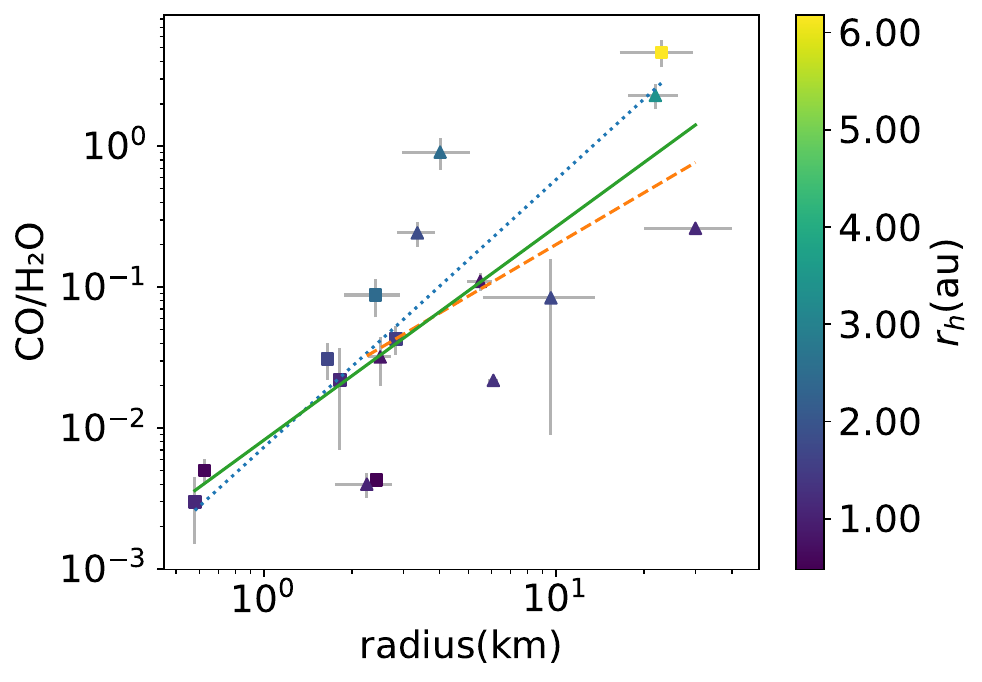} \\   
    \includegraphics[width=0.3\textwidth]{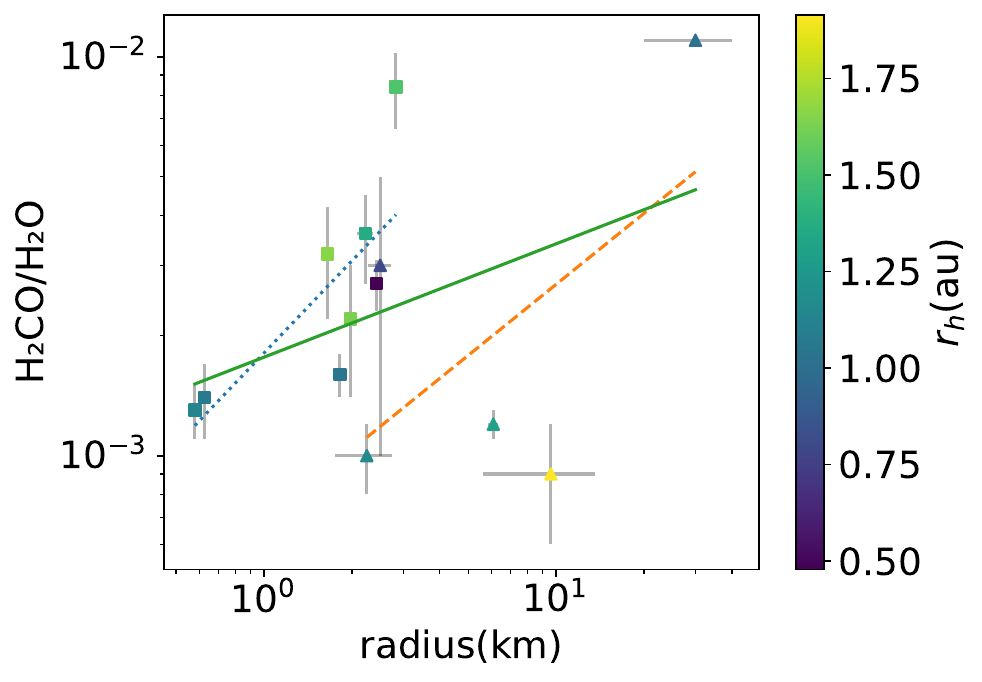} &
	\includegraphics[width=0.3\textwidth]{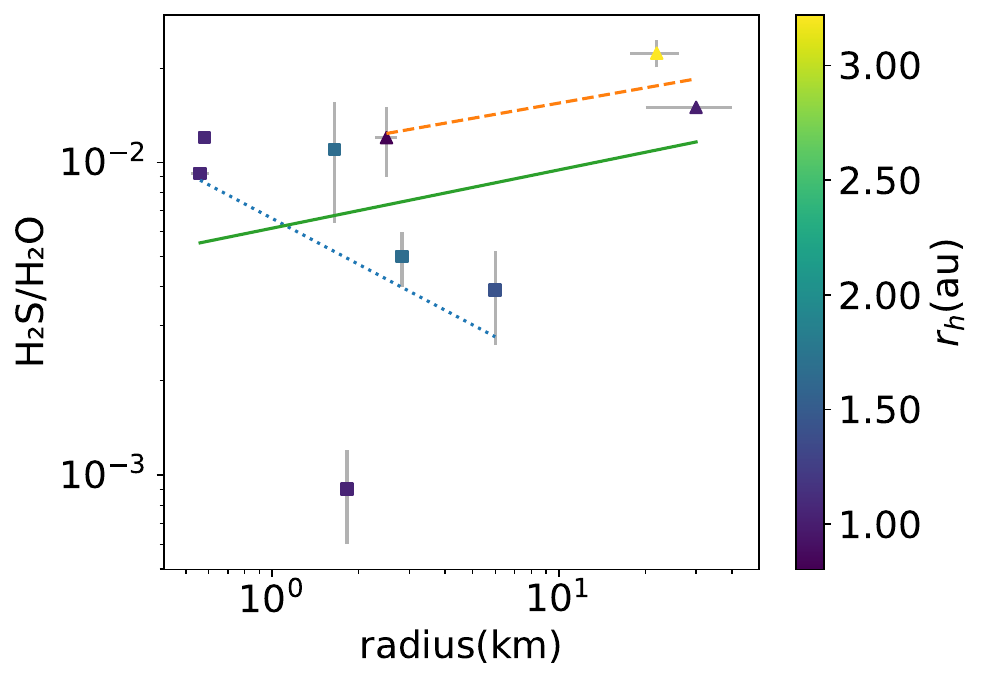} &
	\includegraphics[width=0.3\textwidth]{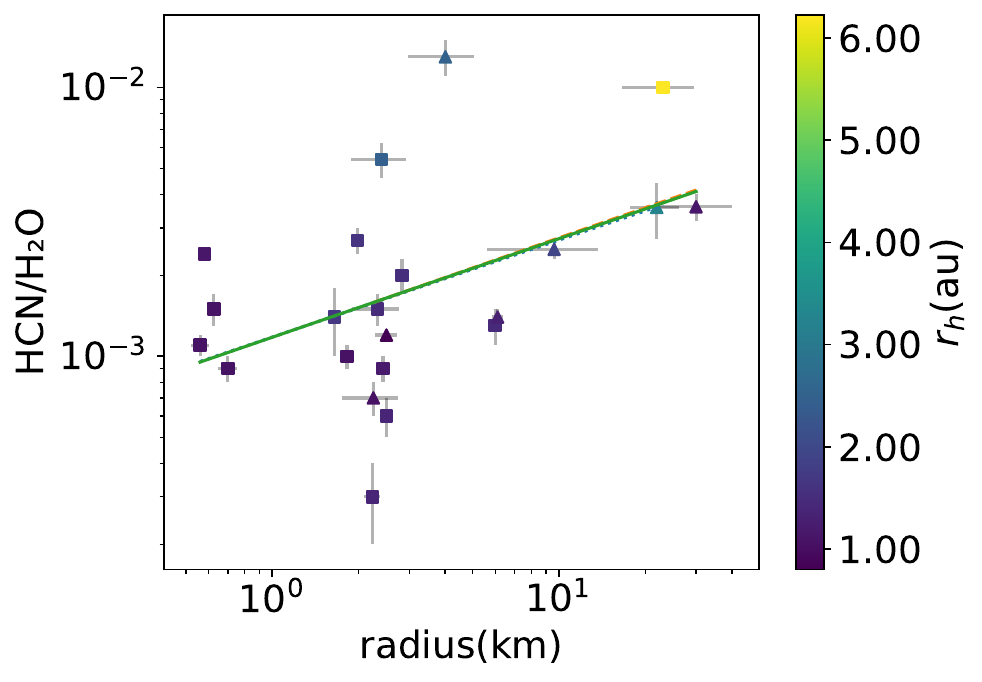} \\
	\includegraphics[width=0.3\textwidth]{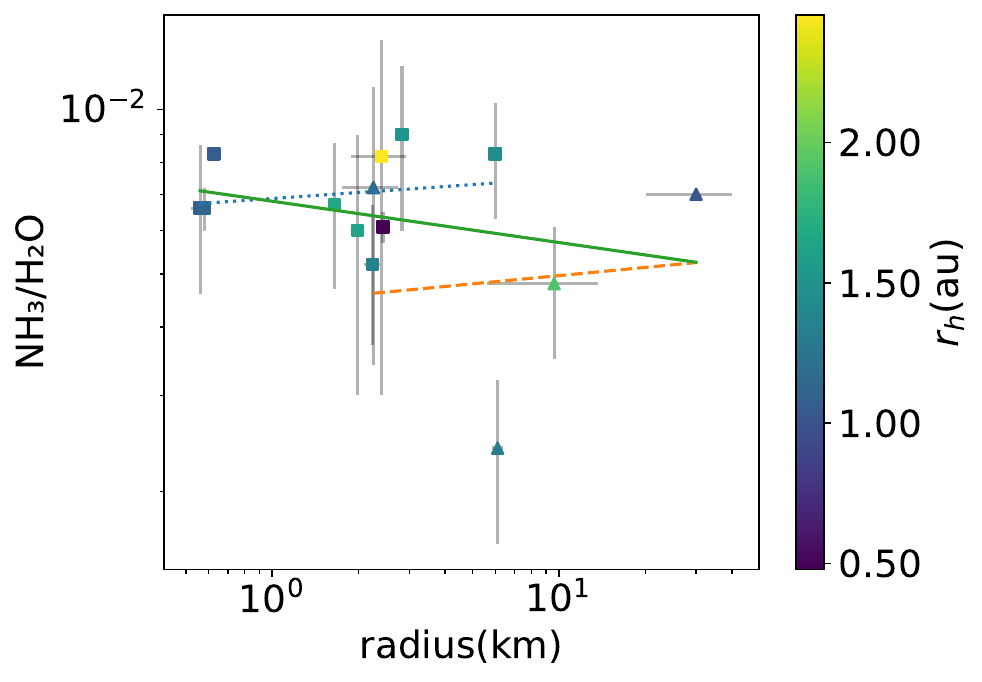} &
 	\includegraphics[width=0.3\textwidth]{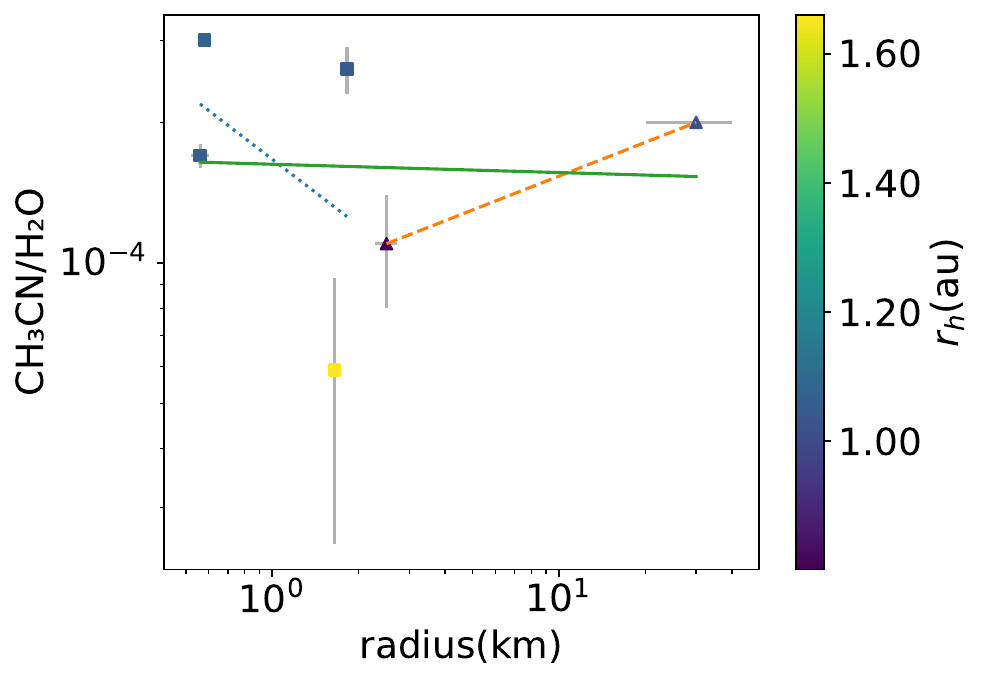} &
\end{tabular}
\caption{Log scale plots showing the relation between comet composition (of various parent species relative to \ch{H2O}) and radius of the nucleus.
Marker shape denotes the dynamical class of each comet, either an Ecliptic Comet (EC, square markers) or Nearly Isotropic Comet (NIC, triangular markers).
Marker colour indicates the heliocentric distance of the comet when the composition was measured.
The error bars denote the uncertainty in the measured composition or size (when this was available).
We indicate the correlation of the radius-composition data with a linear fit (in log-log space) for the whole dataset (solid line), ECs (dotted line) and NICs (dashed line).
The Pearson correlation coefficients for each parent species are given in Table \ref{tab:correlations1}.
}
\label{fig:comp_size_H2O_parents}
\end{figure*}

Figure \ref{fig:comp_size_H2O_parents} presents the abundance relative to water of a range of species commonly considered to be parent species as a function of the nucleus radius, for our entire data set. Each point represents a single comet. Different symbols are used for the ecliptic comets (ECs) and nearly isotropic comets (NICs), and the heliocentric distance of the comet when the abundance measurement was performed is indicated by the colour scale.
Figure \ref{fig:comp_size_H2O_daughters} shows the same information but for daughter species (and the proxy for dust production rate $\mathrm{Af\rho}$). 
Only species for which a significant number of measurements were available are shown. 
Additional figures displaying abundances relative to CN are presented in Figure \ref{fig:comp_size_CN}. 
For each species we assess the possible presence of a correlation between the relative production rate of that species and the comet nuclear size. 
In order to do so we calculated the Pearson correlation coefficient ($\gamma$) of these data (in log-log space) to measure the degree of linear correlation. 
When the data has strong linear correlation $\gamma$ has values approaching $\pm 1$ (signifying positive or negative correlation). 
In order to test the statistical significance of $\gamma$, a $p$-value is also calculated, which represents the probability of obtaining a result assuming that the null hypothesis (no correlation) is true. 
Therefore, when we measure large values of $\gamma$ with a corresponding small value of $p$ we can assume that the correlation is statistically significant, where typically a significance level of $p<0.05$ is the de facto threshold often used in literature.
Table \ref{tab:correlations1} presents the values of the Pearson correlation and $p$-values for all species for ECs and NICs separately as well as for the full sample. 
The exact $p$-value for a significant result can be a somewhat arbitrary choice, therefore in Table \ref{tab:correlations1} we have highlighted different ranges of $p$-value.
We select thresholds that are analogous to 3-, 2-, 1-sigma significances, i.e.\ $p \leq 0.003$ (strong significance), $0.003<p \leq 0.05$ (moderate significance), and $0.05<p \leq 0.32$ (marginal significance), respectively.

The strongest correlation by far is seen for CO at the 3-sigma level. CO is one of the most volatile ices in cometary nuclei. 
The trend is stronger for ECs than NICs.
One potential bias to keep in mind is that short period comets tend to have lower CO abundances due to repeated passages close to the Sun \citep{dellorussoEmergingTrendsComet2016}. 
The effect of the heliocentric distance on the CO abundance measurements in comets and how it could affect these results is discussed below in Section \ref{s:effect_of_heliocentric_distance}. 
For more detail on the compilation of \ch{CO}/\ch{H2O} abundances from the literature, please refer to Appendix \ref{Appendix:CO/H2O}.

We also see a trend at the 1-sigma level for HCN for ECs and no significant trend for the NICs, however, the correlation significance increases to 2-sigma when the whole sample is considered.
As noted by \cite{biverObservationsComet20202022}, HCN production rates derived from millimetre observations can differ from production rates derived from infrared observations by typically a factor two, which complicates the interpretation of the trend for HCN. We do not see any trend for CN but this is not entirely surprising. While CN was originally thought to be produced by the photo-dissociation of HCN, evidence indicates that another source is needed to account for the CN abundance and morphology in comets. This other source could be another parent species (C$_2$N$_2$, HC$_3$N, or CH$_3$CN), sublimation of dust grain, salts, or macro-molecules \citep{biverObservationsComet20202022}. The importance of this additional source could vary from comet to comet and explain the different trends seen for CN and HCN. 

We see a trend at the 2-sigma level for H$_2$CO for the  ECs and at 1-sigma for the sample as a whole.
For CO$_2$, we see a 1-sigma level trend for the ECs and the full sample. 
We do not see any correlation for any of the other typical parent species. However, this does not mean that the correlation is not present, but most likely that the current data available are insufficient to draw strong conclusions. This should improve in the future when full composition and size measurements become available for a larger number of comets. The only exception might be methanol for which the number of data points available are similar to CO, CO$_2$, and HCN, but no trend can be seen. 

With the possible exception of \ch{CS}, we generally do not see strong correlations for daughter species. However, we do note a moderately significant anti-correlation for NH which is driven by the ECs.
This correlation is surprising given the lack for correlation for ECs for NH$_2$ and NH$_3$. However, the scale-length used to compute NH production rates using the Haser model are difficult to constrain, which could influence the results. We thus disregard the NH trend for the rest of the discussion.

In Figure \ref{fig:comp_size_H2O_daughters}, we also see a 2-sigma correlation for the $\mathrm{Af\rho}$ for NICs, with higher dust to gas ratios for larger comets. An increase of the $\mathrm{Af\rho}$/OH ratio at large heliocentric distances has been noticed by \cite{ahearnEnsemblePropertiesComets1995} and \cite{langland-shulaCometClassificationNew2011}, which has been explained either by a selection effect (high dust to gas ratio comets have higher visual magnitudes), by the presence of large grains less volatile than water, or the build-up of a crust on the surface of the nucleus. Since larger comets tend to be more active, and thus brighter, they can be observed farther from the Sun. This is particularly true for NICs, which are more likely to be observed far from the Sun. The trend of higher dust-to-gas ratio at larger distances from the Sun could thus bias our correlation between $\mathrm{Af\rho}$/H$_2$O and the nucleus size.

\begin{table*}  
\begin{tabular}{l|rrr|rrr|rrr}
\toprule
 & \multicolumn{3}{l|}{Ecliptic Comets} & \multicolumn{3}{l|}{Nearly Isotropic Comets} & \multicolumn{3}{l|}{All Comets} \\
Species & Number & Correlation & $p$-value & Number & Correlation & $p$-value & Number & Correlation & $p$-value \\
\midrule
\ch{C2H2/H2O} & \cellcolor{lightgray}6 & \cellcolor{lightgray}0.7796 & \cellcolor{lightgray}0.0675 & \cellcolor{gray}\color{white}4 & \cellcolor{gray}\color{white}0.9537 & \cellcolor{gray}\color{white}0.0463 & 10 & 0.3421 & 0.3332 \\
\ch{C2H6/H2O} & \cellcolor{lightgray}11 & \cellcolor{lightgray}-0.3479 & \cellcolor{lightgray}0.2944 & 5 & 0.0177 & 0.9775 & 16 & 0.0862 & 0.7508 \\
\ch{CH3CN/H2O} & 4 & -0.4120 & 0.5880 & 2 & 1.0000 & 1.0000 & 6 & -0.0430 & 0.9356 \\
\ch{CH3OH/H2O} & 14 & -0.2295 & 0.4300 & 7 & -0.1155 & 0.8053 & 21 & 0.0933 & 0.6874 \\
\ch{CH4/H2O} & 5 & -0.5520 & 0.3347 & 5 & 0.0376 & 0.9521 & 10 & 0.2667 & 0.4564 \\
\ch{CO2/H2O} & \cellcolor{lightgray}19 & \cellcolor{lightgray}0.2628 & \cellcolor{lightgray}0.2770 & 10 & 0.2205 & 0.5405 & \cellcolor{lightgray}29 & \cellcolor{lightgray}0.3003 & \cellcolor{lightgray}0.1135 \\
\ch{CO/H2O} & \cellcolor{darkgray}\color{white}8 & \cellcolor{darkgray}\color{white}0.9143 & \cellcolor{darkgray}\color{white}0.0015 & \cellcolor{lightgray}9 & \cellcolor{lightgray}0.5740 & \cellcolor{lightgray}0.1060 & \cellcolor{darkgray}\color{white}17 & \cellcolor{darkgray}\color{white}0.7880 & \cellcolor{darkgray}\color{white}0.0002 \\
\ch{H2CO/H2O} & \cellcolor{gray}\color{white}8 & \cellcolor{gray}\color{white}0.7575 & \cellcolor{gray}\color{white}0.0295 & \cellcolor{lightgray}5 & \cellcolor{lightgray}0.5971 & \cellcolor{lightgray}0.2877 & \cellcolor{lightgray}13 & \cellcolor{lightgray}0.3832 & \cellcolor{lightgray}0.1962 \\
\ch{H2S/H2O} & 6 & -0.4590 & 0.3599 & 3 & 0.6947 & 0.5111 & 9 & 0.2737 & 0.4761 \\
\ch{HCN/H2O} & \cellcolor{lightgray}15 & \cellcolor{lightgray}0.4108 & \cellcolor{lightgray}0.1283 & 7 & 0.3957 & 0.3796 & \cellcolor{gray}\color{white}22 & \cellcolor{gray}\color{white}0.4678 & \cellcolor{gray}\color{white}0.0281 \\
\ch{NH3/H2O} & 10 & 0.1611 & 0.6565 & 4 & 0.1050 & 0.8950 & 14 & -0.2537 & 0.3815 \\
\bottomrule
\end{tabular}

\caption{Table showing the results of the Pearson correlation tests between abundance of parent species and nucleus size, described in Section \ref{S:Findings}.
For each species abundance (with respect to \ch{H2O}) we present the correlations for dynamical subsets of the data, ecliptic comets and nearly isotropic comets, as well as the results for all comets.
For each group we state the number of comets analysed, the Pearson correlation coefficient and the associated $p$-value of the correlation test.
We highlight the species and dynamical groups in order of Pearson correlation significance, i.e.\ by $p$-value.
The strongest significance correlations ($p \leq 0.003$, equivalent to a 3-sigma threshold) are indicated by dark grey cells with white text.
Moderate significance ($0.003 < p \leq 0.05$, 2-sigma) results are indicated by grey cells with white text.
Results with only marginal significance ($ 0.05 < p \leq 0.32$, 1-sigma) are shaded light grey.
All other results have been deemed to be statistically insignificant in this analysis (plain white cells).
}
\label{tab:correlations1} 
\end{table*}

\subsection{The critical influence of heliocentric distance}
\label{s:effect_of_heliocentric_distance}

Cometary activity is driven by solar heating and is therefore strongly correlated with the heliocentric distance. This is why we have assessed the abundance ratio of each species relative to a common volatile such as \ch{H2O} rather than considering production rates directly. However, more volatile species are able to drive activity at lower temperatures and greater heliocentric distances. As such we must assess whether the correlations reported above are driven primarily by comet size or by heliocentric distance of the measurements. It is for example expected that the abundances of CO/H$_2$O and CO$_2$/H$_2$O increase for comets past 2-3 au as the water sublimation becomes less efficient \citep{dellorussoEmergingTrendsComet2016,ootsuboAKARINEARINFRAREDSPECTROSCOPIC2012}. For other species, \cite{langland-shulaCometClassificationNew2011} reported a trend of decreasing C$_2$/CN ratio as comets moved away from the Sun whereas \cite{ahearnEnsemblePropertiesComets1995,cochranThirtyYearsCometary2012,finkTaxonomicSurveyComet2009} did not report a similar trend. 
\cite{dellorussoEmergingTrendsComet2016} report increases in the abundance of H$_2$CO, NH$_3$, and C$_2$H$_2$ within heliocentric distances of 0.8 au compared to the abundances measured between 1 and 2 au, which might be caused by an additional contribution from extended sources. 

\begin{figure*}
\begin{tabular}{c c c}
	\includegraphics[width=0.3\textwidth]{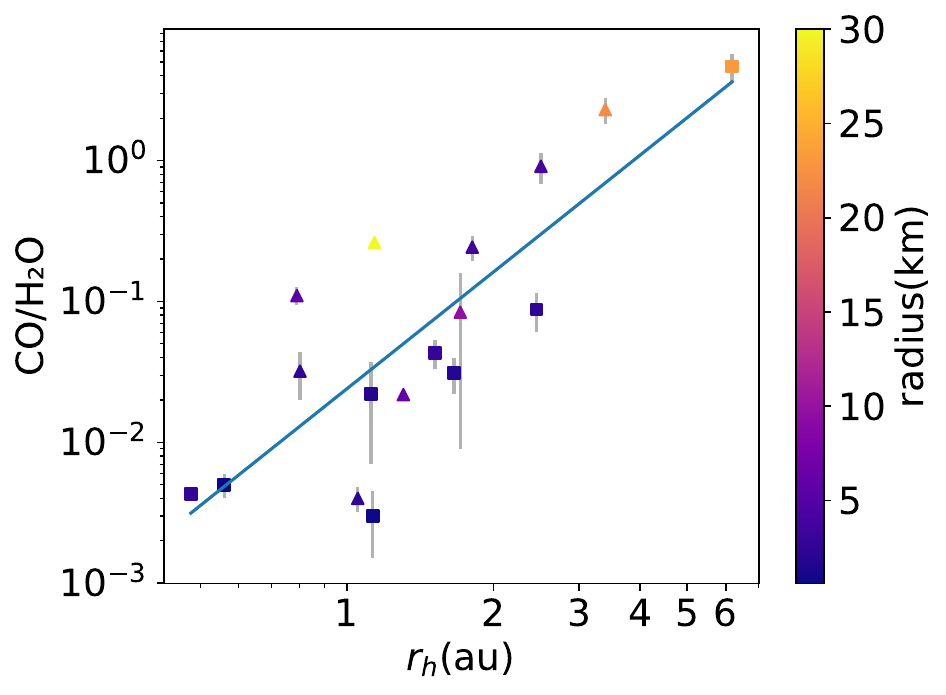} &
 	\includegraphics[width=0.3\textwidth]{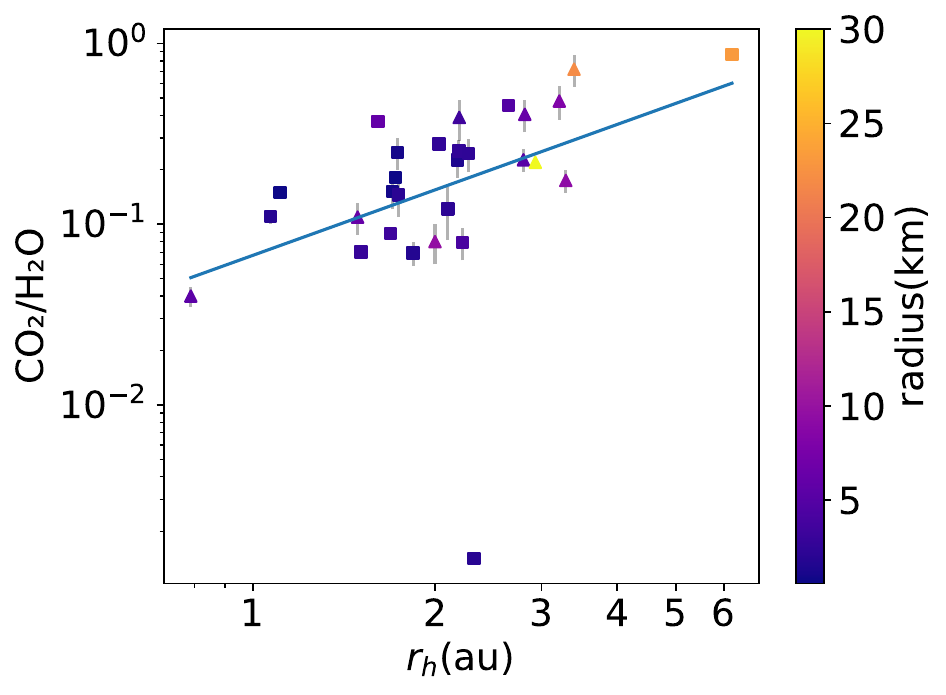} &
  	\includegraphics[width=0.3\textwidth]{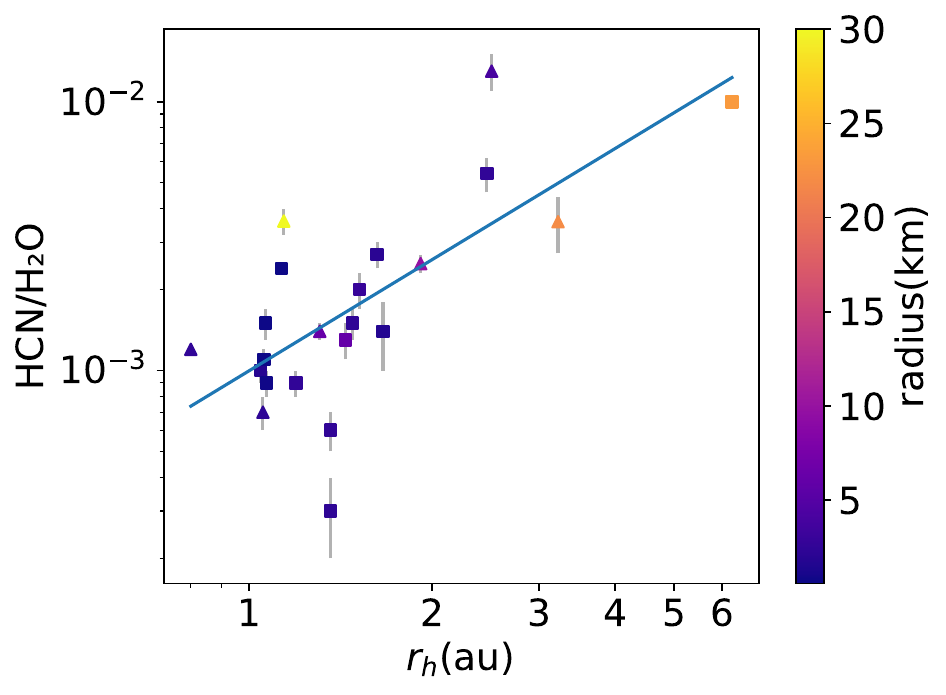}
\end{tabular}
\caption{
Plots of the \ch{CO/H2O}, \ch{CO2/H2O}, \ch{HCN/H2O} abundance as a function of the mean heliocentric distance of the observations for each comet.
The datasets consists of the same comets as the corresponding panels in Figure \ref{fig:comp_size_H2O_parents}
(except for \ch{HCN/H2O} where we have excluded an outlying measurement for C/2002 X5 at $r_h = 0.21\ \si{au}$). 
As in Figure \ref{fig:comp_size_H2O_parents}, square and triangle markers denote ECs and NICs, respectively, and here the marker colour indicates the comet nucleus size.
The Pearson correlation coefficients for abundance vs $r_h$ are $\gamma=0.810$ ($p = 8\times10^{-5}$), $\gamma=0.410$ ($p = 2.7\times10^{-2}$), $\gamma=0.709$ ($p = 2\times10^{-4}$) for  \ch{CO/H2O}, \ch{CO2/H2O}, \ch{HCN/H2O} respectively.
}
\label{fig:comp_size_dist_check}
\end{figure*}

On the plots shown in Figure \ref{fig:comp_size_H2O_parents} the colour of the data points reflects the heliocentric distance of the observation. For most species, there is no obvious heliocentric trend in the plots.
However, for some of the species which show the strongest correlations between abundance and size (\ch{CO}, \ch{CO2}, \ch{HCN}) there are strong indications of a compositional dependence on heliocentric distance. 
In order to test this additional correlation, in Figure \ref{fig:comp_size_dist_check} we plot the abundance of these species (relative to \ch{H2O}) against heliocentric distance. 
This is the same data as shown in Figure \ref{fig:comp_size_H2O_parents}, with the exception of an outlying \ch{HCN/H2O} measurement for C/2002 X5 made at $r_h = 0.21\ \si{au}$, which we have excluded as an outlier in terms of heliocentric distance.
There is strong positive correlation between abundance of these species and heliocentric distance as indicated by the corresponding values of $\gamma$ and $p$ provided in Figure \ref{fig:comp_size_dist_check}.
In addition, we highlight these trends with a linear fit in log-log space. 

\begin{figure*}
	\begin{tabular}{ccc} 
		\includegraphics[width=0.3\textwidth]{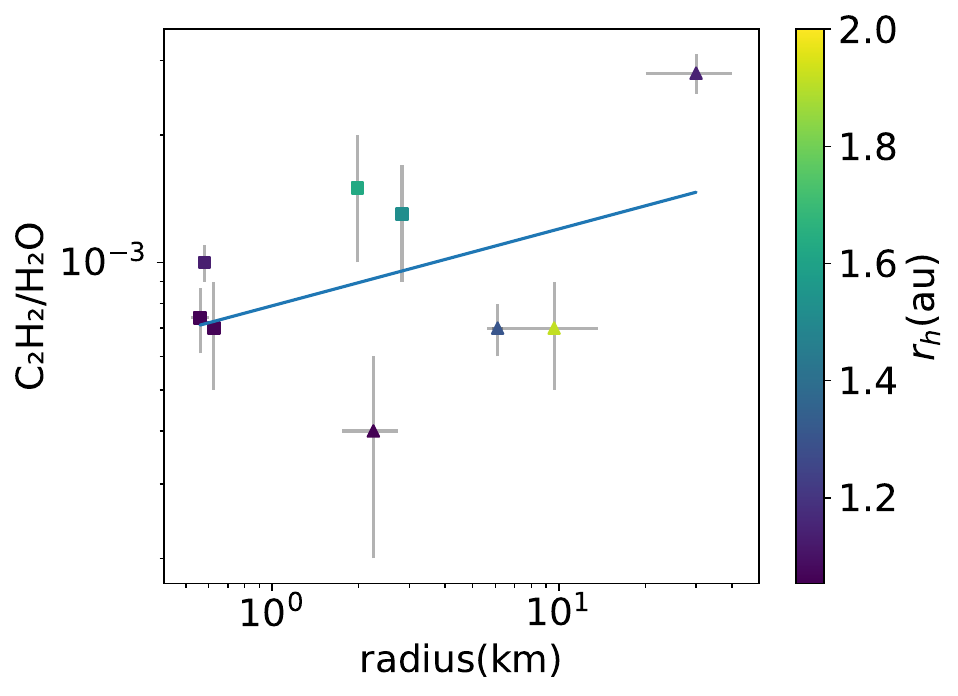} &
		\includegraphics[width=0.3\textwidth]{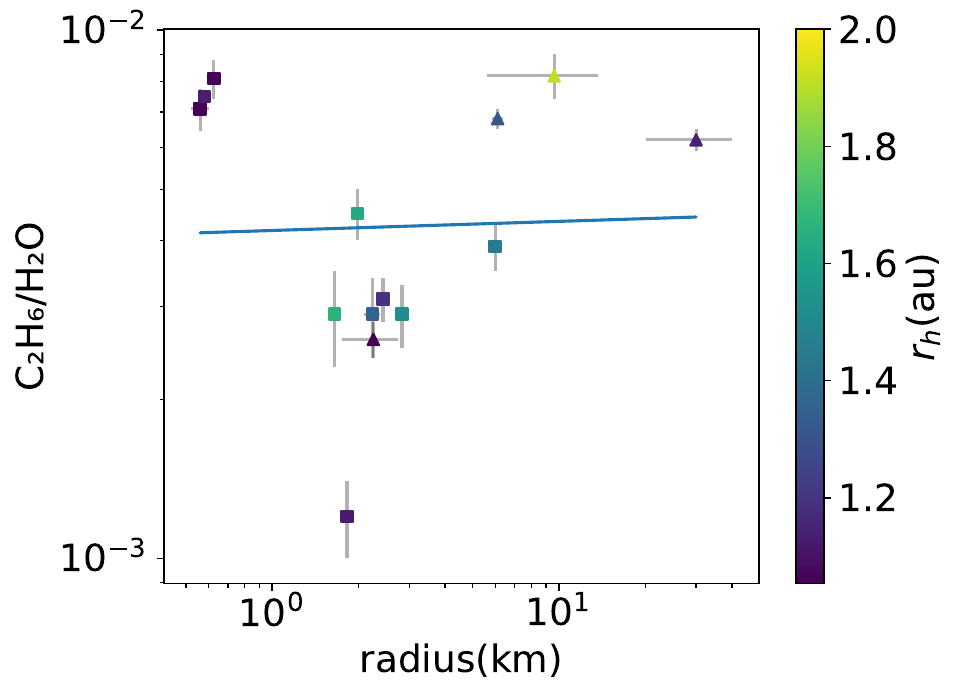} &
		\includegraphics[width=0.3\textwidth]{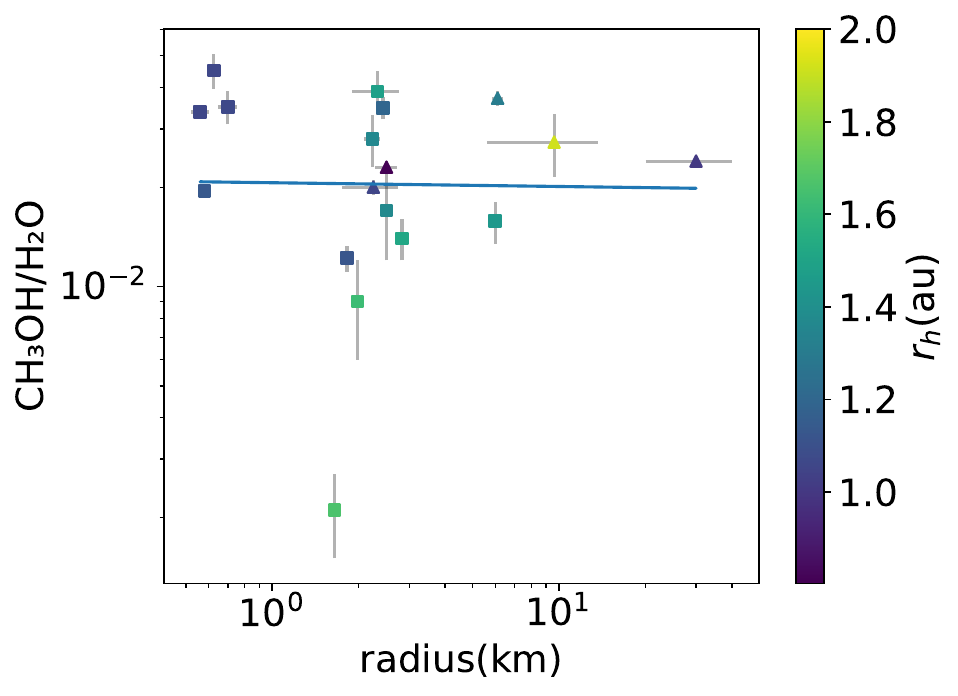} \\
		\includegraphics[width=0.3\textwidth]{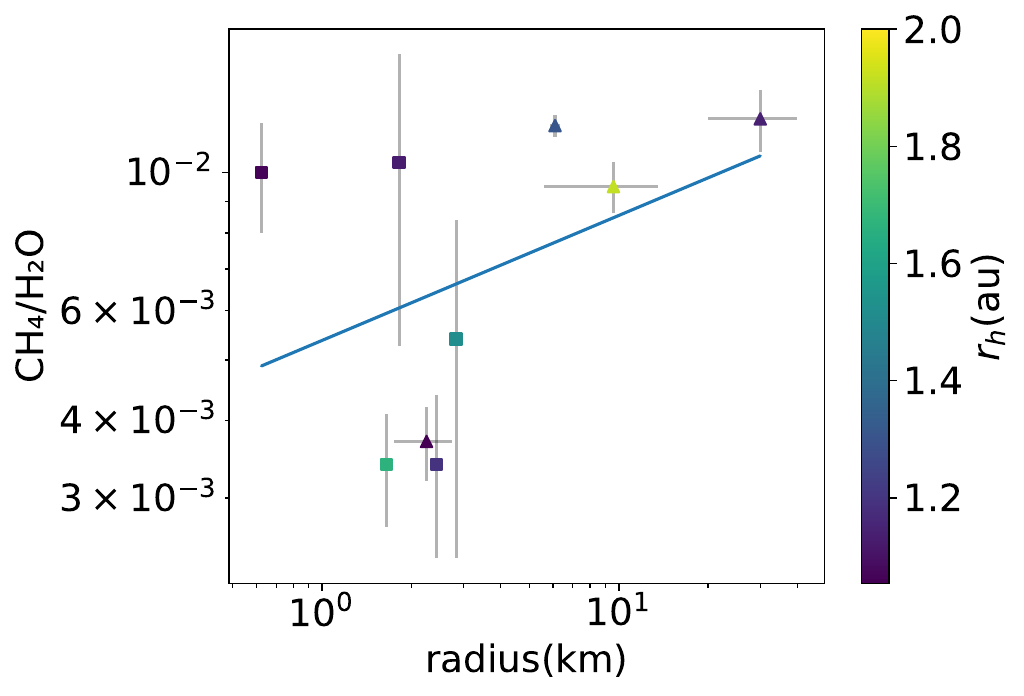} &   
		\includegraphics[width=0.3\textwidth]{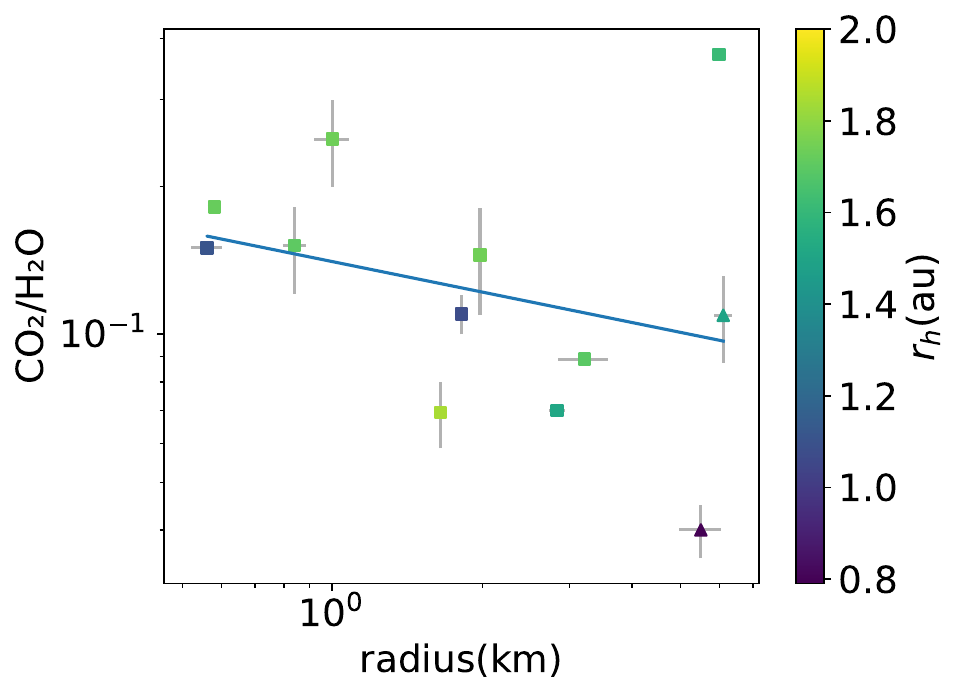} &
		\includegraphics[width=0.3\textwidth]{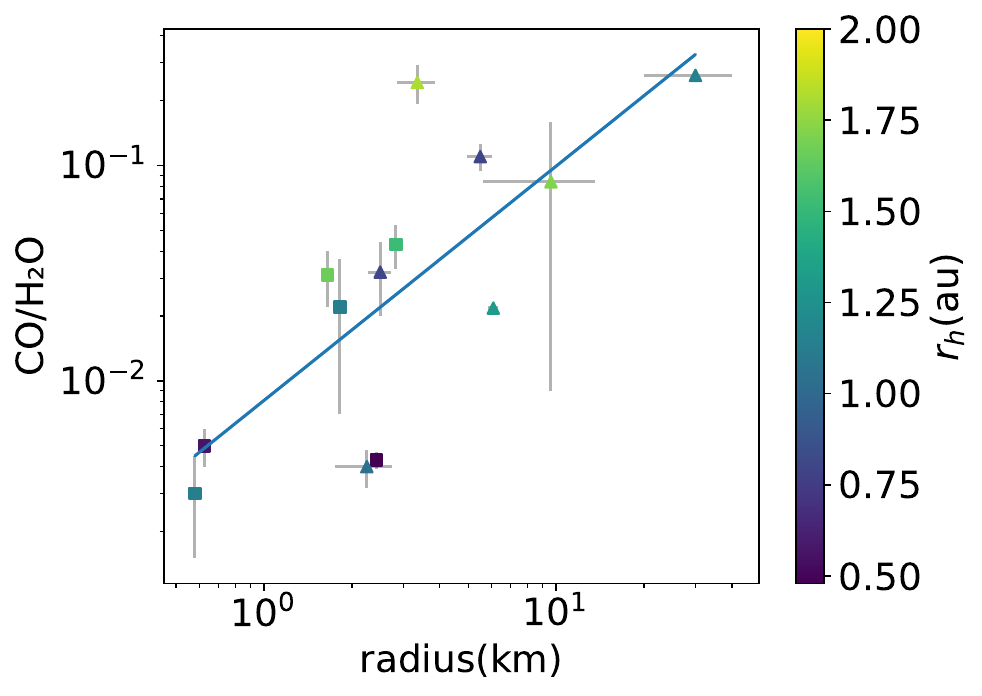} \\   
		\includegraphics[width=0.3\textwidth]{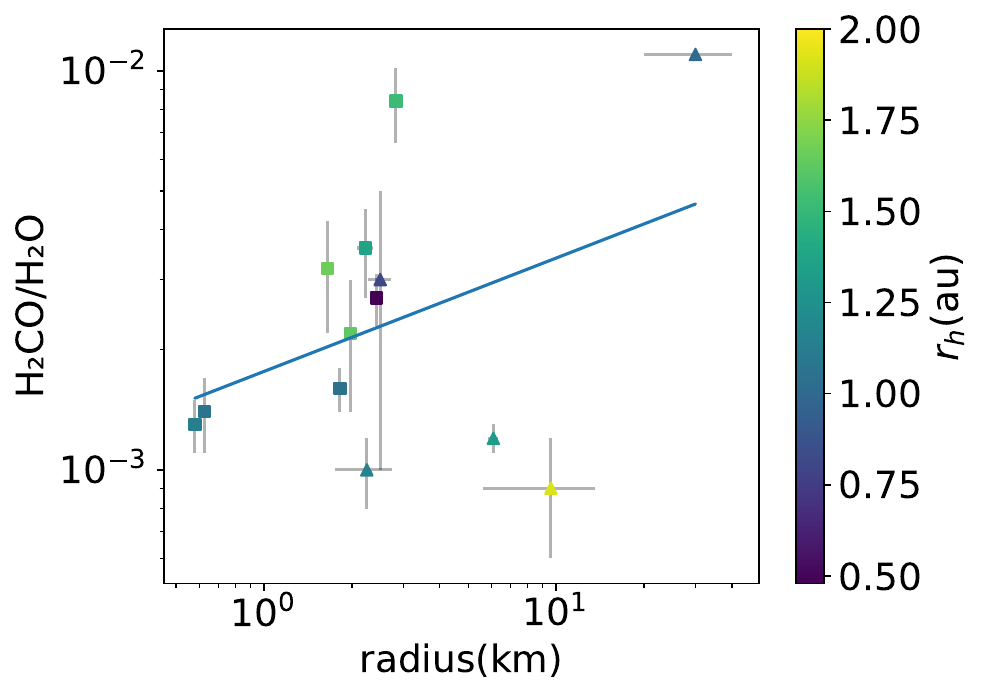} &
		\includegraphics[width=0.3\textwidth]{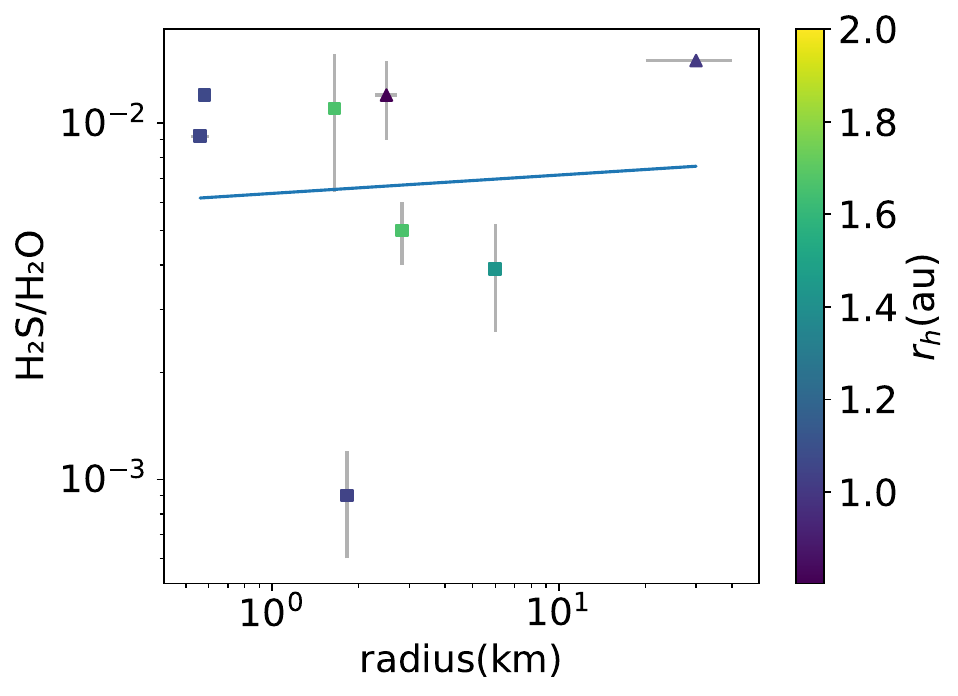} &
		\includegraphics[width=0.3\textwidth]{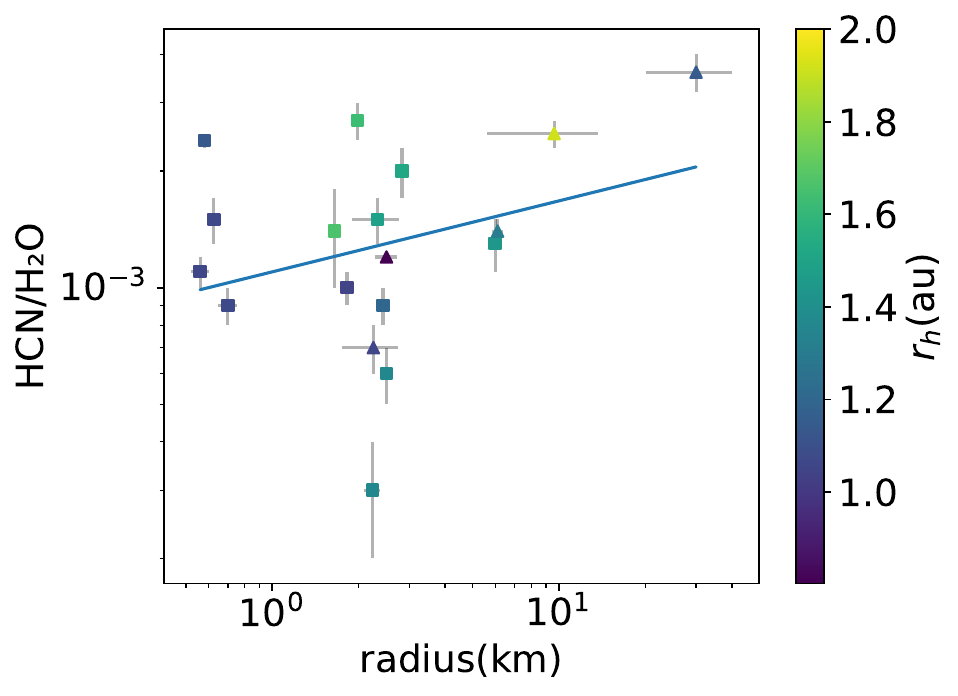} \\
		\includegraphics[width=0.3\textwidth]{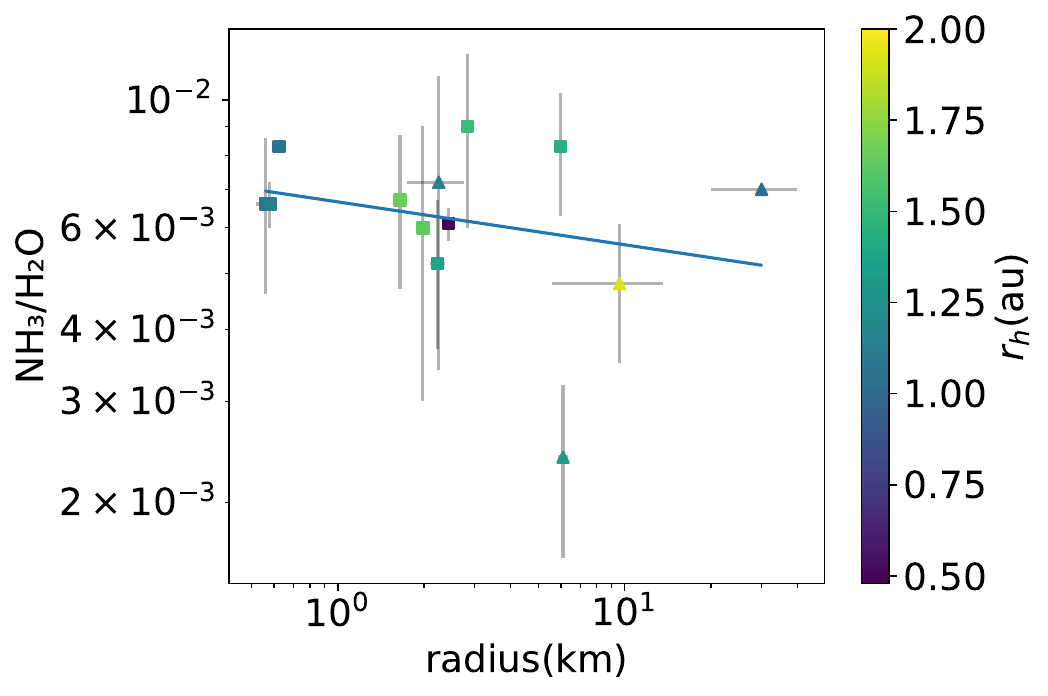} &
		\includegraphics[width=0.3\textwidth]{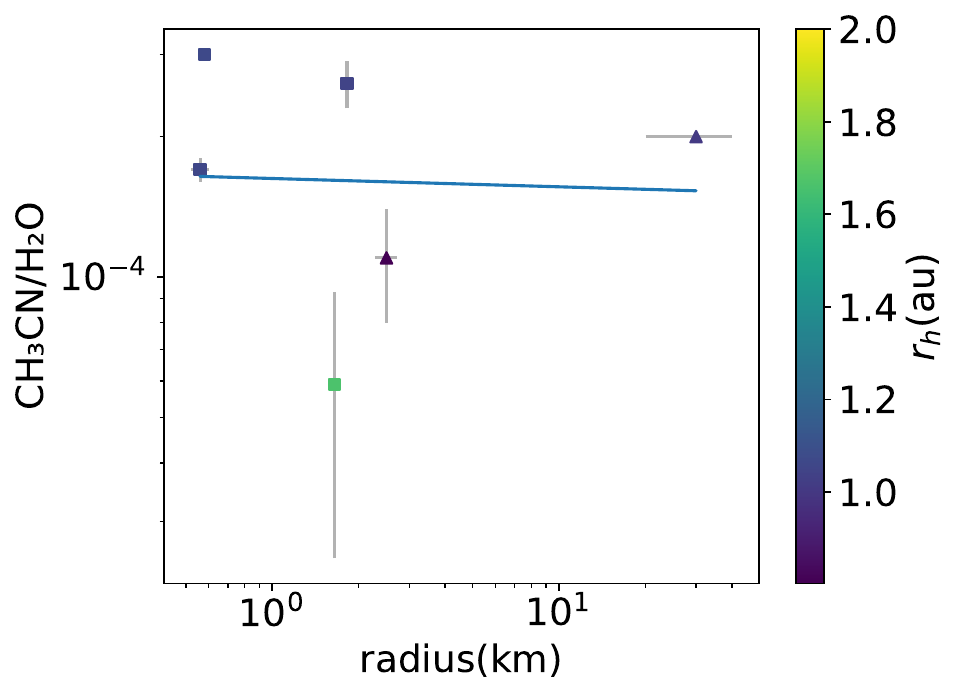} &
	\end{tabular}
	\caption{
		Log-scale plots of parent species abundance (relative to \ch{H2O}) as a function of radius of the nucleus.
		These plots are similar to Figure \ref{fig:comp_size_H2O_parents}, but consider only compositional measurements with heliocentric distances $r_h < 2\ \si{au}$ as observations at large heliocentric distances may not accurately reflect the comet composition.
		The radius-composition correlation for all comets in the sample is indicated by a linear trend line fit (solid line), the corresponding Pearson correlation coefficients are provided in Table \ref{tab:correlations1_rh_lim}.
	}
	\label{fig:comp_size_dist_check_bin_parents}
\end{figure*}

In order to test the strength of the heliocentric distance dependence we considered the composition - size correlations for subsets of the data that have been limited to measurements made with heliocentric distances of $r_h < 2\ \si{au}$. 
This restricted range is selected to remove observations at large $r_h$ where the changes in the production rate of \ch{H2O} due to reduced solar heating may skew the measured abundance of a particular species (see further discussion in Section \ref{S:Discussion}).
However, it must be noted that this selection criteria preferentially excludes some of the largest comets, primarily due to observational biases. 
Figure \ref{fig:comp_size_dist_check_bin_parents} and Table \ref{tab:correlations1_rh_lim} show the results of this test and we see that the subset of \ch{CO}/\ch{H2O} abundances still displays a statistically significant Pearson correlation coefficient ($\gamma = 0.750$, $p = 0.0031$) at the 2-sigma level (albeit on the 3-sigma boundary).
This implies that the correlation between \ch{CO}/\ch{H2O} and comet size dominates over the $r_h$ dependence. 
In contrast, the positive correlations of \ch{CO2}/\ch{H2O} and \ch{HCN}/\ch{H2O} disappear or are reduced in significance ($\gamma=-0.295$, $p = 0.352$ and $\gamma=0.312$, $p = 0.207$ respectively), implying that the correlations seen in Figure \ref{fig:comp_size_H2O_parents} are driven primarily by heliocentric distance effects for these species.

To further test the robustness of this correlation we repeated the analysis with statistical resampling of the 13 comets measured at $r_h<2\ \si{au}$ in our \ch{CO}/\ch{H2O} dataset.
We conducted a bootstrap resampling, i.e.\ sampling with replacement, and found that over the course of 10,000 repeats 52\% of the resulting correlations were of 3-sigma significance.
93\% of tests had a significance of 2-sigma or stronger.
We also conducted a jack-knife resampling, where the test is repeated with a given data point dropped in turn.
In this test the correlation had 3-sigma significance 23\% of the time and all permutations resulted in at least a 2-sigma correlation.
The weakest correlation occurred when C/1995 O1 was excluded, as would be expected given that this is the largest comet in the dataset, however the overall correlation was still moderate ($\gamma = 0.68$, $p = 0.016$).
Overall these resampling tests show that the correlation between \ch{CO}/\ch{H2O} and nucleus size is relatively robust for the given dataset and is not overly dominated by a particular comet.
However we acknowledge that our dataset is limited by its small size and the inherent difficulties in accurately determining the size and composition of cometary nuclei.
The veracity of this correlation would be greatly strengthened with more measurements and improved estimates of the \ch{CO} abundances in particular, given the variation in literature values for some comets.

\begin{table*}  
	\begin{tabular}{l|rrr|rrr|rrr}
\toprule
 & \multicolumn{3}{l|}{Ecliptic Comets} & \multicolumn{3}{l|}{Nearly Isotropic Comets} & \multicolumn{3}{l|}{All Comets} \\
Species & Number & Correlation & $p$-value & Number & Correlation & $p$-value & Number & Correlation & $p$-value \\
\midrule
\ch{C2H2/H2O} & \cellcolor{lightgray}5 & \cellcolor{lightgray}0.8573 & \cellcolor{lightgray}0.0633 & \cellcolor{gray}\color{white}4 & \cellcolor{gray}\color{white}0.9537 & \cellcolor{gray}\color{white}0.0463 & \cellcolor{lightgray}9 & \cellcolor{lightgray}0.4411 & \cellcolor{lightgray}0.2347 \\
\ch{C2H6/H2O} & \cellcolor{lightgray}10 & \cellcolor{lightgray}-0.6085 & \cellcolor{lightgray}0.0619 & \cellcolor{lightgray}4 & \cellcolor{lightgray}0.6838 & \cellcolor{lightgray}0.3162 & 14 & 0.0340 & 0.9083 \\
\ch{CH3CN/H2O} & 4 & -0.4120 & 0.5880 & 2 & 1.0000 & 1.0000 & 6 & -0.0430 & 0.9356 \\
\ch{CH3OH/H2O} & 13 & -0.2809 & 0.3526 & 5 & 0.2691 & 0.6616 & 18 & -0.0162 & 0.9493 \\
\ch{CH4/H2O} & 5 & -0.5520 & 0.3347 & \cellcolor{lightgray}4 & \cellcolor{lightgray}0.8068 & \cellcolor{lightgray}0.1932 & \cellcolor{lightgray}9 & \cellcolor{lightgray}0.4084 & \cellcolor{lightgray}0.2752 \\
\ch{CO2/H2O} & 10 & -0.0354 & 0.9227 & 2 & 1.0000 & 1.0000 & 12 & -0.2947 & 0.3524 \\
\ch{CO/H2O} & \cellcolor{lightgray}6 & \cellcolor{lightgray}0.6573 & \cellcolor{lightgray}0.1560 & \cellcolor{lightgray}7 & \cellcolor{lightgray}0.5990 & \cellcolor{lightgray}0.1553 & \cellcolor{gray}\color{white}13 & \cellcolor{gray}\color{white}0.7503 & \cellcolor{gray}\color{white}0.0031 \\
\ch{H2CO/H2O} & \cellcolor{gray}\color{white}8 & \cellcolor{gray}\color{white}0.7575 & \cellcolor{gray}\color{white}0.0295 & \cellcolor{lightgray}5 & \cellcolor{lightgray}0.5971 & \cellcolor{lightgray}0.2877 & \cellcolor{lightgray}13 & \cellcolor{lightgray}0.3832 & \cellcolor{lightgray}0.1962 \\
\ch{H2S/H2O} & 6 & -0.4590 & 0.3599 & 2 & 1.0000 & 1.0000 & 8 & 0.0712 & 0.8669 \\
\ch{HCN/H2O} & 13 & -0.1625 & 0.5958 & \cellcolor{gray}\color{white}5 & \cellcolor{gray}\color{white}0.9446 & \cellcolor{gray}\color{white}0.0155 & \cellcolor{lightgray}18 & \cellcolor{lightgray}0.3124 & \cellcolor{lightgray}0.2069 \\
\ch{NH3/H2O} & 9 & 0.1157 & 0.7669 & 4 & 0.1050 & 0.8950 & 13 & -0.2562 & 0.3981 \\
\bottomrule
\end{tabular}

	\caption{
		Similar to Table \ref{tab:correlations1}, here we show the Pearson correlations coefficients for the composition-radius relations for the parent species shown in Figure \ref{fig:comp_size_H2O_parents}. 
		As described in Section \ref{s:effect_of_heliocentric_distance} only compositions with $r_h < 2\ \si{au}$ where considered.
	}
	\label{tab:correlations1_rh_lim} 
\end{table*}

\begin{table*}  
	\begin{tabular}{l|rrr|rrr|rrr}
\toprule
 & \multicolumn{3}{l|}{Ecliptic Comets} & \multicolumn{3}{l|}{Nearly Isotropic Comets} & \multicolumn{3}{l|}{All Comets} \\
Species & Number & Correlation & $p$-value & Number & Correlation & $p$-value & Number & Correlation & $p$-value \\
\midrule
\ch{C2H2/H2O} & 2 & -1.0000 & 1.0000 & \cellcolor{gray}\color{white}4 & \cellcolor{gray}\color{white}0.9537 & \cellcolor{gray}\color{white}0.0463 & 6 & 0.4825 & 0.3324 \\
\ch{C2H6/H2O} & 7 & 0.3974 & 0.3774 & \cellcolor{lightgray}4 & \cellcolor{lightgray}0.6838 & \cellcolor{lightgray}0.3162 & \cellcolor{gray}\color{white}11 & \cellcolor{gray}\color{white}0.7194 & \cellcolor{gray}\color{white}0.0126 \\
\ch{CH3CN/H2O} & 2 & 1.0000 & 1.0000 & 2 & 1.0000 & 1.0000 & 4 & 0.4053 & 0.5947 \\
\ch{CH3OH/H2O} & 9 & 0.3317 & 0.3832 & 5 & 0.2691 & 0.6616 & \cellcolor{lightgray}14 & \cellcolor{lightgray}0.3563 & \cellcolor{lightgray}0.2112 \\
\ch{CH4/H2O} & 4 & -0.1136 & 0.8864 & \cellcolor{lightgray}4 & \cellcolor{lightgray}0.8068 & \cellcolor{lightgray}0.1932 & \cellcolor{lightgray}8 & \cellcolor{lightgray}0.6835 & \cellcolor{lightgray}0.0616 \\
\ch{CO2/H2O} & \cellcolor{lightgray}6 & \cellcolor{lightgray}0.7001 & \cellcolor{lightgray}0.1214 & 2 & 1.0000 & 1.0000 & 8 & 0.1737 & 0.6809 \\
\ch{CO/H2O} & 4 & -0.1458 & 0.8542 & \cellcolor{lightgray}7 & \cellcolor{lightgray}0.5990 & \cellcolor{lightgray}0.1553 & \cellcolor{gray}\color{white}11 & \cellcolor{gray}\color{white}0.6136 & \cellcolor{gray}\color{white}0.0447 \\
\ch{H2CO/H2O} & \cellcolor{lightgray}6 & \cellcolor{lightgray}0.7043 & \cellcolor{lightgray}0.1182 & \cellcolor{lightgray}5 & \cellcolor{lightgray}0.5971 & \cellcolor{lightgray}0.2877 & 11 & 0.2525 & 0.4537 \\
\ch{H2S/H2O} & 4 & 0.0417 & 0.9583 & 2 & 1.0000 & 1.0000 & 6 & 0.4129 & 0.4158 \\
\ch{HCN/H2O} & 9 & 0.0371 & 0.9245 & \cellcolor{gray}\color{white}5 & \cellcolor{gray}\color{white}0.9446 & \cellcolor{gray}\color{white}0.0155 & \cellcolor{gray}\color{white}14 & \cellcolor{gray}\color{white}0.5512 & \cellcolor{gray}\color{white}0.0411 \\
\ch{NH3/H2O} & \cellcolor{lightgray}6 & \cellcolor{lightgray}0.5948 & \cellcolor{lightgray}0.2131 & 4 & 0.1050 & 0.8950 & 10 & -0.1333 & 0.7135 \\
\bottomrule
\end{tabular}

	\caption{Similar to the results shown in Table \ref{tab:correlations1_rh_lim} we display the Pearson correlation coefficients for the relation between parent species abundance and nucleus radius.
		In this analysis we have considered only measurements with $r_h <2\ \si{au}$ and made a further cut to exclude comets with radii $<1\ \si{km}$; these smaller objects are more likely to be collisional fragments as opposed to primordial nuclei.
	}
	\label{tab:correlations1_rad_lim} 
\end{table*}

\subsection{Potential significance of a minimum size cutoff}

\subsubsection{Dependence on unknown fragmentation history}

An important assumption in this work is that the cometary nuclei in the dataset have not had their internal composition and structure significantly altered since their formation, e.g.\ by collisional or tidal events.
However it has been proposed that many small Solar System bodies could be fragments of larger, primordial parent bodies, either through past collisions \citep{morbidelliCometsCollisionalFragments2015}, or tidal/rotational breakup \citep{Boehnhardt-2002}.
Due to the limited size of our dataset, and the difficulties in definitively determining the history of small bodies without detailed \textit{in situ} analysis, we have thus far only excluded the comets with a known fragmentation history from our analysis.

Here we perform a brief test to determine the influence of potential \emph{unknown} fragmentation history in the data. 
This could distort the abundances displayed by the smallest comets in our dataset and therefore skew the results, altering the real composition-size correlation by reflecting the abundance of the larger primordial parent objects instead.
In order to check for this possibility, we have repeated the Pearson correlation analysis from the previous section, but in addition to the $r_h <2\ \si{au}$ restriction we also removed any comets with radius $<1\ \si{km}$. The 1 km cutoff was determined arbitrarily, representing exceptionally small nuclei.

The results of this test are displayed in Table \ref{tab:correlations1_rad_lim}.
In comparison to Table \ref{tab:correlations1_rh_lim} we see that the removal of the smallest comets in the dataset leads to more significant correlations for some abundances, e.g.\ \ch{C2H6}/\ch{H2O} with a decrease in $p$-value of $0.908 \rightarrow 0.012$.
However, the significance of the correlation for other species is reduced when these objects are excluded. E.g., For \ch{CO}/\ch{H2O}, the $p$-value increased from $0.003 \rightarrow 0.045$.
Given that the sample now consists of a smaller number of comets a reduction of statistical significance is generally to be expected. Despite this, the correlation became highly significant (3-sigma) for ethane, which is among the species with the lowest sublimation temperature beyond that of methane.

\subsubsection{Dependence on volatility}
For species that are much less volatile than \ch{CO}, e.g.\ \ch{CO2} and \ch{HCN}, the thermophysical models of \cite{MalamudEtAl-2022} predict that significant migration and differentiation requires higher temperatures and thus larger cometary nuclei.
As such, a correlation between these abundances and size might only be relevant beyond a certain size threshold.

\begin{figure}
\begin{tabular}{c}
     \includegraphics[width=\columnwidth]{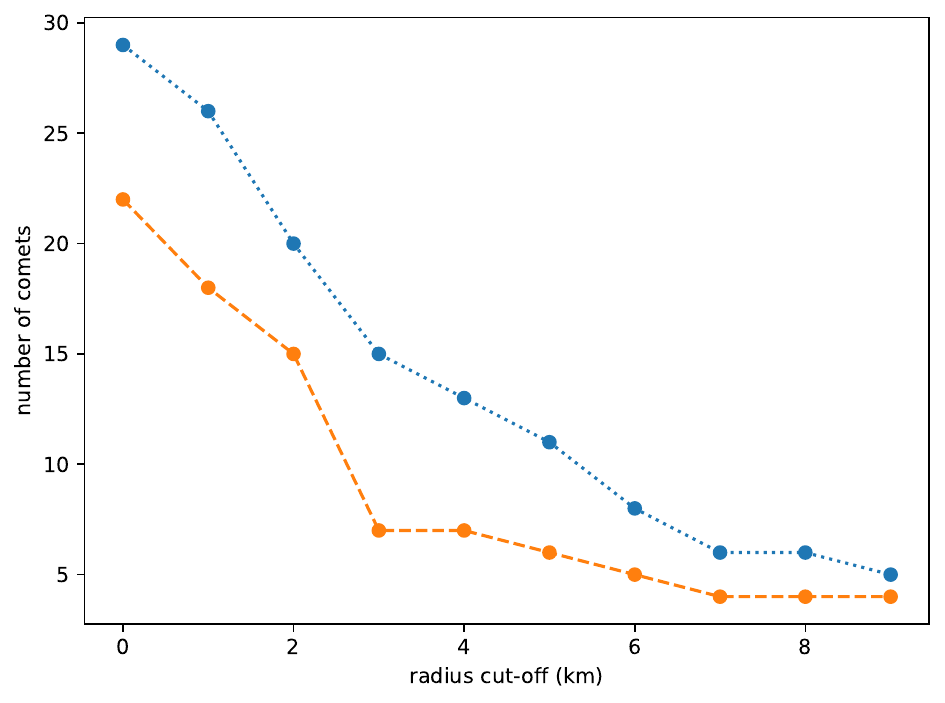}
 \\
      \includegraphics[width=\columnwidth]{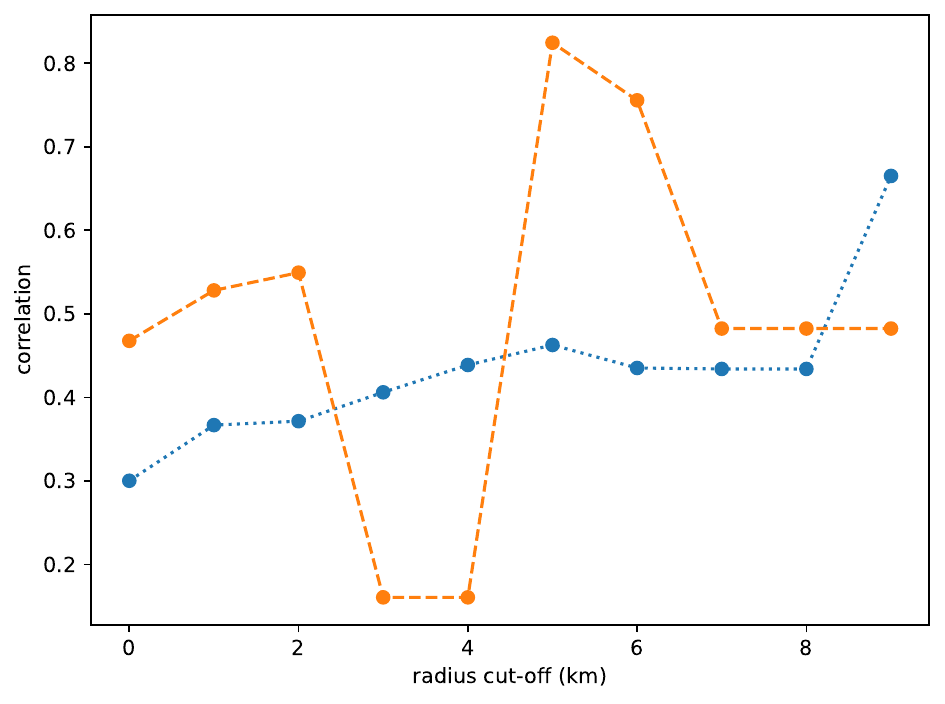}
 \\
      \includegraphics[width=\columnwidth]{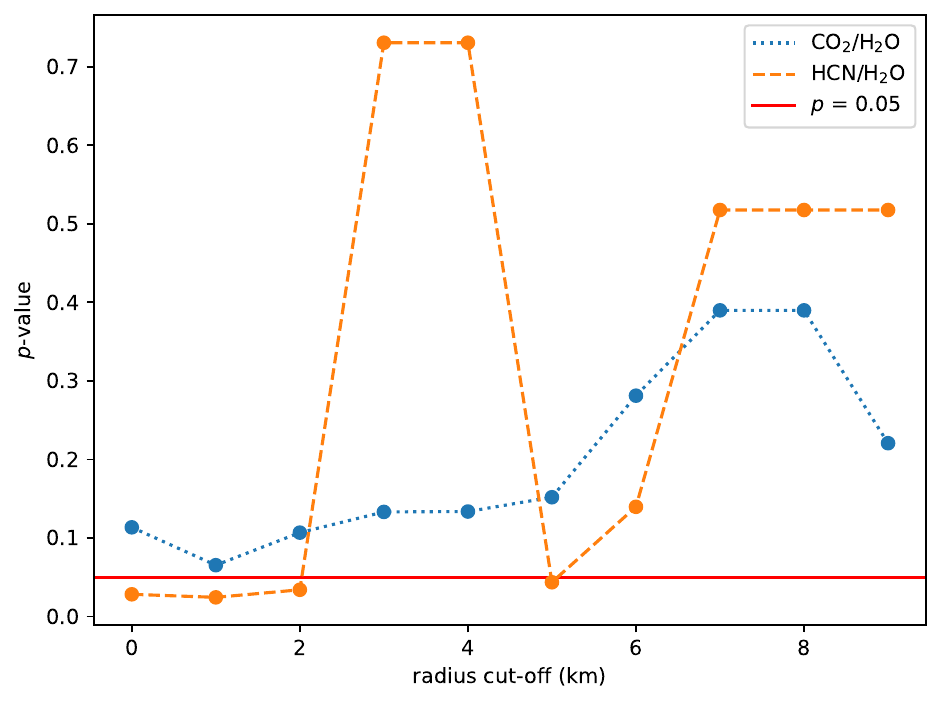} 
\end{tabular}
\caption{
Testing the composition - size correlation as a function of comet sizes for the abundances of \ch{CO2/H2O} and \ch{HCN/H2O}.
Upper panel: The inverse cumulative distribution of number of comets greater than some radius.
Middle panel: The Pearson correlation coefficient of comet composition and size (in log-log space) as a function of the smallest comet radius included in the data subset.
Lower panel: The $p$-value corresponding to each correlation test in the middle panel.
The results for \ch{CO2/H2O} and \ch{HCN/H2O} are indicated by dotted and dashed lines respectively.}
\label{fig:comp_size_limit}
\end{figure}

In order to test this hypothesis we consider the correlation of subsets of the dataset limited by comet radius for \ch{CO2/HCN} and \ch{HCN/H2O}. 
In each test we select only the comets in the dataset with radius $> 0,1,2,...9\ \si{km}$ and determine the Pearson correlation coefficient as before, the results of which are shown in Figure \ref{fig:comp_size_limit}.
For \ch{CO2}/\ch{H2O} the correlation stays approximately the same and the $p$-value increases as more data points are removed; the Pearson correlation test on fewer data points produces less significant results as one would expect.
Therefore we cannot easily determine if this correlation is driven primarily by the comet nuclear size, or if it is being influenced by the more complicated size-heliocentric distance observational bias (Figure \ref{fig:comp_size_dist_check}). 
For \ch{HCN/H2O} the results of this correlation test fluctuate significantly, indicating that this dataset is strongly influenced by the specific objects being considered.
This test implies that both datasets would benefit greatly from being larger and having improved coverage across the range of comet sizes, with more accurate measurements taken in a uniform manner and preferentially from observations at distances less than 2 au.

\section{Discussion}\label{S:Discussion}

\subsection{A model for explaining the correlation between CO and size}
\label{SS:co_size_correlation}
The findings in Section \ref{S:Findings} show that the clearest indication of correlation between size and composition exists within the activity of CO, confirming our first prediction in Section \ref{S:intro}.
In what follows we attempt to reconcile the composition-size trend with the theoretical model of \cite{MalamudEtAl-2022}.

In Section \ref{S:intro} we described the likely mode of transport of hyper-volatile gases within the nucleus. In particular, internal CO might have initially existed as either a pure ice condensate, or as a trapped gas within an amorphous ice host. In most comets the second option might be the more likely, as also newly indicated by the activity of comet 67P/C-G \citep{rubinVolatilesH2OCO22023}, but regardless of which of the two options is correct, internal radiogenic heating would drive migration outwards. Beyond a certain temperature threshold, CO would either sublimate or be released due to the phase transitions of amorphous hosts such as CO$_2$ or H$_2$O. While migrating out, it would encounter cold matrices of pristine amorphous ice. It can therefore become re-incorporated in the amorphous ice host as trapped gas. Multiple lab experiments have demonstrated this to occur. Unlike co-deposition, which is the entrapment of high-volatility gases during the deposition of the amorphous host itself, the process we refer to above is often called sequential deposition (sometimes also entrapment via gas-flow or gas-streaming). It relates to gas which was streamed into an already existing, pre-deposited amorphous ice host. Seminal lab studies have shown that it is an effective way to trap high-volatility species \citep{BarNunEtAl-1985,LauferEtAl-1987,BarNunEtAl-1987,BarNunEtAl-1988}.

Unfortunately, the \cite{MalamudEtAl-2022} code was not explicitly designed to treat the incorporation or release of high-volatility gases into or out of amorphous water ice. It can currently only handle the sublimation and deposition of some species such as CO or CH$_4$, but only as pure condensates. We shall therefore only employ an approximate calculation, giving us a rough estimation of how much CO should be re-incorporated into amorphous ice and in turn quantify the degree of near-surface amorphous ice CO-enrichment, as a consequence.

For the initial state of the comet, prior to any heating, we assume that the internal composition is uniform. Consider that some small fraction of CO is trapped within the amorphous ice hosts -- CO$_2$ or H$_2$O -- and is released during their respective crystallisation phase transitions. For simplicity, here we neglect the possibility of CO initially present as pure ice condensate, because these two scenarios are related. Then, following the aforementioned phase transitions, released CO migrates out towards the surface, and it can become re-trapped within the ice in its path when the temperature is sufficiently cold, but still higher than the temperature of its deposition as pure condensate. Sequential deposition of CO leads to enrichment of the CO fraction stored in the amorphous ice.

Assuming that the comets we observe probe the CO/H$_2$O ratio of trapped hyper-volatiles as envisioned above, we can attempt to interpret the enrichment pattern of larger comet nuclei - using the theoretical model of \cite{MalamudEtAl-2022}. In their work, Figures 5 through 13 showed the distribution of internal temperatures in the comet as a function of various realisations of the model parameters. The two most important model parameters were the nucleus size and the comet formation time. The temperatures correlated with the nucleus size and anti-correlated with formation time. The internal temperatures within different volume fractions in the nucleus were depicted according to a colour scheme. The black colour in those figures represented relatively pristine material heated below 70 K (all amorphous ices are stable against phase transitions); red depicted temperatures in the range 70K<$T$<100K (allowing CO$_2$ amorphous ice to release trapped CO gas when undergoing a phase transition); orange depicted temperatures in the range 100K<$T$<170K (allowing H$_2$O amorphous ice to likewise release its trapped gases); yellow and white correspond to even higher temperature thresholds (also corresponding to full release of all trapped CO). We assume that the released CO gas flows toward the pristine (black coloured) volume fraction, which is closer to the surface. As already pointed out, this layer is sufficiently cold to keep its amorphous ices in their pristine state, but now allowing the excess CO gas to become re-entrapped there, enriching the CO abundance. For simplicity, we assume that these layers are enriched with CO uniformly. Given the aforementioned dependence on model parameters, it immediately follows that the degree of enrichment is greater for large comets and/or comets with a smaller formation time.

\begin{figure*}
	\begin{tabular}[b]{c}
		\includegraphics[scale=0.56]{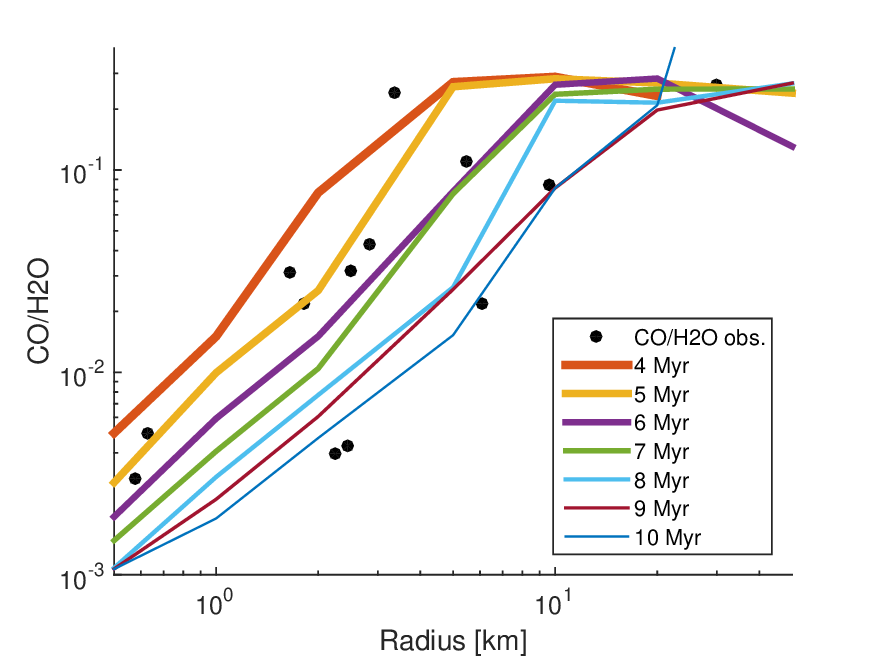}\label{fig:0_1-1-1}\\
		\small (a) Pebble radius $=0.1$ cm; permeability $b=1$
	\end{tabular}
	\begin{tabular}[b]{c}
		\includegraphics[scale=0.56]{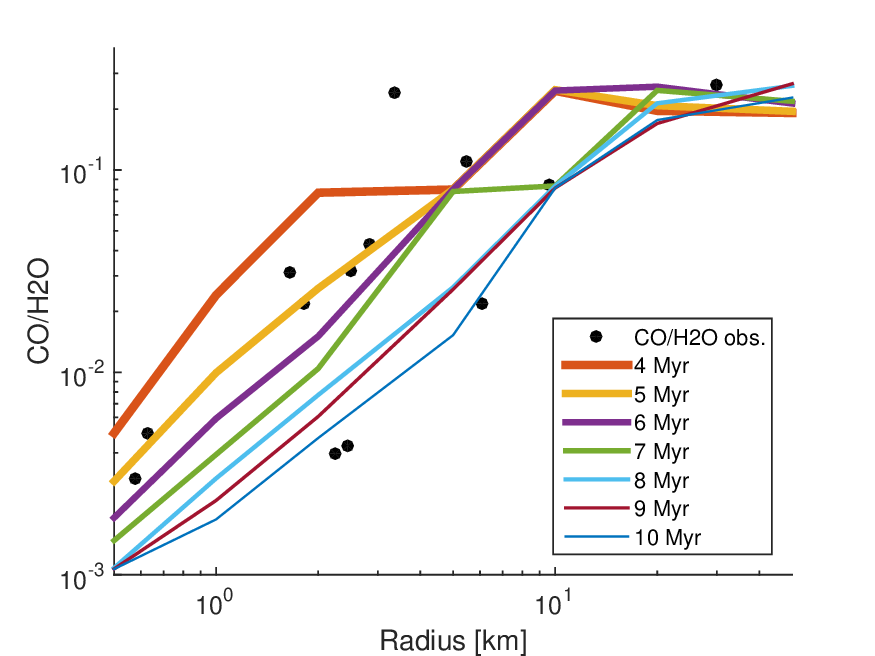}\label{fig:0_1-1-7}\\ 
		\small (b) Pebble radius $=0.1$ cm; permeability $b=7$
	\end{tabular}
	\begin{tabular}[b]{c}
		\includegraphics[scale=0.56]{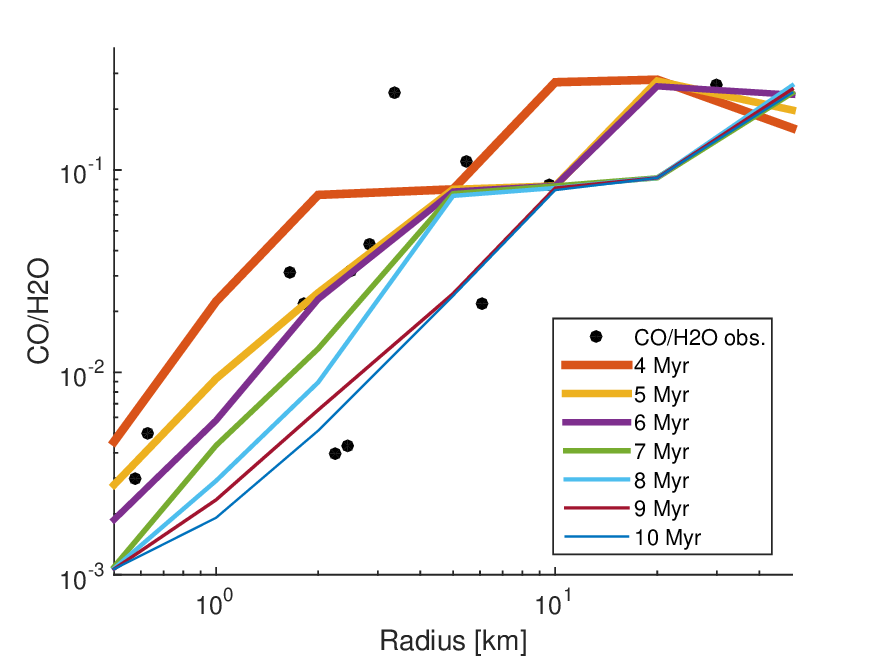}\label{fig:1-1-1}\\
		\small (c) Pebble radius $=1$ cm; permeability $b=1$
	\end{tabular}
	\begin{tabular}[b]{c}
		\includegraphics[scale=0.56]{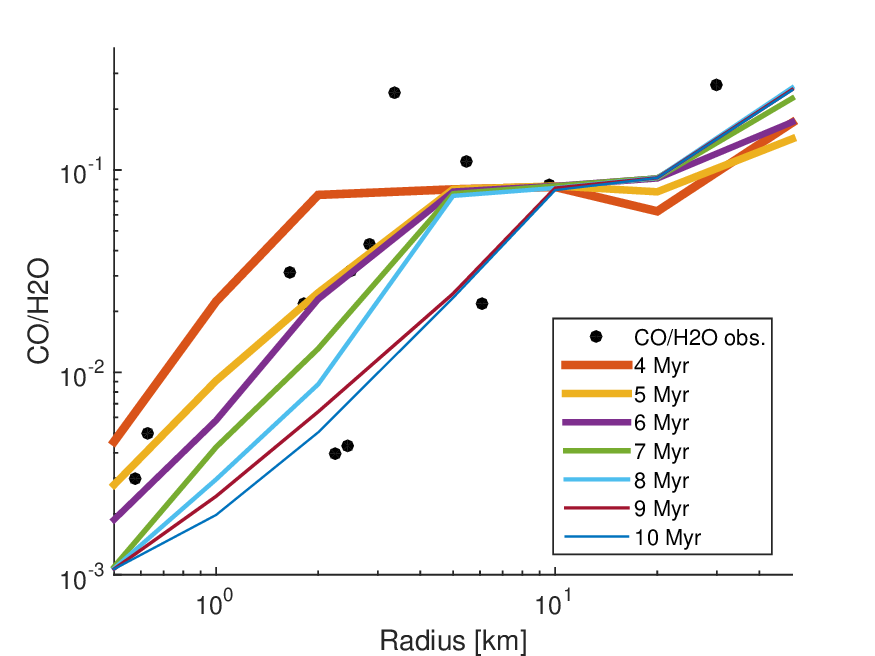}\label{fig:1-1-7}\\
		\small (d) Pebble radius $=1$ cm; permeability $b=7$
	\end{tabular}
	
	\caption{CO/H$_2$O ratio: observed data points (full circles) versus theoretical curves predicted by \protect\cite{MalamudEtAl-2022} (different lines depict different formation times - see legend). The mineral fraction is 1. Other model parameters vary as indicated in the sub-caption. Parameters are explained in the main text.}
	\label{fig:comp_size_theory_frac1}
\end{figure*}

\begin{figure*}
	\begin{tabular}[b]{c}
		\includegraphics[scale=0.56]{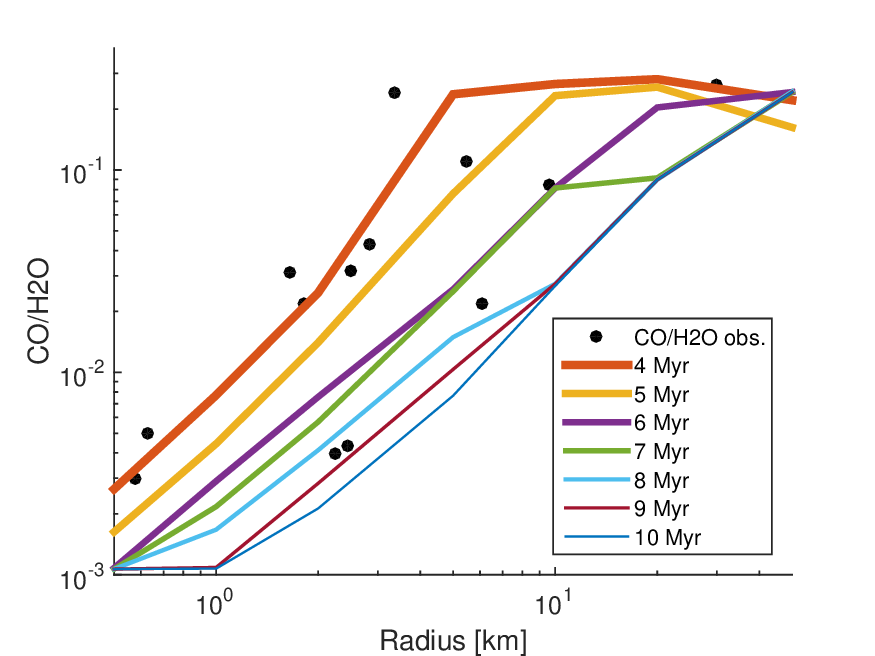}\label{fig:0_1-0_5-1}\\
		\small (a) Pebble radius $=0.1$ cm; permeability $b=1$
	\end{tabular}
	\begin{tabular}[b]{c}
		\includegraphics[scale=0.56]{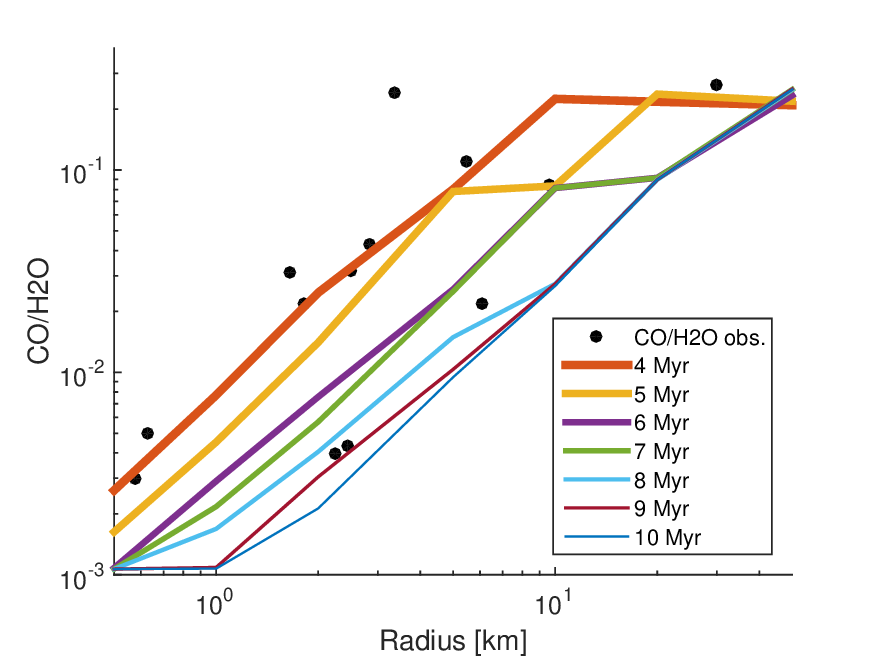}\label{fig:0_1-0_5-7}\\ 
		\small (b) Pebble radius $=0.1$ cm; permeability $b=7$
	\end{tabular}
	\begin{tabular}[b]{c}
		\includegraphics[scale=0.56]{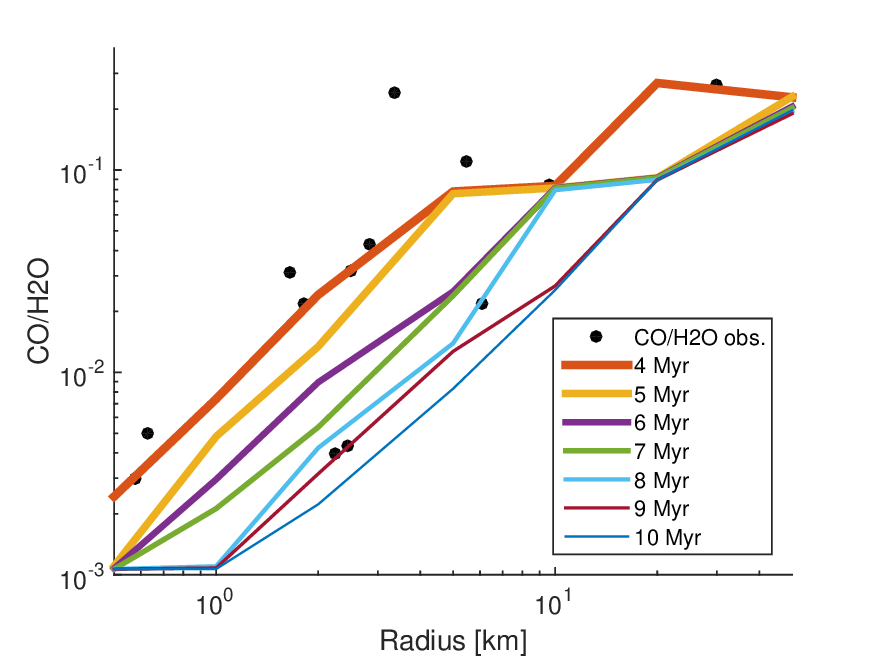}\label{fig:1-0_5-1}\\
		\small (c) Pebble radius $=1$ cm; permeability $b=1$
	\end{tabular}
	\begin{tabular}[b]{c}
		\includegraphics[scale=0.56]{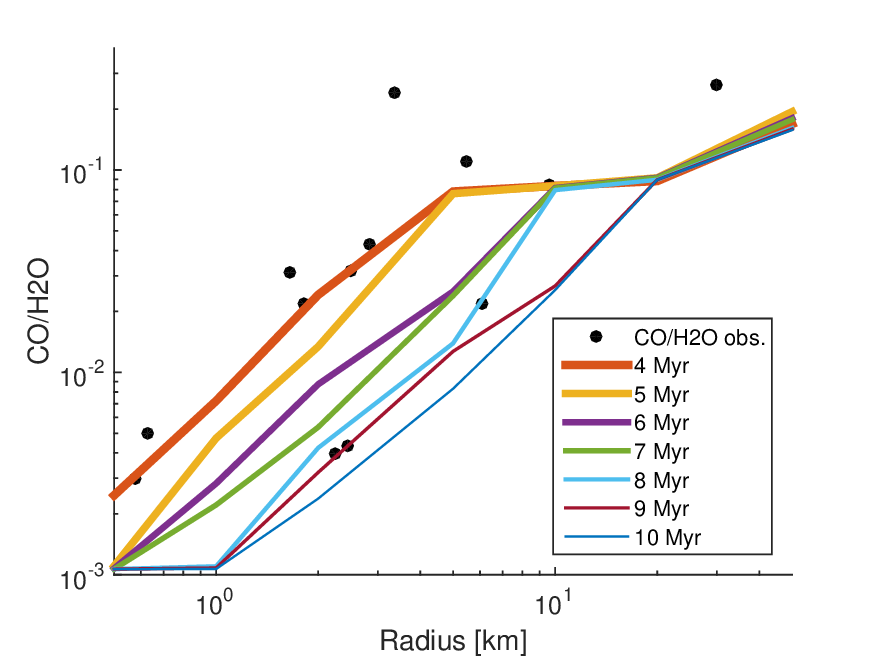}\label{fig:1-0_5-7}\\
		\small (d) Pebble radius $=1$ cm; permeability $b=7$
	\end{tabular}
	
	\caption{CO/H$_2$O ratio: observed data points (full circles) versus theoretical curves predicted by \protect\cite{MalamudEtAl-2022} (different lines depict different formation times - see legend). The mineral fraction is 0.5. Other model parameters vary as indicated in the sub-caption. Parameters are explained in the main text.}
	\label{fig:comp_size_theory_frac0_5}
\end{figure*}

\begin{figure*}
	\begin{tabular}[b]{c}
		\includegraphics[scale=0.56]{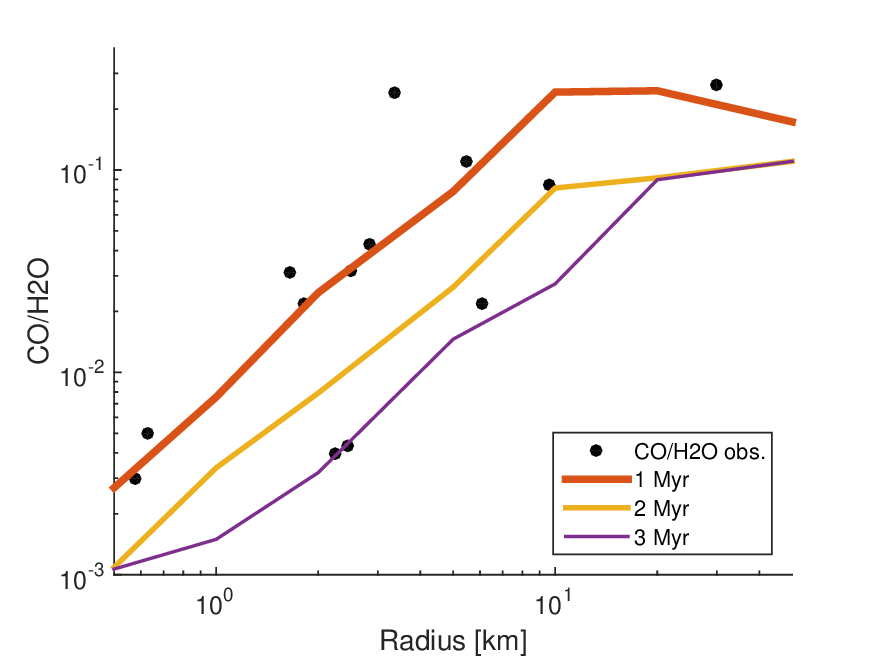}\label{fig:0_1-0_05-1}\\
		\small (a) Pebble radius $=0.1$ cm; permeability $b=1$
	\end{tabular}
	\begin{tabular}[b]{c}
		\includegraphics[scale=0.56]{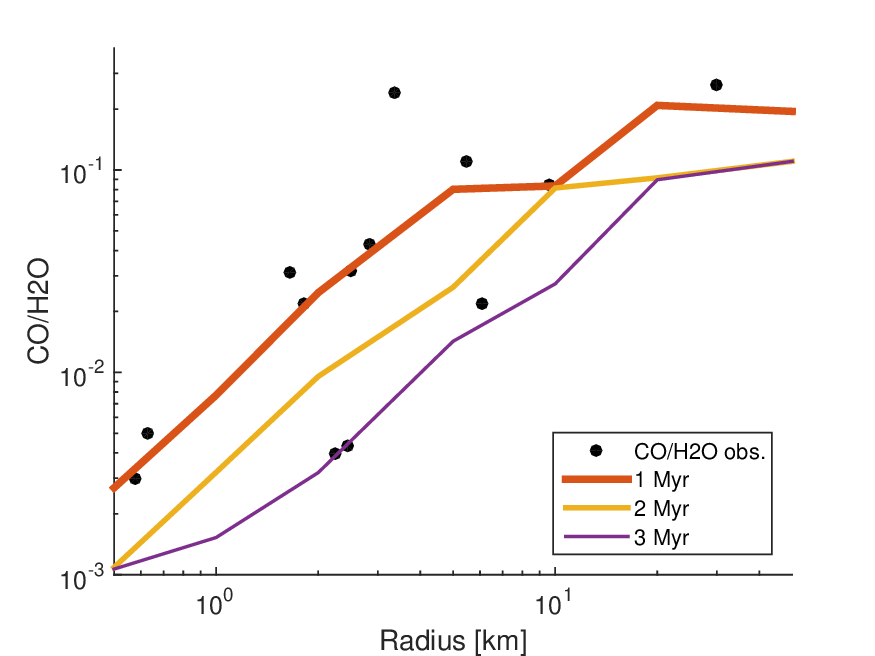}\label{fig:0_1-0_05-7}\\ 
		\small (b) Pebble radius $=0.1$ cm; permeability $b=7$
	\end{tabular}
	\begin{tabular}[b]{c}
		\includegraphics[scale=0.56]{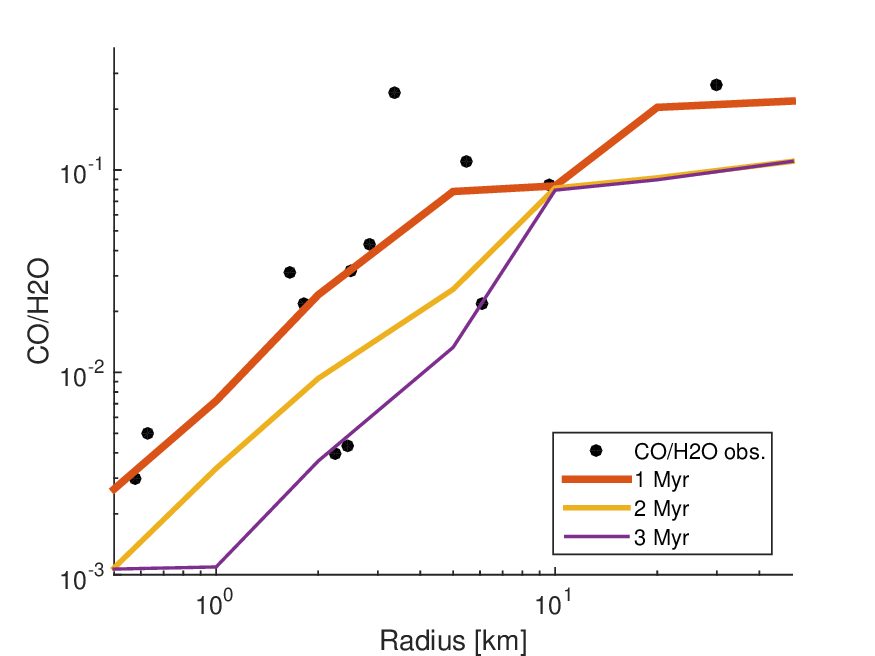}\label{fig:1-0_05-1}\\
		\small (c) Pebble radius $=1$ cm; permeability $b=1$
	\end{tabular}
	\begin{tabular}[b]{c}
		\includegraphics[scale=0.56]{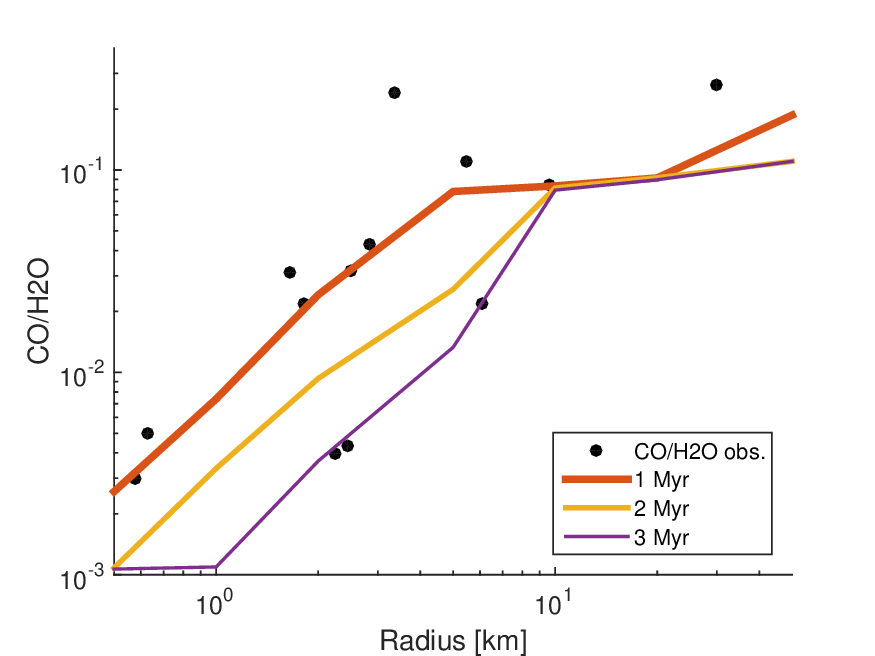}\label{fig:1-0_05-7}\\
		\small (d) Pebble radius $=1$ cm; permeability $b=7$
	\end{tabular}
	
	\caption{CO/H$_2$O ratio: observed data points (full circles) versus theoretical curves predicted by \protect\cite{MalamudEtAl-2022} (different lines depict different formation times - see legend). The mineral fraction is 0.05. Other model parameters vary as indicated in the sub-caption. Parameters are explained in the main text.}
	\label{fig:comp_size_theory_frac0_05}
\end{figure*}

In order to compare the model enrichment to the observations, we will derive a simple mathematical formula based on the assumptions above. In Figures \ref{fig:comp_size_theory_frac1}, \ref{fig:comp_size_theory_frac0_5} \& \ref{fig:comp_size_theory_frac0_05} we plot the model predictions alongside the observed data points. For the comparison we use observed data points from Figure \ref{fig:comp_size_dist_check_bin_parents}. It must be noted that Figure \ref{fig:comp_size_H2O_parents} considers the CO/H$_2$O mass ratio of comets up to an observed distance of 6 au. However, comets that are observed beyond 2 au are not able to directly sublimate significant amounts of water ice. In contrast, they are certainly able to expel trapped CO gas, released through crystallisation of the amorphous ice hosts. While water can still be expelled to some extent as a byproduct of hyper-volatile activity, we can expect the CO/H$_2$O ratio to be enhanced. Therefore, beyond 2 au, this ratio should not be indicative of the \emph{intrinsic} mass fraction of entrapped CO within the amorphous host ice at the surface. Indeed, Figure \ref{fig:comp_size_H2O_parents} shows that the peak observed CO/H$_2$O ratios are in excess of 1. This is only possible due to the large observed distance, because the amorphous ice host cannot contain more trapped gas than matrix \citep{CarMackEtAl-2023}. Figure \ref{fig:comp_size_dist_check_bin_parents} on the other hand, shows only the comets whose distance from the Sun at the time of observation is less than 2 au. It should therefore be more indicative of the intrinsic properties of the amorphous ices, which is why we have chosen to use it for the comparison.

In order to plot the theoretical enrichment curves, we first express Figures 5-13 of \cite{MalamudEtAl-2022} in terms of mass rather than volume. 
We then define the following free parameters with respect to mass: $f_{\rm H2O}$ - the mass fraction of amorphous H$_2$O ice in the nucleus; $f_{\rm CO2}$ - the mass fraction of amorphous CO$_2$ ice in the nucleus; and $f_{\rm CO}$ - the initial mass fraction of trapped CO in the amorphous H$_2$O or CO$_2$ host ices (assumed to be equal for simplicity). 

Comets are presently regarded to be highly refractory-rich bodies, having a refractory to ice mass ratios in approximately the range of 3-5 \citep{RotundiEtAl-2015,FulleEtAl-2016,FulleEtAl-2017,FulleEtAl-2019,ChoukrounEtAl-2020}, and comet 67P/C-G, the most well-studied cometary archetype \citep{FulleEtAl-2016, FilacchioneEtAl-2019, GroussinEtAl-2019} has a refractory to ice mass ratio of about 4. We therefore use a combined ice mass fraction of $f_{\rm H2O}+f_{\rm CO2} = 0.2$ to comply with these estimates. For the mass ratio between H$_2$O and CO$_2$ we also rely on estimates from comet 67P/C-G, with a respective ratio of $\sim$15 \citep{rubinVolatilesH2OCO22023}. For our fiducial parameter set we thus have: $f_{\rm H2O}=0.1875$ and $f_{\rm CO2}=0.0125$.

The choice of $f_{\rm trapCO}$ is motivated by the raw observed data.
Figure \ref{fig:comp_size_dist_check_bin_parents} shows that the smallest CO/H$_2$O number fraction, in the smallest comet nucleus, is around 0.003. Were comets to be completely pristine, this would have given us the approximate value of $f_{\rm trapCO}$ (amorphous water ice near the surface sublimates along with its entrapped CO, which is uniformly distributed throughout the whole nucleus). However, Figures 5-13 in \cite{MalamudEtAl-2022} show that even comets with radii as small as 0.5 km can still attain temperatures in excess of 70 K deep beneath the surface, despite their small size (but only when minimising their formation time). Therefore, even in small nuclei some degree of migration and enrichment of CO is possible, and in such cases the incipient CO/H$_2$O could be slightly smaller than 0.003. To account for this, we choose a round value of  CO/H$_2$O=0.001, slightly lower than yet characteristic of the 0.003 minimum observed. From this we obtain the mass fraction $f_{\rm trapCO}$ (multiplying by the molecular weight ratio - see below), capturing the right order of magnitude based on the minimum observed CO fraction.

Using $F1$, $F2$ and $F3$ to denote the mass fractions of layers within comet nuclei that have $T<70$ K, $70<T<100$ K, $T>100$ K, obtained from \cite{MalamudEtAl-2022}, we can calculate the fraction of CO released from the bulk of the comet, denoted as

\begin{equation}
    {\rm CO}_{\rm bulk}=F3 \cdot f_{\rm H2O} \cdot f_{\rm trapCO}+(F2+F3) \cdot f_{\rm CO2} \cdot f_{\rm trapCO}
\label{eq:CO-bulk}
\end{equation}

Using Eq. \ref{eq:CO-bulk}, the fraction of CO newly trapped inside the pristine amorphous water ice, i.e. the degree of its enrichment, denoted by $f_{\rm enrichCO}$, is approximately given by

\begin{equation}
    f_{\rm enrichCO} \cong \left(\frac{F1 \cdot (f_{\rm H2O}+f_{\rm CO2}) \cdot f_{\rm trapCO}+{\rm CO}_{\rm bulk}}{F1 \cdot f_{\rm H2O}} \right) \frac{m_{\rm H2O}}{m_{\rm CO}}
\label{eq:CO-to-H2O}
\end{equation}

\noindent where $m_{\rm H2O}=18$ and $m_{\rm CO}=28$ are the molecular weights of H$_2$O and CO molecules. The molecular weight ratio is required in order to go from mass fraction to number fraction, as in our reported values from observations. Note that the actual ratios in the coma also depend on the relative life times of the molecules in the coma, an effect which we do not consider here. Therefore, Eq. \ref{eq:CO-to-H2O} has to be taken only as a first-order approximation, but this approximation is good enough to capture the trend in the data. 
Recall again that $f_{\rm trapCO}$ denotes the \emph{initial} uniform fraction of trapped CO within the amorphous ice matrices (assumed equal for H$_2$O and CO$_2$), whereas $f_{\rm enrichCO}$ denotes the \emph{final} enriched ratio in the remaining H$_2$O amorphous ice.

Figures \ref{fig:comp_size_theory_frac1}-\ref{fig:comp_size_theory_frac0_05} show the CO/H$_2$O abundances predicted by Eq. \ref{eq:CO-to-H2O}.
Different lines depict different formation times for each comet (quicker formation corresponds to greater heating by short-lived radionuclides), and the observations are marked by the full circles, for comparison. A detailed explanation of all the model parameters is given in \cite{MalamudEtAl-2022}. Here we provide a brief explanation. The mineral fraction is introduced to the model since radionuclides are only incorporated into refractory silicate minerals, and not organics. The former might not be present in comets in the same proportion as they are in meteorites, based on which the radionuclide information is derived (we also consider 50\% and 5\% of meteoritic fraction). The pebble radius controls heat and mass transport inside the comet, and we take a binary selection for the pebble radii of 1 mm and 1 cm. This choice roughly represents the lower and upper limits expected in the literature. The permeability $b$ coefficient is related to the Knudsen diffusivity and in turn gas permeability and flow within the comet. We also consider lower and upper limit values. 

It is encouraging that a physical interpretation, however approximated, nicely captures the observed CO/H$_2$O trend. If some comets were to form earlier than others, the various curves span the desired range of the observations. We note that only one set of parameters is adopted for $f_{\rm H2O}$, $f_{\rm CO2}$ and $f_{\rm trapCO}$ in these plots, however our choice of parameters was physically motivated by 67P and explained above. We had also experimented with changing the $f_{\rm H2O}$:$f_{\rm CO2}$ mass ratio considerably, and we always find that as long as H$_2$O is the dominant amorphous ice host, the model keeps capturing the trend with some small variations. It is indeed expected that water is much more prevalent than carbon dioxide in (the bulk of) comet nuclei.

Based on Figures \ref{fig:comp_size_theory_frac1}-\ref{fig:comp_size_theory_frac0_05} one might also speculate further that the peak CO/H$_2$O mass ratio found for mid-to-large sized comets can be more readily explained (a) by early formation; or (b) if the mineral fraction is not as small as 0.05. In addition, the spread in CO/H$_2$O ratios at each size bin may indicate that comets are formed over an extended period of time and/or have a large variation in mineral fraction. Current data does not provide an easy way to differentiate between various options, but there is indeed some indication that comet nuclei have varied mineral fractions. Recently, \textit{Spitzer} remote observations of cometary dust revealed a wide range of amorphous carbon mass fractions spanning 10-90 \%, based on a large set of a few dozen comets \citep{harkerDustPropertiesComets2023}. While the mean value of 54\% indicates that the mass ratio of silicate minerals to organics is, on average, around 1:1, i.e. very similar to comets 67P/C-G \citep{BardynEtAl-2017}, C/2013 US$_{10}$ (Catalina) \citep{WoodwardEtAl-2021} and Halley \citep{FomenkovaChang-1993}, a spread in the mineral fraction is currently supported. We think that this point is very important and we strongly advocate for future study of the ratio of silicates/organics in cometary dust.

An additional point is that low CO/H$_2$O ratios in certain comets should simply be a sign that the CO-enriched layers have been largely removed already, indicating that these comets are more dynamically evolved than their high CO/H$_2$O ratio counterparts. A prominent example is the comet 2P/Encke, which we know has been active for at least a few centuries \citep{MarsdenSekanina-1974}. Thus, at each size bin, if a comet was observed to have a CO/H$_2$O ratio in the upper part of the spread, it might also be considered a sign of having a relatively fresh dynamical origin.

These plots also reveal that the pebble size and the $b$ coefficient are of lesser importance compared to other parameters, echoing the conclusions already suggested by \cite{MalamudEtAl-2022}.

\subsection{Other species less volatile than CO}
An intriguing result is the presence of a significant correlation between size and CO/H$_2$O ratio, while observing a lack of correlation, or a weaker correlation among the ratios of some other volatiles. There are two possible explanations. The first might be a simple lack of observations, given that we have a rather small statistical sample for many species. The second explanation is much more fundamental. 

We hypothesise that most volatiles released from their amorphous ice hosts would be buried deeper inside the comet, hiding from our sight as only the outermost surface layers are being eroded through activity. Of the many hyper- and super-volatiles that we consider in this work (incorporated into our sample because it is possible to observe them via telescope surveys), only CO and CH$_4$ have extremely low sublimation/deposition temperatures \citep{womackCOOtherVolatiles2017}. That being the case, the other hyper-volatiles would encounter temperatures that should lead to their re-incorporation into amorphous ice before CO and CH$_4$, and the super-volatiles could even deposit as pure ice condensates, when they have characteristic sublimation temperatures that are higher than the crystallisation temperatures of the host amorphous ice. The resulting outcome is that they are buried deeper within the comet, somewhere between the surface and the location of their release by radiogenic heating.

The explanation is qualitatively straightforward and depends on temperature. The surface temperature of an active comet in the inner Solar System is determined by its exact heliocentric distance. However, below the skin depth the temperature is much colder -- and is a relic of its previous location before it was perturbed into the inner Solar System. For example, if it came from the Kuiper belt, this temperature is certainly lower than the temperature of crystallisation of amorphous ice, but often not lower than the deposition temperature of CO and CH$_4$ \citep{LisseEtAl-2021,ParhiPrialnik-2023}. We therefore expect the main volatiles which are not CO or CH$_4$ to be buried deeper below the surface of the comet. Their release location within the comet corresponds to the inner cubic-amorphous ice boundary. In this context, inner refers to the boundary that forms as a result of an internal temperature gradient due to radiogenic heating from within. It should not be confused with the (outer) amorphous-cubic boundary that might form externally by a heat wave propagating inwards from the insolated comet surface (triggering crystallisation). Their exact burial location is a function of the characteristic deposition temperature (e.g., HCN would be buried deeper than C$_2$H$_6$).

One interpretation of our results therefore might be that most active comets are still eroding their outermost layers, which are not significantly enriched by any of these less volatile gases. It should be noted that some species, such as HCN (see Figure \ref{fig:comp_size_dist_check_bin_parents}) do exhibit a much more moderate slope, in contrast to the steep slope we obtained for CO. This could still be reconciled with our hypothesis, since comet nuclei are neither spheri-symmetric nor is their surface eroded homogeneously. In reality only small fractions of the comet surface might be eroded to expose deeper layers, so the integrated result for the entire comet circumference gives the outcome that larger nuclei also release more of these gases, but the slope is moderated by these geometrical factors.

It begs the question however, why is CH$_4$ not giving us the same slope as CO? While CH$_4$ indeed sublimates at a slightly higher temperature, the small difference is not a likely explanation. A more robust explanation involves the trapping efficiency of these two gases in amorphous ice. \cite{BarNunEtAl-1988} have shown that when both CO and CH$_4$ gases are streamed into a pre-deposited amorphous ice, the entrapment of CH$_4$ is 150 times more efficient than that of CO \footnote{This result is for a temperature of 50 K and when ample CO and CH$_4$ gas is used in the experiment (CH$_4$ and CO molecules differ in both their size and energy of interaction with the host ice due to their polarity, leading to the greater ease of trapping of CH$_4$).}. For comets this would mean that sequential deposition of these two gases when they are released together from the underlying crystallising ice, leads to the preferential trapping of CH$_4$, while CO continues to flow until it sees no competition from CH$_4$. The expectation is therefore that CO lies closer to the surface, while CH$_4$ is buried deeper. A final and trivial explanation is that it is simply due to the small amount of data available for CH$_4$, only 10 comets in total, and the larger dispersion of the measurements. This latter comment is however true, in general, for virtually all the other species as well.

\subsection{A note about cometary outbursts}
We briefly note that cometary outbursts at large heliocentric distances, post perihelion passage, are often associated with a heat wave propagating inwards from an insolated surface, to a deeply buried crystalline-amorphous ice boundary. Upon reaching this boundary with sufficient energy, crystallisation releases latent heat as well as entrapped gases. Latent heat is important because it can trigger further crystallisation, a process however which cannot continue indefinitely, since the eventual sublimation of ice absorbs a large amount of energy. Trapped gases are important since they are effectively the cause of the outbursts. On release, these gases lead to build-up of pressure, and only when this pressure exceeds the tensile strength of the ambient solid materials, it leads to cracking and possibly more rapid expulsion of gas \citep{PrialnikBarNun-1987,PrialnikBarNun-1992}.

The current study does not alter this basic picture in any way. However, we have envisioned here the formation of localised spots of amorphous ice, highly enriched with various high-volatility species. These are buried at various distances from the surface, based on the sublimation properties of each particular volatile, and after having concentrated them from the bulk of the comet. For a spherically-symmetric nucleus, an onion-like stratification might be expected. Yet in reality comet nuclei have irregular shapes, which means that these pockets might be rather more sporadically placed. The significance in relation to our study, is simply to justify a large fraction of high-volatility species required for an enhanced outburst.

\subsection{Other predictions}
In Section \ref{S:intro} we also presented two additional predictions besides the general size-composition correlation. We suggested that different dynamical classes of comets might exhibit significant differences in their size-composition correlation, as a result of differences in their erosional state. Our findings in Section \ref{S:Findings} however cannot confirm this hypothesis. We believe that this could certainly be due to insufficient statistics to drive a conclusion.

We additionally speculated that dynamically evolved active comets are smaller than their long period counterparts. Figure \ref{fig:comp_size_dist_check_bin_parents} shows that this is strictly correct (the triangle positions tend to be located more to the right, wheres the squares are positioned more to the left, within each sub-plot). However, we caution that this result might also simply be an observational bias -- ECs are generally less active and can be observed inactive closer than NICs and therefore it is preferentially easier to work out the size for them. At this time we require more data to confirm all our other predictions.

\subsection{Future observations}

In order to improve comparisons between models and comet composition in the future, more data are necessary. In particular, we need larger homogeneous datasets for comet sizes but also comet composition information where the abundance of species is measured simultaneously (or as close to that as possible), at different distances from the Sun, and in a consistent way in terms of observational techniques and models used to derive production rates. 
With CO and CO$_2$ being the most abundant volatiles in comets, abundance measurements for these elements for a larger number of comets for which we have size information is particularly critical. Additional measurements of CH$_4$ and N$_2$ abundances would also be very valuable, as these species have sublimation temperatures close to that of CO. N$_2$ abundances were not presented in this work as they are extremely difficult to measure. This species was only detected \textit{in situ} in the coma of comet 67P/C-G by the ROSINA instrument onboard the Rosetta mission \citep{Rubin2015}. While N$_2$ itself is difficult to detect, N$_2^+$ can be observed at optical wavelengths from the ground and has been observed in a handful of comets \citep{Korsun2008,Ivanova2018,Cochran2018,Opitom2019}. This can then be used to infer the abundance of N$_2$ in comets. More N$_2$ or N$_2^+$ measurements would be very valuable in the future. In general, this type of work would particularly benefit from a more substantial sample of large comets for which composition information is available. Indeed, this database only contains a handful of comets larger than 5~km with composition information.

\section{Conclusions}\label{S:Conclusions}
In this manuscript, following predictions of a model from \cite{MalamudEtAl-2022}, we gathered a large number of literature data to search for correlations between the size and composition of comets.

\begin{itemize}
    \item For the dataset we have gathered we found  a statistically significant correlation between the CO/H$_2$O abundance ratios and the sizes of both ecliptic and nearly isotropic comets.
     This trend persists even when selecting for comets observed within 2 au from the Sun, indicating it is not driven by changes in the abundance ratios with heliocentric distance. 
    \item A weaker correlation was also observed for some other volatile species, however further tests indicate that our analysis would critically benefit from obtaining a bigger statistical sample in the future.
    \item We do not see any strong correlations for daughter species.
    \item We do not see a similarly strong correlation for CH$_4$, in spite of having a comparable sublimation temperature to that of CO.
\end{itemize}

We develop a simple theoretical framework based on the \cite{MalamudEtAl-2022} model, with which we rather accurately obtain the CO/H$_2$O abundance-to-size trend in our observed data. In this framework we consider CO to migrate from the bulk of the nucleus outwards, becoming entrapped within its outer amorphous ice layers, and in turn enhancing their CO-enrichment as a function of the nucleus size.

		We emphasise that the correlation between \ch{CO}/\ch{H2O} abundance and size appears to be robust for the dataset we have presented, where we have gathered together a wide range of measurements from the available literature. 
		However this dataset is ultimately limited by its size and also by the intrinsic difficulties in accurately determining the physical properties of cometary nuclei from a variety of observations and techniques.
This study would have benefited from, and therefore strongly motivates, a larger homogeneous set of composition measurements in the future, in particular for highly volatile species like CO, CH$_4$, or N$_2$.
State of the art observatories, e.g.\ JWST and the upcoming Vera C. Rubin Observatory and ELT, could provide more opportunities to characterise the physical properties of cometary nuclei, especially the sizes of long period comets which are otherwise sparse in the literature.

\section*{Acknowledgements}
The authors would like to thank the referee for a thorough review that helped to improve the work.
We wish to thank Diana Laufer for providing information about the relative entrapment efficiency of CO versus CH$_4$ in amorphous water ice. 
We also thank Rosita Kokotanekova for valuable input on the selection of comet sizes from literature sources.
UM and HBP acknowledge support by the Niedersächsisches Vorab in the framework of the research cooperation between Israel and Lower Saxony under grant ZN 3630 and grant by MOST-space.
CO and JR ackowledge the support of the Royal Society. 
This work made use of the NASA SBDB service and PDS datasets \verb|ear-c-phot-5-rdr-lowell-comet-db-pr-v1.0| \citep{osipLowellObservatoryCometary2003} and \verb|urn:nasa:pds:compil-comet:nuc_properties::1.0| \citep{barnesPropertiesCometNuclei2010}.
The following software packages were used in this work: 
\verb|matplotlib| \citep{hunterMatplotlib2DGraphics2007}, 
\verb|numpy| \citep{harrisArrayProgrammingNumPy2020}, 
\verb|scipy| \citep{virtanenSciPyFundamentalAlgorithms2020}, 
\verb|pandas| \citep{mckinneyDataStructuresStatistical2010}, 
\verb|pds3| \citep{kelley_pds3},
\verb|pds4_tools| \citep{nagdimunov_pds4_tools},
\verb|astropy| \citep{astropycollaborationAstropyProjectSustaining2022},
\verb|astroquery| \citep{ginsburgAstroqueryAstronomicalWebquerying2019},
\verb|sbpy| \citep{mommertSbpyPythonModule2019} and
\verb|camelot| \citep{mehta_camelot}.
For the purpose of open access, the authors have applied a Creative Commons Attribution (CC BY) licence to any Author Accepted Manuscript version arising from this submission. 
\section*{Data Availability}

The dataset constructed in this work is available online as Supple-
mentary data and also from the University of Edinburgh DataShare
repository (\url{https://doi.org/10.7488/ds/7723}).
The data compiled in this work may also be obtained via reasonable e-mail request to the lead authors.



\bibliographystyle{mnras}
\bibliography{zotero_library,extra_refs} 



\section*{Supporting Information}

Supplementary data are available at \href{https://academic.oup.com/mnras/article-lookup/doi/10.1093/mnras/stae881#supplementary-data}{\textit{MNRAS}} online.

\noindent Please note: Oxford University Press is not responsible for the content
or functionality of any supporting materials supplied by the authors.
Any queries (other than missing material) should be directed to the
corresponding author for the article.


\appendix

\section{Comet Radius Sources}
\label{Appendix:Radius_Sources}
\begin{table}
\begin{tabular}{llp{2.5cm}}
\toprule
Source & Number & Method \\
\midrule
\cite{lamySizesShapesAlbedos2004} & 22 & Compilation (Photometric/Thermal) \\
\cite{bauerDebiasingNEOWISECryogenic2017} & 20 & Thermal \\
\cite{fernandezThermalPropertiesSizes2013} & 15 & Thermal \\
\cite{tancrediCatalogObservedNuclearmagnitudes2000} & 8 & Photometric \\
\cite{lamyPropertiesNucleiComae2009} & 7 & Photometric \\
\cite{lamyPropertiesNucleiComae2011} & 2 & Photometric \\
\cite{rosserBehavioralCharacteristicsCO2018} & 2 & Thermal \\
\cite{Scotti1994} & 1 & Photometric \\
\cite{Weissman2008} & 1 & Photometric \\
\cite{boehnhardtNucleiComets26P1999} & 1 & Photometric \\
\cite{boehnhardtPhotometryPolarimetryNucleus2008} & 1 & Photometric \\
\cite{boehnhardtVLTObservationsComet2002} & 1 & Photometric \\
\cite{eisnerPropertiesBareNucleus2019} & 1 & Photometric \\
\cite{mazzottaepifaniDistantActivityShort2008} & 1 & Photometric \\
\cite{harmonComet8PTuttle2008} & 1 & Radar \\
\cite{harmonRadarDetectionNucleus1997} & 1 & Radar \\
\cite{lejolyRadialDistributionDust2022} & 1 & Radar \\
\cite{burattiDeepSpacePhotometry2004} & 1 & Spacecraft \\
\cite{farnhamResolvedNucleusComet2017} & 1 & Spacecraft \\
\cite{jordaGlobalShapeDensity2016} & 1 & Spacecraft \\
\cite{sekaninaModelingNucleusJets2004a} & 1 & Spacecraft \\
\cite{thomasNucleusComet9P2013a} & 1 & Spacecraft \\
\cite{thomasShapeDensityGeology2013} & 1 & Spacecraft \\
J. Bauer (unpubl. data) & 1 & Thermal \\
\cite{boissierMillimetreContinuumObservations2013} & 1 & Thermal \\
\cite{fomenkovaMidInfraredObservationsNucleus1995} & 1 & Thermal \\
\cite{groussinNucleusComet19832010} & 1 & Thermal \\
\cite{pittichovaGROUNDBASEDOPTICALSPITZER2008} & 1 & Thermal \\
\bottomrule
\end{tabular}

\caption{Table summarising the sources used for comet radii in the combined composition-size dataset.
	For each source we state the number of comet size measurements used in this study and also the general method by which these sizes were obtained.
These counts include the objects with only limits on size.
    }
\label{tab:sizes} 
\end{table}

In general it was a simple procedure to apply our guidelines for comet size selection from a range of literature sources (Section \ref{SS:size_selection_criteria}).
However for a small number of objects there were conflicting measurements or more detailed circumstances that complicated the size selection, which we describe in more detail here.

For comets 7P, 17P, 37P, 64P, 109P and 116P we rejected smaller nucleus sizes measured by photometric methods in favour of larger sizes from thermal observations \citep[primarily from][]{fernandezThermalPropertiesSizes2013,bauerDebiasingNEOWISECryogenic2017}.
With the exception of the thermal size of 109P by \cite{fomenkovaMidInfraredObservationsNucleus1995} these measurements made use of more modern data than the earlier photometric observations.
Furthermore our selected sources provided uncertainties on the nucleus size whereas this was not always the case for the photometric measurements \citep[most of which were from][]{lamySizesShapesAlbedos2004}.
Often the lower photometric nucleus size was consistent with the larger thermal estimate when the uncertainties were taken into account.
Furthermore, for most of these objects our selected size is the same as that selected by the literature compilation of \cite{knightPhysicalSurfaceProperties2023}.

Likewise, C/2009 P1 has a radius measurement of $r=13.5 \pm 2.5\ \si{km}$ \citep{bauerDebiasingNEOWISECryogenic2017} which is in conflict with an upper limit of $r<5.6\ \si{km}$ from a non-detection with IRAM on 04/03/2012 by \cite{boissierMillimetreContinuumObservations2013}.
These measurements could be explained if the nucleus of C/2009 P1 is elongated and presented a smaller cross-section during the IRAM observations.
As such we combine the two measurements into a single size estimate by taking the mean and using the range to define the uncertainty such that $r = 9.6 \pm 4\ \si{km}$.

A thermal size of $r = 2.465 \pm 0.135\ \si{km}$ was measured for comet 10P from NEOWISE data \citep{bauerDebiasingNEOWISECryogenic2017}, however, this is smaller than a photometric size of $r = 5.98 \pm 0.04\ \si{km}$ from HST observations \citep{lamyPropertiesNucleiComae2011}.
It would appear that the comet was active and trailed in the NEOWISE data which may have led to a smaller size estimate (R. Kokotanekova, personal communication).
Therefore we followed \cite{knightPhysicalSurfaceProperties2023} in selecting the larger photometric size.

The nucleus of 45P was imaged by radar and found to have a radius in the range $r = 0.6 - 0.65\ \si{km}$ \citep{disantiComet45PHondaMrkosPajdusakova2017,lejolyRadialDistributionDust2022}.
In order to include this object in our analysis we used the centroid of this range for the radius and the lower/upper bounds for the uncertainty, resulting in $r = 0.625 \pm 0.025\  \si{km}$.
In addition, a personal communication mentioned in \cite{lejolyRadialDistributionDust2022} describes a radar diameter of $1.4\ \si{km}$ for 46P.
Although we consider radar measurements to be preferable to other remote observations we could find no further details of this measurement in the available literature.
Therefore we used the radius of $r = 0.56 \pm 0.04\ \si{km}$ from \cite{boehnhardtVLTObservationsComet2002} which was also selected by \cite{knightPhysicalSurfaceProperties2023}.

In the compilation of \cite{lisTerrestrialDeuteriumtohydrogenRatio2019} the nucleus radius of comet 73P is given as $r = 1.10 \pm 0.03\ \si{km}$.
This measurement was ultimately derived from photometric observations by \cite{boehnhardtNucleiComets26P1999} which were made before fragmentation of the comet nucleus in 1995.
However, in this work it is clearly shown that comet 73P was active at the time of these observations and \cite{boehnhardtNucleiComets26P1999} stated that the nuclear radius of 73P must be $<1.1\ \si{km}$.
We did not account for size limits in our methods, and given the likelihood of an earlier fragmentation event \citep{schullerCometSchwassmannWachmann1930} this object was not included in our analysis.

\begin{figure}
\includegraphics[width=\columnwidth]{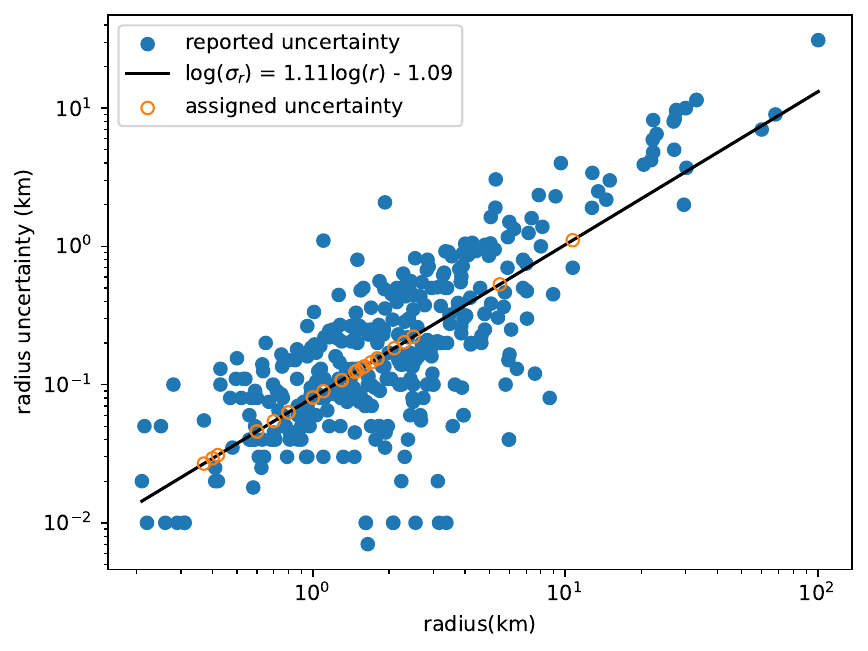}   
\caption{Relation between the reported comet nucleus radius and associated uncertainty from the sources searched in this work.
We have fit a linear relation in log-log space to allow us to assign approximate uncertainties to radius measurements missing them.
}
\label{fig:radius_uncertainty}
\end{figure}

In Table \ref{tab:sizes} we present a summary of the different sources used to obtain comet radii for this analysis.
Several sources present comet radii without formal uncertainties, however there is a clear power law relation between radius and the associated uncertainty (Figure \ref{fig:radius_uncertainty}).
We use this relation to assign approximate uncertainties to radius measurements without them in the compiled dataset.

\section{Composition - Size Data Table}
\label{Appendix:composition_data_table}

Here we present a sample of the complete dataset used in this study, compiled as described in Section \ref{S:DataSet}.
Each row contains a single abundance measurement of species X for a particular comet, where abundance is given with respect to either \ch{H2O} or \ch{CN}.
The circumstances of the compositional observation are provided, and the literature sources of both the composition and size measurement are stated.
The dataset contains 909 unique species measurements for 96 unique comets with sizes; this includes measurements of composition/size with limits and comets with a known fragmentation history.
In our analysis we rejected limits and split comets and were left with 710 composition measurements for 69 comets.
Table \ref{tab:composition_size_sample_table} gives a sample of selected rows and columns from the full dataset, which is available online as Supplementary data and \href{https://doi.org/10.7488/ds/7723}{at this link}.

\begin{landscape}

\begin{table}
\begin{tabular}{llrlrrlrrlrrl}
\toprule
Type & Designation & Number & Name & Date(MJD) & $r_h$(au) & X & X/\ch{H2O} & $\sigma_\textrm{X/H2O}$ & Composition Source & $r$(km) & $\sigma_r$(km) & Radius Source \\
\midrule
P &  & 49 & Arend-Rigaux & 46035.0 & 1.56 & \ch{Af$\rho$} & 5.881e-26 & 4.7e-27 & \cite{ahearnEnsemblePropertiesComets1995} & 3.21 & 0.37 & \cite{bauerDebiasingNEOWISECryogenic2017} \\
P &  & 49 & Arend-Rigaux & 46035.0 & 1.56 & \ch{C2} & 1.817e-03 & 1.8e-04 & \cite{ahearnEnsemblePropertiesComets1995} & 3.21 & 0.37 & \cite{bauerDebiasingNEOWISECryogenic2017} \\
P &  & 49 & Arend-Rigaux & 46035.0 & 1.56 & \ch{C3} & 2.567e-04 & 2.6e-05 & \cite{ahearnEnsemblePropertiesComets1995} & 3.21 & 0.37 & \cite{bauerDebiasingNEOWISECryogenic2017} \\
P &  & 49 & Arend-Rigaux & 46035.0 & 1.56 & \ch{CN} & 2.087e-03 & 1.3e-04 & \cite{ahearnEnsemblePropertiesComets1995} & 3.21 & 0.37 & \cite{bauerDebiasingNEOWISECryogenic2017} \\
P &  & 49 & Arend-Rigaux & 46035.0 & 1.56 & \ch{NH} & 1.620e-03 & 2.6e-03 & \cite{ahearnEnsemblePropertiesComets1995} & 3.21 & 0.37 & \cite{bauerDebiasingNEOWISECryogenic2017} \\
P &  & 59 & Kearns-Kwee & 45599.0 & 1.00 & \ch{C2} & 3.626e-03 & 3.1e-03 & \cite{cochranThirtyYearsCometary2012} & 0.79 & 0.03 & \cite{lamyPropertiesNucleiComae2009} \\
P &  & 59 & Kearns-Kwee & 45599.0 & 1.00 & \ch{C3} & 4.461e-04 & 4.0e-04 & \cite{cochranThirtyYearsCometary2012} & 0.79 & 0.03 & \cite{lamyPropertiesNucleiComae2009} \\
P &  & 59 & Kearns-Kwee & 45599.0 & 1.00 & \ch{NH} & 1.046e-02 & 7.0e-03 & \cite{cochranThirtyYearsCometary2012} & 0.79 & 0.03 & \cite{lamyPropertiesNucleiComae2009} \\
P &  & 59 & Kearns-Kwee & 46578.0 & 2.23 & \ch{Af$\rho$} & 3.710e-26 & 3.3e-27 & \cite{ahearnEnsemblePropertiesComets1995} & 0.79 & 0.03 & \cite{lamyPropertiesNucleiComae2009} \\
P &  & 59 & Kearns-Kwee & 46578.0 & 2.23 & \ch{CN} & 1.735e-03 & 1.7e-04 & \cite{ahearnEnsemblePropertiesComets1995} & 0.79 & 0.03 & \cite{lamyPropertiesNucleiComae2009} \\
P &  & 65 & Gunn & 46547.0 & 2.64 & \ch{Af$\rho$} & 9.537e-27 & 1.7e-27 & \cite{ahearnEnsemblePropertiesComets1995} & 4.80 & 1.02 & \cite{bauerDebiasingNEOWISECryogenic2017} \\
P &  & 65 & Gunn & 46547.0 & 2.64 & \ch{C2} & 1.735e-04 & 4.7e-05 & \cite{ahearnEnsemblePropertiesComets1995} & 4.80 & 1.02 & \cite{bauerDebiasingNEOWISECryogenic2017} \\
P &  & 65 & Gunn & 46547.0 & 2.64 & \ch{C3} & 2.135e-04 & 1.1e-04 & \cite{ahearnEnsemblePropertiesComets1995} & 4.80 & 1.02 & \cite{bauerDebiasingNEOWISECryogenic2017} \\
P &  & 65 & Gunn & 46547.0 & 2.64 & \ch{CN} & 4.780e-04 & 8.6e-05 & \cite{ahearnEnsemblePropertiesComets1995} & 4.80 & 1.02 & \cite{bauerDebiasingNEOWISECryogenic2017} \\
P &  & 65 & Gunn & 46547.0 & 2.64 & \ch{NH} & 8.698e-04 & 3.3e-03 & \cite{ahearnEnsemblePropertiesComets1995} & 4.80 & 1.02 & \cite{bauerDebiasingNEOWISECryogenic2017} \\
P &  & 88 & Howell & 44728.0 & 2.09 & \ch{Af$\rho$} & 5.363e-26 & 4.3e-27 & \cite{ahearnEnsemblePropertiesComets1995} & 1.00 &  & \cite{tancrediCatalogObservedNuclearmagnitudes2000} \\
P &  & 88 & Howell & 44728.0 & 2.09 & \ch{C2} & 2.396e-03 & 2.4e-04 & \cite{ahearnEnsemblePropertiesComets1995} & 1.00 &  & \cite{tancrediCatalogObservedNuclearmagnitudes2000} \\
P &  & 88 & Howell & 44728.0 & 2.09 & \ch{C3} & 1.583e-04 & 1.6e-05 & \cite{ahearnEnsemblePropertiesComets1995} & 1.00 &  & \cite{tancrediCatalogObservedNuclearmagnitudes2000} \\
P &  & 88 & Howell & 44728.0 & 2.09 & \ch{CN} & 2.880e-03 & 2.3e-04 & \cite{ahearnEnsemblePropertiesComets1995} & 1.00 &  & \cite{tancrediCatalogObservedNuclearmagnitudes2000} \\
P &  & 88 & Howell & 55015.1 & 1.74 & \ch{CO2} & 2.495e-01 & 5.0e-02 & \cite{ootsuboAKARINEARINFRAREDSPECTROSCOPIC2012} & 1.00 &  & \cite{tancrediCatalogObservedNuclearmagnitudes2000} \\
... & ... & ... & ... & ... & ... & ... & ... & ... & ... & ... & ... & ... \\ 
C & 1983 J1 &  & Sugano-Saigusa-Fujikawa & 45455.0 & 0.74 & \ch{Af$\rho$} & 3.797e-27 & 1.9e-28 & \cite{ahearnEnsemblePropertiesComets1995} & 0.37 &  & \cite{lamySizesShapesAlbedos2004} \\
C & 1983 J1 &  & Sugano-Saigusa-Fujikawa & 45455.0 & 0.74 & \ch{C2} & 5.881e-03 & 1.8e-04 & \cite{ahearnEnsemblePropertiesComets1995} & 0.37 &  & \cite{lamySizesShapesAlbedos2004} \\
C & 1983 J1 &  & Sugano-Saigusa-Fujikawa & 45455.0 & 0.74 & \ch{C3} & 6.018e-05 & 3.6e-06 & \cite{ahearnEnsemblePropertiesComets1995} & 0.37 &  & \cite{lamySizesShapesAlbedos2004} \\
C & 1983 J1 &  & Sugano-Saigusa-Fujikawa & 45455.0 & 0.74 & \ch{CN} & 2.947e-03 & 8.8e-05 & \cite{ahearnEnsemblePropertiesComets1995} & 0.37 &  & \cite{lamySizesShapesAlbedos2004} \\
C & 1983 J1 &  & Sugano-Saigusa-Fujikawa & 45455.0 & 0.74 & \ch{NH} & 2.627e-03 & 1.8e-04 & \cite{ahearnEnsemblePropertiesComets1995} & 0.37 &  & \cite{lamySizesShapesAlbedos2004} \\
C & 2006 W3 &  & Christensen & 54909.9 & 3.40 & \ch{CO2} & 7.204e-01 & 1.5e-01 & \cite{ootsuboAKARINEARINFRAREDSPECTROSCOPIC2012} & 21.88 & 4.20 & \cite{bauerDebiasingNEOWISECryogenic2017} \\
C & 2006 W3 &  & Christensen & 54909.9 & 3.40 & \ch{CO} & 2.296e+00 & 4.6e-01 & \cite{ootsuboAKARINEARINFRAREDSPECTROSCOPIC2012} & 21.88 & 4.20 & \cite{bauerDebiasingNEOWISECryogenic2017} \\
C & 2006 W3 &  & Christensen & 55073.3 & 3.22 & \ch{CH3OH} & 3.355e-02 & 6.7e-03 & \cite{bockelee-morvanStudyDistantActivity2010} & 21.88 & 4.20 & \cite{bauerDebiasingNEOWISECryogenic2017} \\
C & 2006 W3 &  & Christensen & 55073.3 & 3.22 & \ch{CS} & 1.118e-03 & 4.5e-04 & \cite{bockelee-morvanStudyDistantActivity2010} & 21.88 & 4.20 & \cite{bauerDebiasingNEOWISECryogenic2017} \\
C & 2006 W3 &  & Christensen & 55073.3 & 3.22 & \ch{H2S} & 2.237e-02 & 2.2e-03 & \cite{bockelee-morvanStudyDistantActivity2010} & 21.88 & 4.20 & \cite{bauerDebiasingNEOWISECryogenic2017} \\
C & 2006 W3 &  & Christensen & 55073.3 & 3.22 & \ch{HCN} & 3.579e-03 & 8.5e-04 & \cite{bockelee-morvanStudyDistantActivity2010} & 21.88 & 4.20 & \cite{bauerDebiasingNEOWISECryogenic2017} \\
\bottomrule
\end{tabular}

\caption{
	A sample of selected rows and columns from the comet composition-size dataset used in this work.
The columns presented are the comet identifiers (type, designation and number), details of the compositional measurement (date and heliocentric distance $r_h$) and the abundance of species X relative to water (X/\ch{H2O} and the corresponding uncertainty $\sigma_\textrm{X/H2O}$ if available) with the source of the compositional measurement.
Size information is provided as radius, $r$, with uncertainty $\sigma_r$ (if available) alongside the literature source of the measurement.
The full table with all rows and additional columns is available online as Supplementary data and \href{https://doi.org/10.7488/ds/7723}{at this link}.
}
\label{tab:composition_size_sample_table}
\end{table}

\end{landscape}

\section{Additional composition data}
\label{Appendix:additional}

\begin{figure*}
	\begin{tabular}{ccc}
		\includegraphics[width=0.3\textwidth]{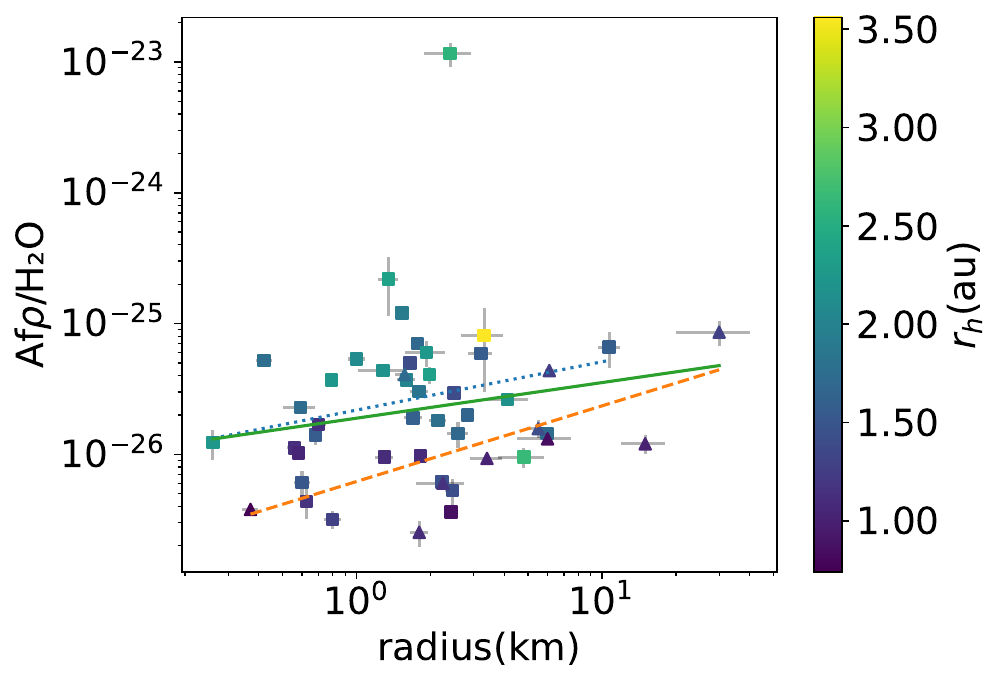} &   
		\includegraphics[width=0.3\textwidth]{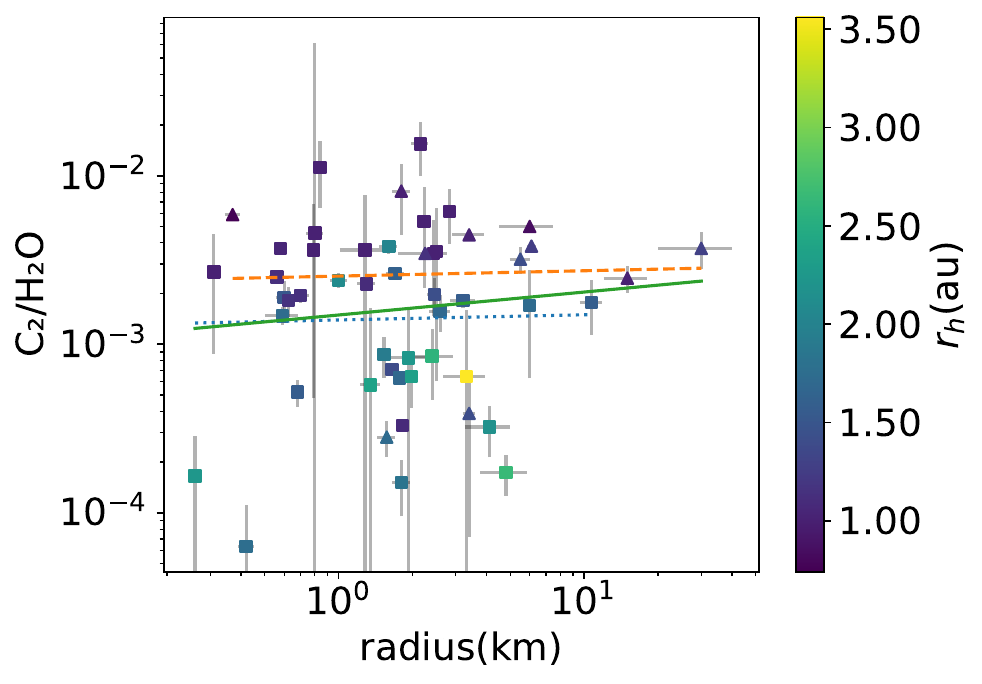} &
		\includegraphics[width=0.3\textwidth]{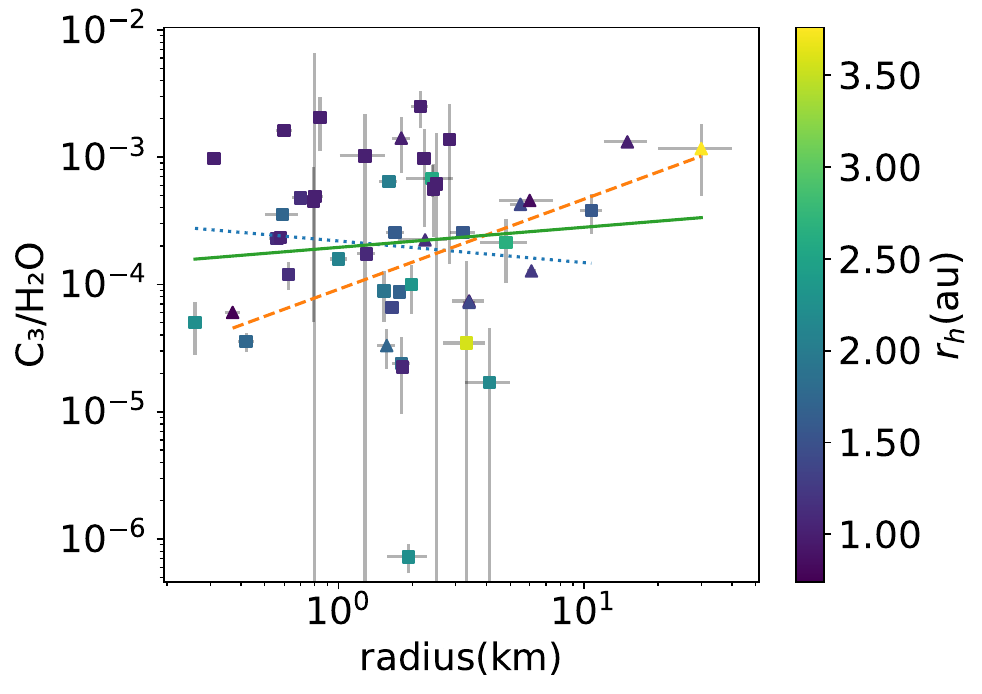} \\   
		\includegraphics[width=0.3\textwidth]{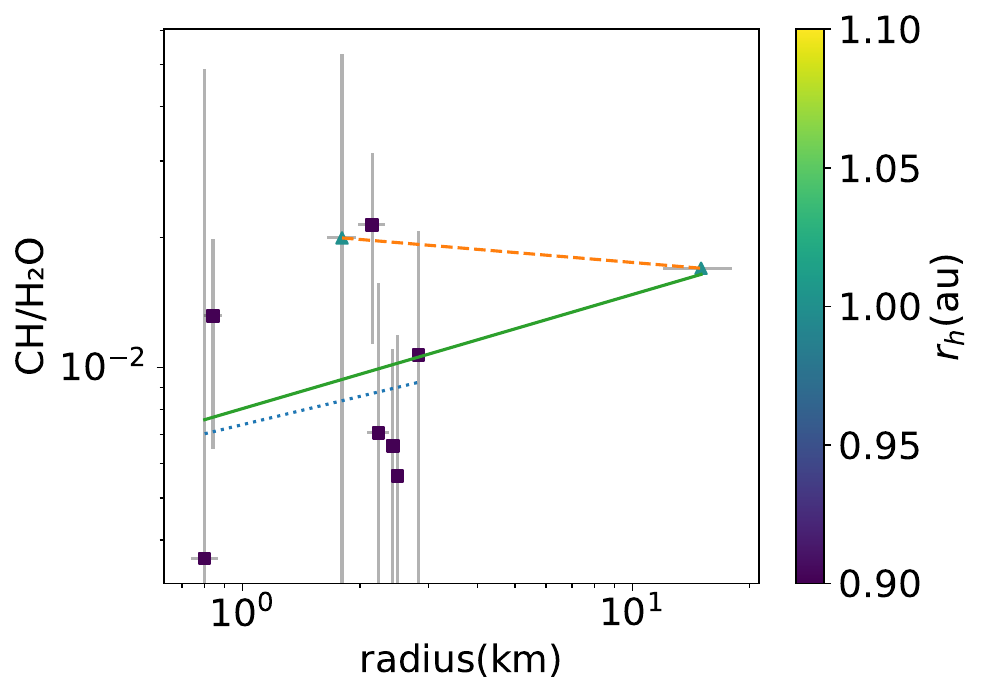} &   
		\includegraphics[width=0.3\textwidth]{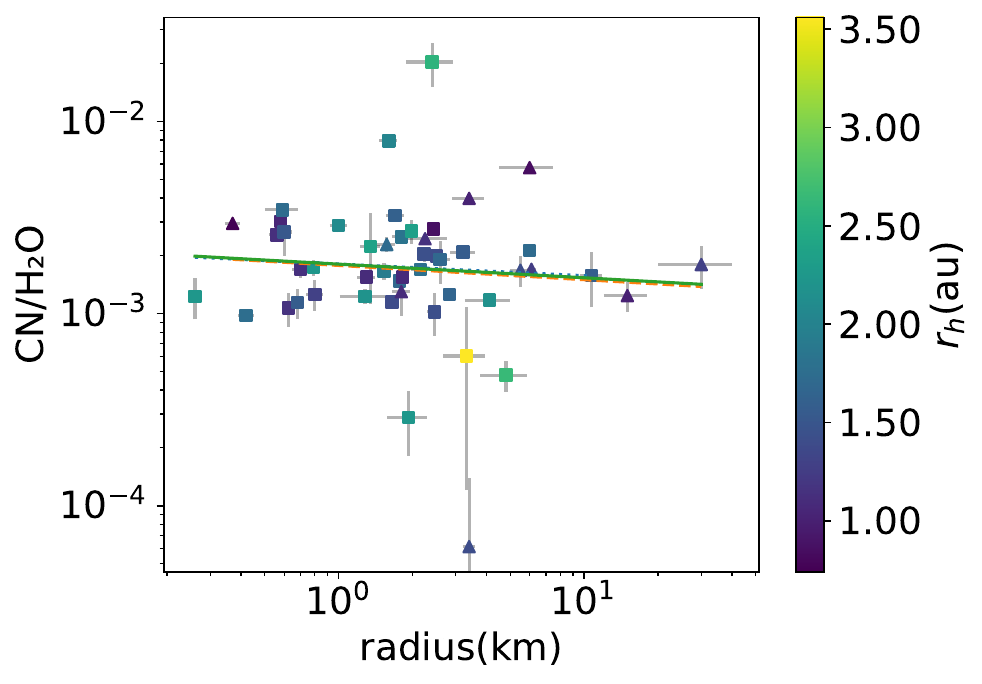} &
		\includegraphics[width=0.3\textwidth]{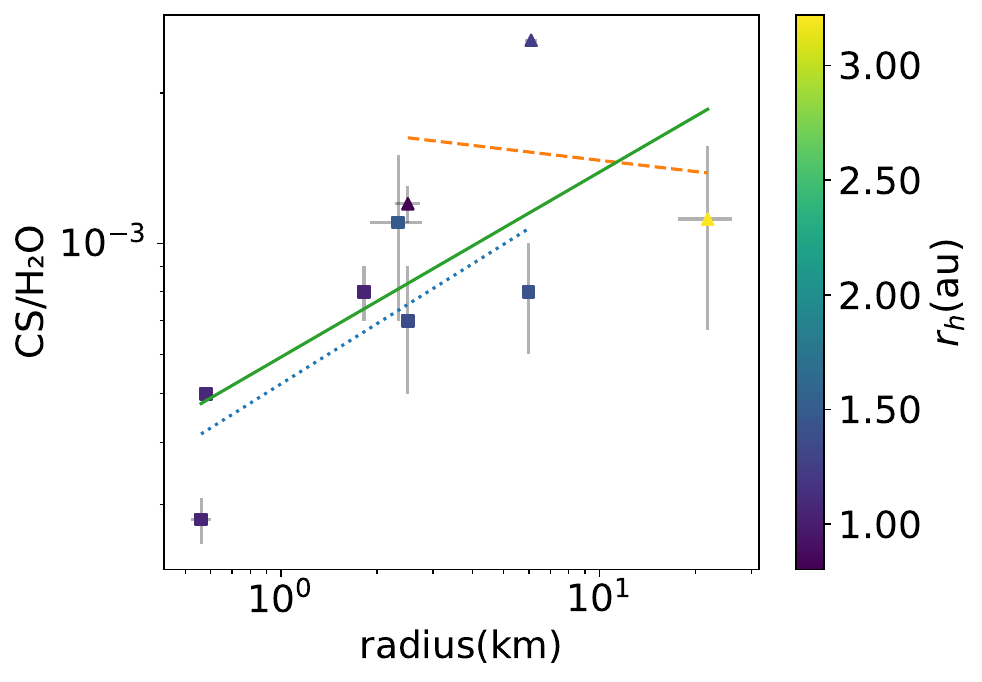} \\
		\includegraphics[width=0.3\textwidth]{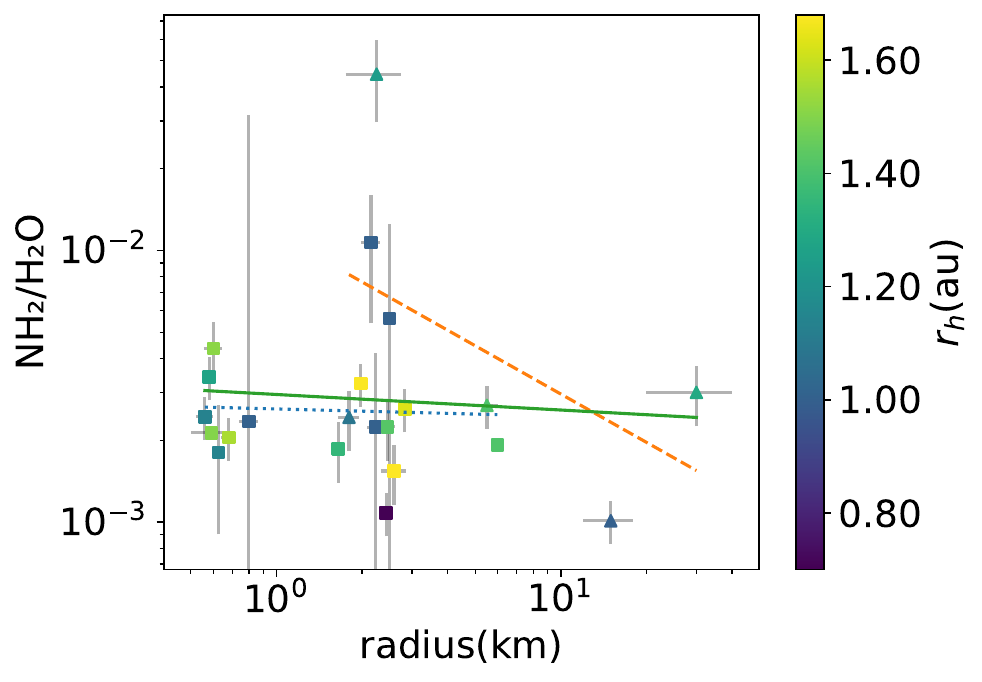} &
		\includegraphics[width=0.3\textwidth]{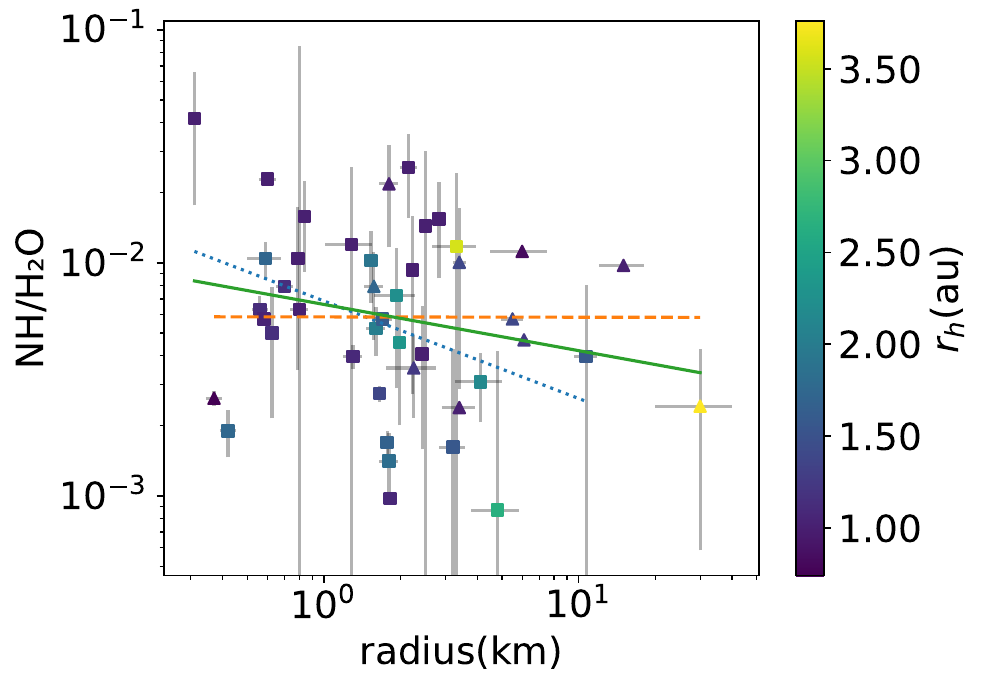} &
	\end{tabular}
	\caption{Log scale plots showing the relation between comet composition (of various daughter species relative to \ch{H2O}) and radius of the nucleus.
		The meanings of the markers and lines are the same as Figure \ref{fig:comp_size_H2O_parents} and the corresponding Pearson correlation coefficients are provided in Table \ref{tab:correlations2}.
	}
	\label{fig:comp_size_H2O_daughters}
\end{figure*}

This annex presents additional figures for the composition vs radius of our sample of comets for daughter species, compared to the parent species considered in the main analysis.
Figure \ref{fig:comp_size_H2O_daughters} shows the daughter species abundance (relative to \ch{H2O}) as a function of comet radius.
The full results of the Pearson correlation tests for each daughter species, and the dynamical sub-populations in the dataset, are provided in Table \ref{tab:correlations2}.

\begin{figure*}
	\begin{tabular}{ccc}
		\includegraphics[width=0.3\textwidth]{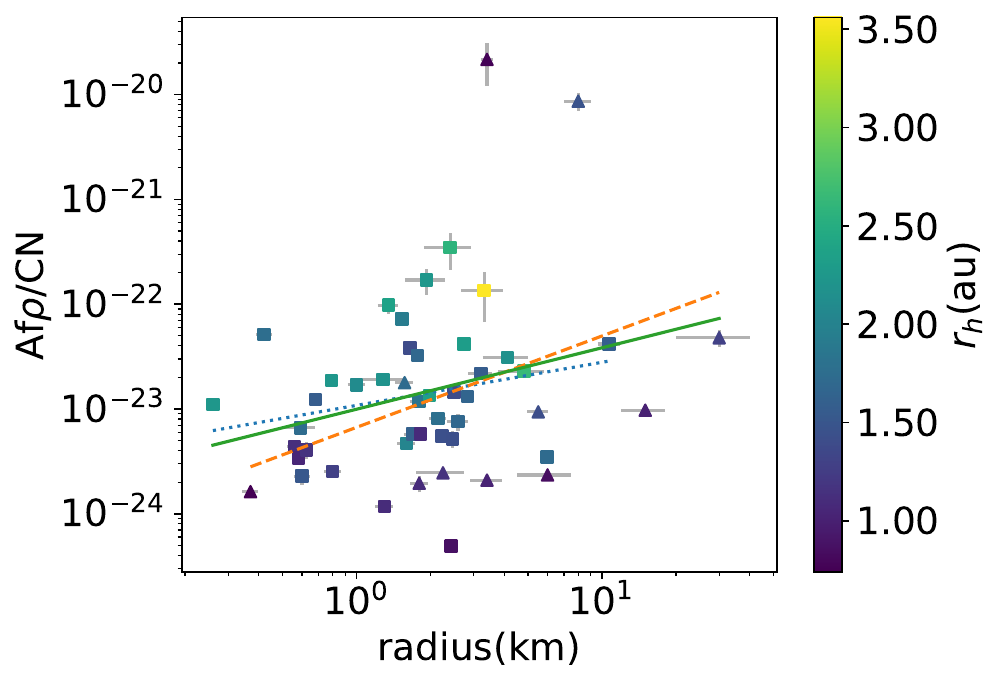} &   \includegraphics[width=0.3\textwidth]{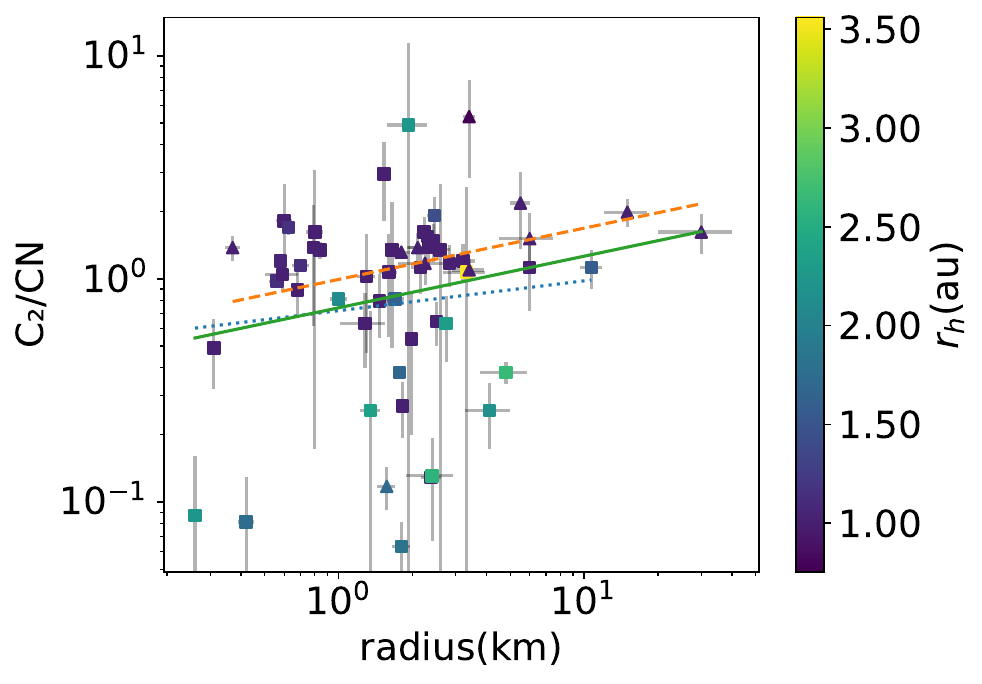} &
		\includegraphics[width=0.3\textwidth]{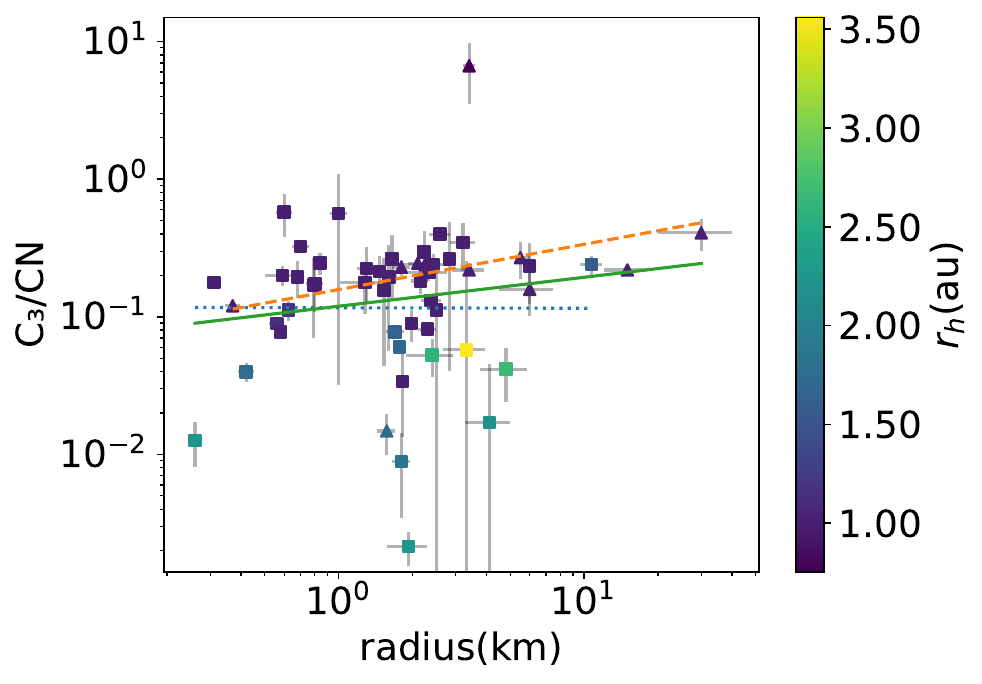} \\
		\includegraphics[width=0.3\textwidth]{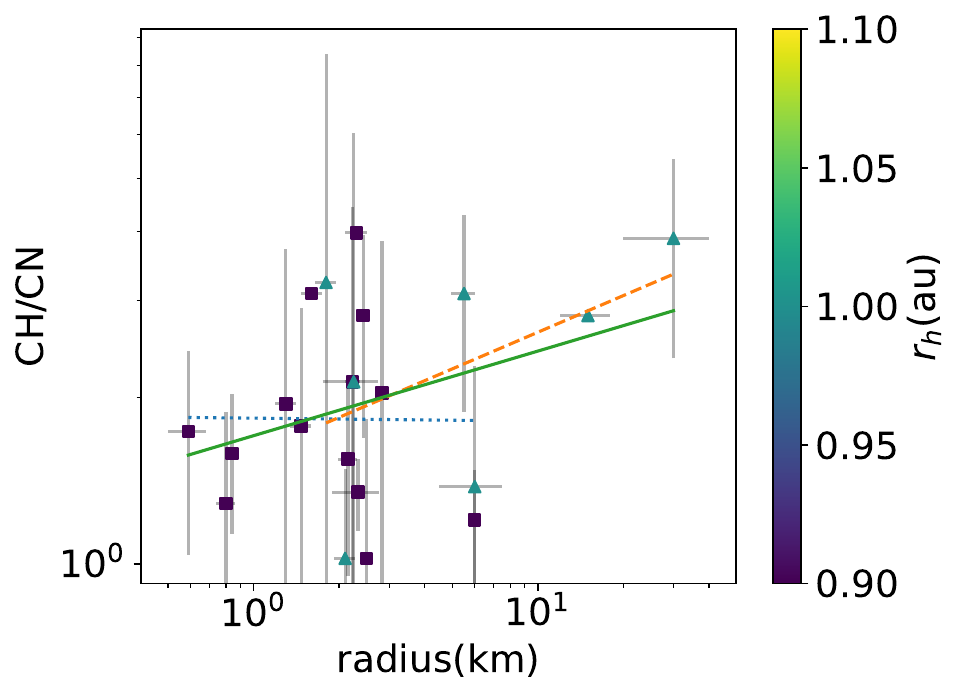} &
		\includegraphics[width=0.3\textwidth]{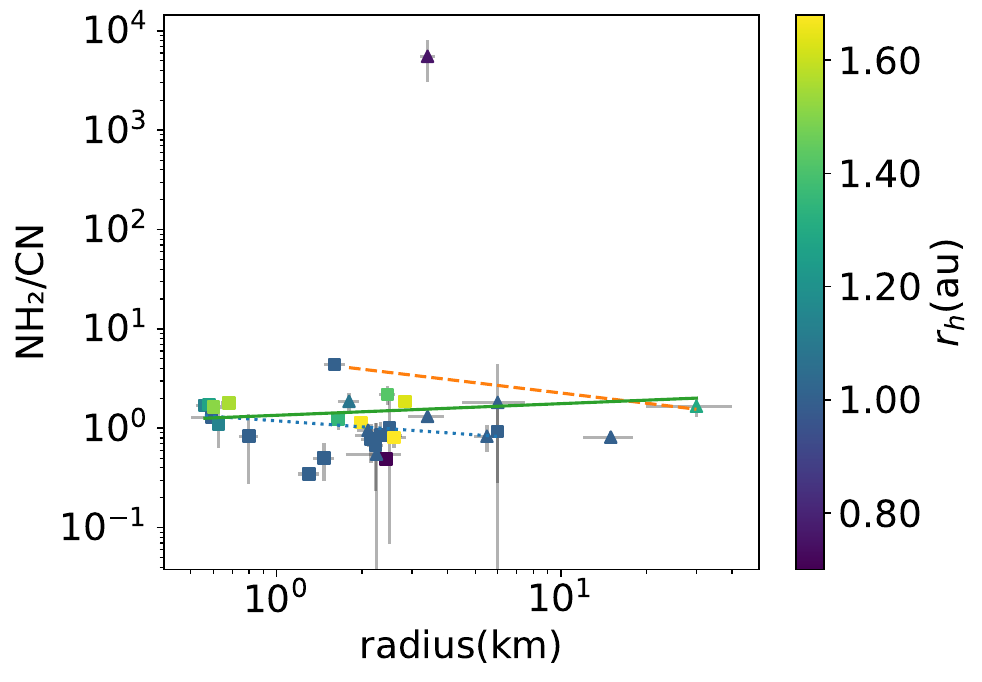} &   
        \includegraphics[width=0.3\textwidth]{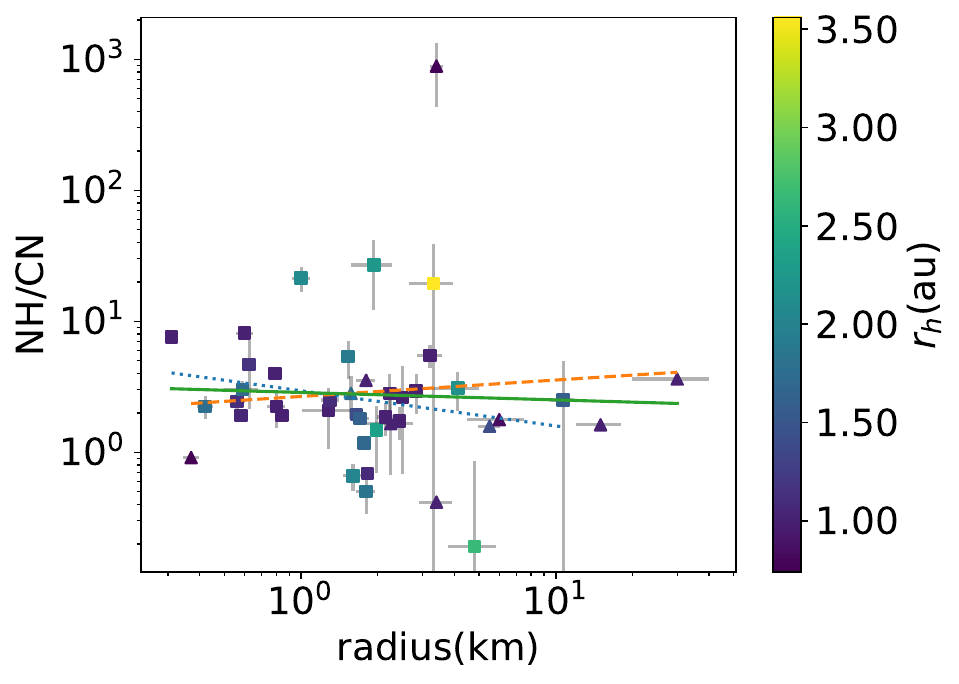} \\
		\includegraphics[width=0.3\textwidth]{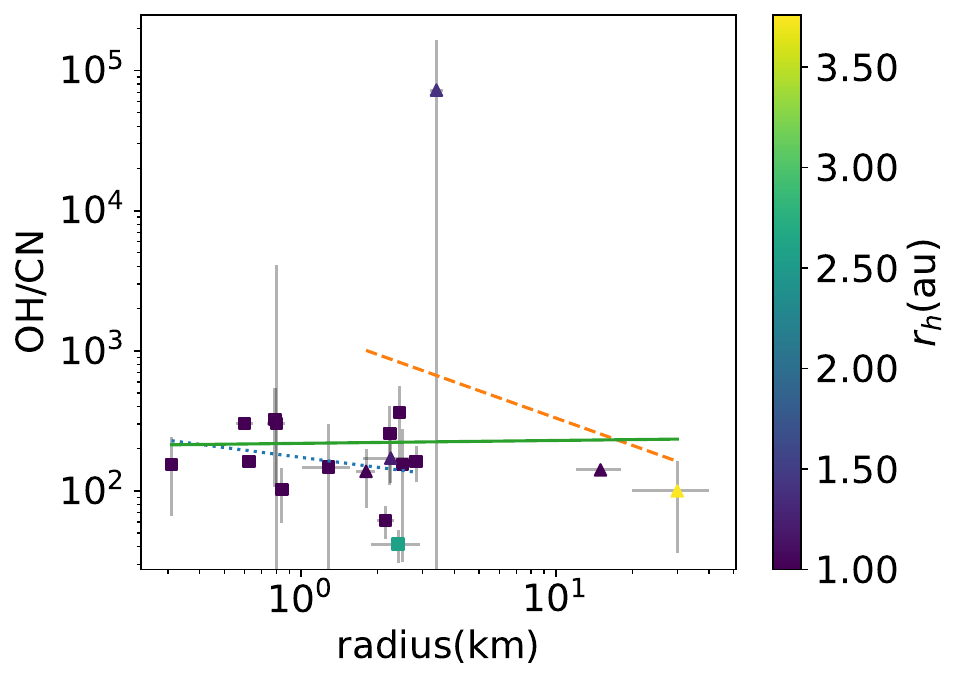} &
	\end{tabular}
	\caption{Log scale plots showing the relation between comet composition (of various daughter species relative to \ch{CN}) and radius of the nucleus.
	Marker shape, colour and plotted lines have the same meanings as Figure \ref{fig:comp_size_H2O_parents}.
The results of the Pearson correlation tests are provided in Table \ref{tab:correlations3}.
}
	\label{fig:comp_size_CN}
\end{figure*}

In addition we have considered the daughter species abundance relative to \ch{CN} instead of \ch{H2O} (Figure \ref{fig:comp_size_CN}). 
The results of the Pearson correlation analysis for this dataset are provided in \ref{tab:correlations3}.
We note that a moderately significant correlation for \ch{CH}/\ch{CN} is visible in these data, while no correlation was seen for \ch{CH}/\ch{H2O}. 
However, this is likely a small number statistics effect as we have only a handful of measurements for \ch{CH}/\ch{H2O}. 

\begin{table*}  
	\begin{tabular}{l|rrr|rrr|rrr}
\toprule
 & \multicolumn{3}{l|}{Ecliptic Comets} & \multicolumn{3}{l|}{Nearly Isotropic Comets} & \multicolumn{3}{l|}{All Comets} \\
Species & Number & Correlation & $p$-value & Number & Correlation & $p$-value & Number & Correlation & $p$-value \\
\midrule
\ch{Af$\rho$/H2O} & \cellcolor{lightgray}38 & \cellcolor{lightgray}0.2028 & \cellcolor{lightgray}0.2221 & \cellcolor{gray}\color{white}10 & \cellcolor{gray}\color{white}0.6396 & \cellcolor{gray}\color{white}0.0464 & \cellcolor{lightgray}48 & \cellcolor{lightgray}0.1885 & \cellcolor{lightgray}0.1994 \\
\ch{C2/H2O} & 40 & 0.0209 & 0.8981 & 11 & 0.0345 & 0.9198 & 51 & 0.1114 & 0.4366 \\
\ch{C3/H2O} & 35 & -0.0829 & 0.6360 & \cellcolor{gray}\color{white}11 & \cellcolor{gray}\color{white}0.6139 & \cellcolor{gray}\color{white}0.0445 & 46 & 0.0995 & 0.5107 \\
\ch{CH/H2O} & 7 & 0.1965 & 0.6728 & 2 & -1.0000 & 1.0000 & 9 & 0.3612 & 0.3395 \\
\ch{CN/H2O} & 38 & -0.0668 & 0.6902 & 11 & -0.0737 & 0.8294 & 49 & -0.0819 & 0.5757 \\
\ch{CS/H2O} & \cellcolor{lightgray}6 & \cellcolor{lightgray}0.7663 & \cellcolor{lightgray}0.0755 & 3 & -0.1783 & 0.8859 & \cellcolor{gray}\color{white}9 & \cellcolor{gray}\color{white}0.6955 & \cellcolor{gray}\color{white}0.0375 \\
\ch{NH2/H2O} & 17 & -0.0381 & 0.8846 & 5 & -0.4985 & 0.3926 & 22 & -0.0758 & 0.7373 \\
\ch{NH/H2O} & \cellcolor{lightgray}32 & \cellcolor{lightgray}-0.3465 & \cellcolor{lightgray}0.0520 & 11 & -0.0019 & 0.9955 & \cellcolor{lightgray}43 & \cellcolor{lightgray}-0.2190 & \cellcolor{lightgray}0.1583 \\
\bottomrule
\end{tabular}

	\caption{Table showing the Pearson test results for daughter species abundance relative to \ch{H2O}, including the number of comets tested, correlation coefficients and associated $p$-values
		Similar to Table \ref{tab:correlations1} results are shown for the ecliptic comets, nearly isotropic comets and all objects when considered together.}
	\label{tab:correlations2} 
\end{table*}

\begin{table*}  
\begin{tabular}{l|rrr|rrr|rrr}
\toprule
 & \multicolumn{3}{l|}{Ecliptic Comets} & \multicolumn{3}{l|}{Nearly Isotropic Comets} & \multicolumn{3}{l|}{All Comets} \\
Species & Number & Correlation & $p$-value & Number & Correlation & $p$-value & Number & Correlation & $p$-value \\
\midrule
\ch{Af$\rho$/CN} & \cellcolor{lightgray}38 & \cellcolor{lightgray}0.2307 & \cellcolor{lightgray}0.1636 & 11 & 0.3078 & 0.3571 & \cellcolor{gray}\color{white}49 & \cellcolor{gray}\color{white}0.2826 & \cellcolor{gray}\color{white}0.0491 \\
\ch{C2/CN} & 45 & 0.1089 & 0.4763 & 11 & 0.2958 & 0.3771 & \cellcolor{lightgray}56 & \cellcolor{lightgray}0.2203 & \cellcolor{lightgray}0.1028 \\
\ch{C3/CN} & 42 & -0.0025 & 0.9872 & 11 & 0.2747 & 0.4136 & \cellcolor{lightgray}53 & \cellcolor{lightgray}0.1623 & \cellcolor{lightgray}0.2456 \\
\ch{CH/CN} & 14 & -0.0085 & 0.9771 & \cellcolor{lightgray}7 & \cellcolor{lightgray}0.4812 & \cellcolor{lightgray}0.2743 & \cellcolor{lightgray}21 & \cellcolor{lightgray}0.3369 & \cellcolor{lightgray}0.1354 \\
\ch{NH2/CN} & \cellcolor{lightgray}21 & \cellcolor{lightgray}-0.2368 & \cellcolor{lightgray}0.3014 & 9 & -0.1134 & 0.7715 & 30 & 0.0669 & 0.7254 \\
\ch{NH/CN} & \cellcolor{lightgray}32 & \cellcolor{lightgray}-0.2019 & \cellcolor{lightgray}0.2678 & 10 & 0.0736 & 0.8399 & 42 & -0.0413 & 0.7953 \\
\ch{OH/CN} & 13 & -0.2573 & 0.3960 & 5 & -0.2853 & 0.6418 & 18 & 0.0146 & 0.9541 \\
\bottomrule
\end{tabular}

\caption{Table showing species abundance relative to \ch{CN}, number of comets for which we have an abundance measurement, Pearson correlation coefficients for the abundance vs nucleus size and associated $p$-values.
Results are shown for the ecliptic comets (EC), nearly isotropic comets (NIC) and all objects for which a radius and composition estimate are available.
Similar to Table \ref{tab:correlations1} the strong, moderate and marginal significance correlations are highlighted.
}
\label{tab:correlations3} 
\end{table*}

\section{\ch{CO}/\ch{H2O} data}
\label{Appendix:CO/H2O}

In Table \ref{tab:CO_H2O_table} we present the complete data for \ch{CO}/\ch{H2O} abundance ratios and sizes of the comets used in our correlation analysis and the detailed discussion in Section \ref{SS:co_size_correlation}.
In this subset of the full composition - size dataset (a sample of which of which is presented in Table \ref{tab:composition_size_sample_table}) we have already rejected measurements that are only limits, and any comets with a known fragmentation history prior to measurement.
We followed the methodology described in Section \ref{SS:comp_selection_criteria} to select a specific source for the abundance ratio when there were multiple measurements available.
These steps preferentially selected the source with the largest number of measurements for unique comets and species.
We did this to try compile as homogeneous a dataset as possible given the range of literature sources available.

Most of the \ch{CO}/\ch{H2O} abundance ratios were selected from the large scale study by \cite{dellorussoEmergingTrendsComet2016}.
This work is a compilation of the abundance of 8 volatile molecules for 30 comets determined from a database of high resolution infrared spectroscopy taken from 1997 - 2013 using a variety of telescopes/instruments.
In our methodology this source was frequently selected due to its large size and number of different species measured in a consistent manner, thus helping to increase the consistency across our literature-complied dataset.
Furthermore, we note that for many comets with multiple sources the abundance ratios are of a similar value to that of \cite{dellorussoEmergingTrendsComet2016}.
In this work we repeated our analysis while selecting different abundances from different sources and found that variation in the value of the logarithm abundance was small and so there was little change to the trends discussed in Section \ref{S:Findings}.

Within the dataset of \cite{dellorussoEmergingTrendsComet2016} we highlight some notable abundance ratios.
The observations from which \ch{CO}/\ch{H2O} was determined for 9P were taken shortly after the collision of the Deep Impact spacecraft with the nucleus of 9P on 04/07/2005.
There were no pre-impact measurements of the \ch{CO} abundance for direct comparison, however
\cite{biverRadioObservationsComet2007} did not observe significant changes in the abundance of \ch{HCN} and only a possible increase for \ch{CH3OH}.
Likewise \cite{mummaParentVolatilesComet2005} found no changes in the abundances of \ch{HCN}, \ch{CH3OH}, but they did observe a significant increase for \ch{C2H6}.
For comet 9P measurements of \ch{CO}/\ch{H2O} were also available from \cite{biverRadioObservationsComet2007,lippiInvestigationOriginsComets2021} and the literature compilation of \cite{harringtonpintoSurveyCOCO2022}.
However these sources either published upper limits on the abundance ratio, or did not include uncertainties, therefore the value from \cite{dellorussoEmergingTrendsComet2016} was selected.
Likewise, for comet 103P additional measurements of \ch{CO}/\ch{H2O} are available from \cite{harringtonpintoSurveyCOCO2022}, but with no uncertainty, and from \cite{lippiInvestigationOriginsComets2021}.
However the latter measurement is identical to that of \cite{dellorussoEmergingTrendsComet2016} as they both obtained this value from UV spectroscopic observation with HST at the time of the NASA EPOXI flyby of 04/11/2010 \citep[][the only non-IR measurements included in this work]{weaverCARBONMONOXIDEABUNDANCE2011}.
In any case, following our methodology the measurement was selected from \cite{dellorussoEmergingTrendsComet2016} as it was the larger study.

The hyperbolic comet C/2009 P1 demonstrated unusual behaviour in the observed production rates of \ch{CO} during its perihelion passage of December 2011.
For most comets volatile production rates are expected to peak sometime around perihelion approach and then decrease.
This was the case for the production of \ch{H2O} by C/2009 P1, however, the observed production of \ch{CO} continued to increase past the perihelion passage \citep[see Figure 9 of][]{feagaUNCORRELATEDVOLATILEBEHAVIOR2013}.
This resulted in a large variation of the measured \ch{CO}/\ch{H2O} abundance across the perihelion passage; as such we assessed the available literature in an attempt to determine a suitable value of \ch{CO}/\ch{H2O} for our investigation.
The \ch{CO}/\ch{H2O} abundance of C/2009 P1 as presented in \cite{dellorussoEmergingTrendsComet2016} is the weighted mean of abundances from the following sources: \cite{paganiniCHEMICALCOMPOSITIONCORICH2012}, \cite{villanuevaMultiinstrumentStudyComet2012}, \cite{disantiPrePostperihelionObservations2014}, and \cite{mckayEvolutionH2OCO2015}.
In addition we retrieved the abundance ratios presented by \cite{gicquelEvolutionVolatileProduction2015}
Furthermore the largest abundance ratio for this comet, \ch{CO}/\ch{H2O} = $0.630 \pm 0.206$, was derived by \cite{feagaUNCORRELATEDVOLATILEBEHAVIOR2013} from remote observations from the Deep Impact Flyby spacecraft when of C/2009 P1 was at $r_h = 2.00 - 2.06\ \si{au}$ \citep[abundance value and uncertainty retrieved from][]{harringtonpintoSurveyCOCO2022}. 
For consistency with our methodology we excluded measurements with $r_h \geq 2\ \si{au}$ and took the mean abundance, getting a similar value to the composition presented in \cite{dellorussoEmergingTrendsComet2016}: \ch{CO}/\ch{H2O} = $0.084 \pm 0.076$, where we reflect the large variation in abundance by assigning an uncertainty derived from the range of measured values\footnote{$\pm$ (max(\ch{CO}/\ch{H2O}) - min(\ch{CO}/\ch{H2O})) / 2}.
It should be noted that we repeated our analysis using the much larger estimate of \cite{feagaUNCORRELATEDVOLATILEBEHAVIOR2013} and we found no significant changes in the overall strength of the composition - size correlation presented in Table \ref{tab:correlations1_rh_lim}.
This is in line with the bootstrap/jack-knife resampling tests described in section \ref{s:effect_of_heliocentric_distance}, which demonstrated that the correlation for this dataset does not depend strongly on any one object.

Abundance ratios for 29P, C/2006 W3 and C/2008/Q3 were selected from \cite{ootsuboAKARINEARINFRAREDSPECTROSCOPIC2012}, a survey of \ch{CO}, \ch{CO2} and \ch{H2O} for 18 comets using NIR spectroscopy from the AKARI spacecraft.
We note that the observations for C/2006 W3 and 29P were taken at a large heliocentric distances of $r_h >3\ \&\ 6\ \si{au}$ respectively.
This is much greater than the typical $r_h \approx 1-2\ \si{au}$ for other comets in the \ch{CO} dataset, which may be another explanation for the higher than average abundance of \ch{CO} relative to less volatile \ch{H2O}, which we attempt to address in our analysis using the additional tests described in Section \ref{s:effect_of_heliocentric_distance}.

We selected the \ch{CO}/\ch{H2O} abundance of periodic comet 17P/Holmes from the population study by \cite{lippiInvestigationOriginsComets2021} as this was the only source available for this comet.
This work reports abundances for 20 comets based on reanalysis of an archive of high resolution infrared spectroscopy from NIRSPEC at the Keck Observatory.

In addition to the large scale surveys described above we searched for literature describing composition of individual comets.
C/2020 F3 was observed by \cite{biverObservationsComet20202022} with IRAM/NOEMA in July/August 2020 with generally poor weather conditions in both runs which limited detection of more complex molecules.
We note that there were relatively few observations of this comet, presumably due pandemic restrictions during its apparition, although similar abundances were also measured by \cite{faggiExtraordinaryPassageComet2021}.
C/2020 F3 has a low \ch{CO}/\ch{H2O} abundance ratio compared to other comets; in the IRAM observations \ch{CO} was only marginally detected.
The reference water production rates were derived from interpolation of SOHO-SWAN observations of Lyman-$\alpha$ Hydrogen emission \citep{combiWaterProductionRate2021} and observations of the 18cm \ch{OH} line at the Green Bank Telescope and Nan\c{c}ay Radio Telescope \citep{drozdovskayaLowNHRatio2023}.
For comet 2P Encke, the \ch{CO}/\ch{H2O} abundance was measured during its 2017 apparition by \cite{rothTaleTwoComets2018} using iSHELL at IRTF.
These observations were made shortly after perihelion passage under favourable conditions, with 2P at geocentric distance of only $\sim 0.75\ \si{au}$.
This allowed the detection of hyper-volatiles, \ch{CO} and \ch{CH4}, which are usually difficult to measure for ecliptic comets from ground-based observations with low geocentric velocities.

67P Churyumov-Gerasimenko was the target of the Rosetta mission and its production rates were measured \textit{in situ} by the ROSINA mass spectrometer instrument in May 2015 \citep{rubinElementalMolecularAbundances2019}.
We selected these \textit{in situ} measurements as we expect them to be more accurate and precise than remote observations.
The Rosetta observations used in this study were taken while 67P was in a period of strong outgassing on the approach to perihelion.
These detailed measurements revealed that the abundance ratios of volatile species varied over the course of the mission, related in a complex way to the heliocentric distance, nucleus spin axis orientation and the relative position of Rosetta to the nucleus.
This highlights that instantaneous measurements of abundance ratios may not necessarily reflect the true abundance ratios within the bulk nucleus, however, such detailed analysis is impossible for remotely observed comets.

The remaining abundance ratios for comets 1P and 45P were selected from the literature compilation by \cite{harringtonpintoSurveyCOCO2022}.
This study gathered production rates for \ch{CO}, \ch{CO2} (and \ch{H2O} where available) for 25 comets from a wide range of published sources using both space and ground-based observations.
They selected sources where \ch{CO} and/or \ch{CO2} production rates were measured contemporaneously with \ch{H2O} and for some comets they have collated multiple measurements for abundance ratio.
Following our methodology we calculated the mean abundance, date and heliocentric distance for each comet to use in our own dataset.
However, as the measurements collected by \cite{harringtonpintoSurveyCOCO2022} are from multiple sources we selected abundance ratios from the larger homogeneous studies \citep[e.g.][]{dellorussoEmergingTrendsComet2016,ootsuboAKARINEARINFRAREDSPECTROSCOPIC2012} where possible.

\begin{landscape}
	
\begin{table}
	\begin{tabular}{llrlrrrrlrrl}
\toprule
Type & Designation & Number & Name & Date(MJD) & $r_h$(au) & \ch{CO}/\ch{H2O} & $\sigma_\textrm{CO/H2O}$ & Composition Source & $r$(km) & $\sigma_r$(km) & Radius Source \\
\midrule
P &  & 1 & Halley & 46495.0 & 0.79 & 0.110 & 0.0160 & \cite{harringtonpintoSurveyCOCO2022} & 5.50 & 0.53 & \cite{lamySizesShapesAlbedos2004} \\
P &  & 2 & Encke & 57834.3 & 0.48 & 0.004 & 0.0004 & \cite{rothTaleTwoComets2018} & 2.43 & 0.06 & \cite{boehnhardtPhotometryPolarimetryNucleus2008} \\
P &  & 8 & Tuttle & 54487.7 & 1.05 & 0.004 & 0.0008 & \cite{dellorussoEmergingTrendsComet2016} & 2.25 & 0.50 & \cite{harmonComet8PTuttle2008} \\
P &  & 9 & Tempel 1 & 53547.5 & 1.52 & 0.043 & 0.0100 & \cite{dellorussoEmergingTrendsComet2016} & 2.83 & 0.10 & \cite{thomasNucleusComet9P2013a} \\
P &  & 17 & Holmes & 54401.5 & 2.46 & 0.088 & 0.0270 & \cite{lippiInvestigationOriginsComets2021} & 2.40 & 0.53 & \cite{bauerDebiasingNEOWISECryogenic2017} \\
P &  & 21 & Giacobini-Zinner & 51455.7 & 1.12 & 0.022 & 0.0150 & \cite{dellorussoEmergingTrendsComet2016} & 1.82 & 0.05 & \cite{pittichovaGROUNDBASEDOPTICALSPITZER2008} \\
P &  & 29 & Schwassmann-Wachmann 1 & 55153.5 & 6.18 & 4.645 & 1.0187 & \cite{ootsuboAKARINEARINFRAREDSPECTROSCOPIC2012} & 23.00 & 6.50 & \cite{bauerDebiasingNEOWISECryogenic2017} \\
P &  & 45 & Honda-Mrkos-Pajdusakova & 57761.0 & 0.56 & 0.005 & 0.0010 & \cite{harringtonpintoSurveyCOCO2022} & 0.62 & 0.03 & \cite{lejolyRadialDistributionDust2022} \\
P &  & 67 & Churyumov-Gerasimenko & 57152.0 & 1.66 & 0.031 & 0.0090 & \cite{rubinElementalMolecularAbundances2019} & 1.65 & 0.01 & \cite{jordaGlobalShapeDensity2016} \\
P &  & 103 & Hartley 2 & 55498.0 & 1.13 & 0.003 & 0.0015 & \cite{dellorussoEmergingTrendsComet2016} & 0.58 & 0.02 & \cite{thomasShapeDensityGeology2013} \\
C & 1995 O1 &  & Hale-Bopp & 50594.6 & 1.14 & 0.262 & 0.0070 & \cite{dellorussoEmergingTrendsComet2016} & 30.00 & 10.00 & \cite{lamySizesShapesAlbedos2004} \\
C & 2006 W3 &  & Christensen & 54909.9 & 3.40 & 2.296 & 0.4648 & \cite{ootsuboAKARINEARINFRAREDSPECTROSCOPIC2012} & 21.88 & 4.20 & \cite{bauerDebiasingNEOWISECryogenic2017} \\
C & 2007 N3 &  & Lulin & 54870.9 & 1.31 & 0.022 & 0.0009 & \cite{dellorussoEmergingTrendsComet2016} & 6.10 & 0.25 & \cite{bauerDebiasingNEOWISECryogenic2017} \\
C & 2008 Q3 &  & Garradd & 55018.0 & 1.81 & 0.243 & 0.0494 & \cite{ootsuboAKARINEARINFRAREDSPECTROSCOPIC2012} & 3.35 & 0.50 & \cite{bauerDebiasingNEOWISECryogenic2017} \\
C & 2009 P1 &  & Garradd & 55943.5 & 1.71 & 0.084 & 0.0750 & See appendix \ref{Appendix:CO/H2O} & 9.60 & 4.00 & \cite{boissierMillimetreContinuumObservations2013} \& \cite{bauerDebiasingNEOWISECryogenic2017} \\
C & 2010 G2 &  & Hill & 55935.5 & 2.50 & 0.910 & 0.2300 & \cite{dellorussoEmergingTrendsComet2016} & 4.01 & 1.04 & \cite{bauerDebiasingNEOWISECryogenic2017} \\
C & 2020 F3 &  & NEOWISE & 59047.3 & 0.80 & 0.032 & 0.0120 & \cite{biverObservationsComet20202022} & 2.50 & 0.22 & J. Bauer (unpubl. data) \\
\bottomrule
\end{tabular}

	\caption{
		The \ch{CO}/\ch{H2O} abundance ratio and nucleus radius for each comet used in our analysis.
		This table is a subset of the full composition - size dataset (which is sampled in table \ref{tab:composition_size_sample_table}) but is presented here for convenience.
	}
	\label{tab:CO_H2O_table}
\end{table}

\end{landscape}


\bsp	
\label{lastpage}
\end{document}